\documentclass[12pt]{article}

\usepackage{graphicx}
\usepackage{epstopdf}
\usepackage[body={17.5cm, 21cm},right=2cm]{geometry}
\usepackage{amssymb}
\usepackage{amsmath}

\newcommand\eea{\end{eqnarray}}
\newcommand\bea{\begin{eqnarray}}

\newcommand\mpl{M_{\rm Pl}}

\def\beq{\begin{equation}}
\def\eeq{\end{equation}}
\def\d{\partial}

\def\d{\partial}

\newcommand{\be}{\begin{equation}}
\newcommand{\ee}{\end{equation}}
\newcommand{\ba}{\begin{align}}
\newcommand{\ea}{\end{align}}
\newcommand{\bg}{\begin{gather}}
\newcommand{\eg}{\end{gather}}
\newcommand{\bseq}{\begin{subequations}}
\newcommand{\eseq}{\end{subequations}}

\begin{document}

\vspace{5mm}
\vspace{1.5cm}
\begin{center}

\def\thefootnote{\fnsymbol{footnote}}
\vspace{1.5cm}
{\Large \bf On Loops in Inflation}
\\[0.5cm]
{\large  Leonardo Senatore and Matias Zaldarriaga}
\\[0.5cm]

{\normalsize { \sl School of Natural Sciences, Institute for Advanced Study, \\Olden Lane, 
Princeton, NJ 08540, USA}}\\
\vspace{.3cm}



\end{center}

\vspace{.8cm}

\hrule \vspace{0.3cm}
{\small  \noindent \textbf{Abstract} \\[0.3cm]
\noindent
We study loop corrections to correlation functions of inflationary perturbations. Previous calculations have found that the two-point function can have a logarithmic running of the form $\log(k/\mu)$, where $k$ is the wavenumber of the perturbation, and $\mu$ is the renormalization scale. We highlight that this result would have profound consequences for both eternal inflation and the predictivity of standard inflation. We find a different result. We consider two sets of theories: one where the inflaton has a large cubic self-interaction and one where the inflaton interacts gravitationally with $N$ massless spectator scalar fields. We find that there is a logarithmic running but of the form $\log(H/\mu)$, where $H$ is the Hubble constant during inflation. We find this result in three independent ways: by performing the calculation with a sharp cutoff in frequency-momentum space, in dimensional regularization and by the simple procedure of making the loop integral dimensionless. For the simplest of our theories we explicitly renormalize the correlation function proving that the divergencies can be reabsorbed and that the correlation function for super-horizon modes does not depend on time (once the tadpole terms have been properly taken into account). We prove the time-independence of the super-horizon correlation function in several additional ways: by doing the calculation of the correlation function at finite time using both the regularizations and by developing a formalism which expresses loop corrections  directly in terms of renormalized quantities at each time. We find this last formalism particularly helpful to develop intuition which we then use to generalize our results to higher loops and different interactions. In particular we argue correlation functions  have no long-term time dependence even if the spectator fields have a potential.

\vspace{0.5cm}  \hrule
\def\thefootnote{\arabic{footnote}}
\setcounter{footnote}{0}




\vspace{.8cm}



 \section{Introduction}
 
 The purpose of this paper is to compute loop corrections to inflationary observables. Why?
 At first there seems to be no good reason.  Let us consider a standard inflaton $\phi$ with action:
 \begin{equation}\label{eq:slow_roll_potential}
S=\int d^4x\; \sqrt{-g}\left[\frac{1}{2}(\partial\phi)^2-V(\phi)\right]\ . 
\end{equation}
In order to have an inflationary solution, all derivatives of the potential need to be small:
\be
\epsilon=\mpl^2\left(\frac{V'}{V}\right)^2\ll1\ ,\quad\quad\quad \eta=\mpl^2\frac{V''}{V}\ll1\ , \quad\quad \ldots \ .
\ee
This means that the inflaton is extremely weakly coupled. In fact Maldacena showed that the leading interactions come from the mixing with gravity ~\cite{Maldacena:2002vr}.
The size of the inflaton two-point function is of order 
\be
\langle\delta\phi_k^2\rangle_{\rm tree}\sim \frac{H^2}{k^3}\ .
\ee
Oservations of the CMB imply that the amplitude of the primordial curvature perturbation $\zeta$ is: 
\be
\langle\zeta^2_k\rangle_{\rm tree}\sim \frac{H^2}{\epsilon\mpl^2}\frac{1}{k^3}\sim 10^{-10}\ ,
\ee
where $\epsilon$ is the slow-roll parameter and thus $H/\mpl\ll 10^{-5}$ during inflation.
Loop corrections will be mediated by gravity, and will therefore be suppressed by $\mpl^2$. By dimensional analysis we expect:
\be
\langle\delta\phi^2_k\rangle_{\rm 1-loop}\sim \frac{H^2}{k^3} \frac{H^2}{\mpl^2}\ .
\ee
This is at least a factor of $10^{-10}$ smaller than the tree level result. Clearly nobody should bother computing it.

This has been the general attitude for about the first twenty-five years after the invention of inflation until Weinberg stressed a different point of view. In the paper where he first studied these effects \cite{Weinberg:2005vy}, he argued that since experiments seem to be providing more and more evidence in favor of the theory it is worth exploring all its predictions, even those that naively appear observationally unverifiable. Perturbation theory ought to be well defined even for inflationary fluctuations, and therefore, given the well known subtleties involved in studying de Sitter space, it is interesting to see how this works in practice. Many times as a result of doing  a non-trivial calculation one can gain new insight into the theory and explore possible subtleties or generalization. For example, this was the case for Maldacena's calculation of inflationary 3-point function in standard slow-roll inflation. Though for standard slow roll inflation the effect was is in practice unobservable it opened the way to formulate alternative theories  which predict much higher levels of non-gaussianities (see for example \cite{Cheung:2007st,Alishahiha:2004eh,ArkaniHamed:2003uz,Senatore:2004rj,Chen:2006nt,Flauger:2009ab,Zaldarriaga:2003my,Lyth:2002my,Green:2009ds,Barnaby:2009mc}) and are already constrained by existing data ~\cite{Smith:2009jr,Senatore:2009gt,Slosar:2008hx}.

The motivation articulated by Weinberg is of course a good one but we find  two other issues to be even more important.  The first is related to eternal inflation. This is a particular regime of inflation that occurs when the potential is flat enough so that quantum fluctuations dominate over the classical motion and there is a finite probability for inflation never to end. This regime of inflation has become more important recently because of Weinberg's proposal to explain anthropically the smallness of the cosmological constant~\cite{Weinberg:1987dv}. Eternal inflation provides a natural and simple mechanism, though not strictly necessary, to populate a number of vacua large enough to explain the tuning of the cosmological constant.
The existent of such a phase was discussed in a more rigorous way only very recently ~\cite{Creminelli:2008es,Dubovsky:2008rf}, following the original works in~\cite{eternal}. It was shown that if we start with a volume of space with the inflaton of (\ref{eq:slow_roll_potential}) up in the potential there is sharp phase transition as we make the potential flatter and flatter:  at a critical value of the slope, the probability of creating an infinite volume goes from being exactly equal to zero to non-zero. It was also shown that the number of $e$-foldings in any finite realization is bounded by $S_{\rm dS}/6$, where $S_{\rm dS}$ is the de Sitter entropy at the end of inflation. This provides additional insight on how to interpret de Sitter space within quantum gravity~\cite{ArkaniHamed:2007ky,Dubovsky:2008rf}. A fundamental ingredient of these proofs is based on the fact that, at zeroth order in the slow roll parameters, the two-point function of the inflaton field at coincidence acquires a calculable logarithmic divergency which leads to a linear time dependence:
\be\label{eq:two-point-coincidence}
\langle\delta\phi(x,t)^2\rangle=\int^{\Lambda a[t]} d^3k \frac{H^2}{k^3}\sim H^2\log (a)\sim H^3 t+\ {\rm const.}\ ,
\ee
where here $\Lambda$ is a physical cutoff, and $a(t)\simeq e^{H t}$ is the scale factor during inflation. Notice that the linear time dependence is in strict correspondence with the scale invariance of the two-point function.
This linear time dependence was essential for the fact that eternal inflation could transition from being eternal to not eternal depending on the slope of the inflationary potential. If the time dependence were to be different, even by the slightest amount, inflation would be either always eternal or never eternal. 
Loop corrections have  the potential of changing the time dependence and is thus interesting to verify  what the dependence is. 

Going beyond eternal inflation, it is well known that de Sitter space is a rather puzzling spacetime in the context of quantum gravity \cite{ArkaniHamed:2007ky}, and inflation offers a natural regularization for the infinities of de Sitter space. Studying in detail inflationary spacetimes can lead to some new insight about its embedding into a theory of quantum gravity.

The first calculation of the one-loop corrections  was done in  \cite{Weinberg:2005vy}. We find that result rather puzzling:
\be\label{eq:two-point-zeta}
\langle\zeta_k^2\rangle_{\rm 1-loop}=N\cdot \frac{H^2}{k^3}\cdot \frac{H^2}{\mpl^2}(c+\frac{\pi}{6}\epsilon \log\left(\frac{k}{\mu}\right))
\ee
where $c$ is a numerical  constant that depends on the renormalization procedure, $\epsilon$ is the slow roll parameter $\epsilon=-\dot H/H^2$, and $\mu$ is a renormalization scale.  $N$ represents the number of massless spectator scalar fields which interact  with the inflaton gravitationally. It is this factor of $N$ that parametrically distinguishes the effect from the spectator fields running in the loop from the effect  of the gravitons. This offers a great simplification to the calculation~\footnote{Here we are using the numerical result of \cite{Adshead:2008gk} which corrected a mistake in \cite{Weinberg:2005vy} associated with the choice of the $i\,\epsilon$ prescription.}.  The same form of the logarithmic running was found in all subsequent related studies ~\cite{Weinberg:2005vy,Adshead:2008gk,literature}~\footnote{It is worth to notice that usually in the just mentioned literature the above logarithmic running was simply stated as $\log(k)+C$, with $C$ a constant containing the logarithm of a quantity with dimensions of mass.}.

Notice that if we take the above result at face value and we neglect the fact that it is slow roll suppressed then the two point function of the inflaton at coincidence would receive a correction of the form:
\be\label{eq:two-point-coincidence-one-loop}
\langle\delta\phi(x,t)^2\rangle_{\rm 1-loop}=-\int^{\Lambda a(t)} d^3k \frac{H^2}{\mpl^2}\cdot N\cdot\frac{H^2}{k^3}\log(k)\sim- N\cdot \frac{H^2}{\mpl^2}\cdot H^2 \left(\log(a)\right)^2 \sim- N \cdot  \frac{H^2}{\mpl^2}\cdot H^4 t^2+\ {\rm const.}
\ee
where we have neglected numerical constants but kept explicit the sign of the $t^2$ correction, which is negative. If we were to use this result to repeat the analysis of \cite{Creminelli:2008es} we would find the very puzzling result that {\it no} model of slow roll inflation is eternal. 

Another puzzle with expressions (\ref{eq:two-point-zeta}) or (\ref{eq:two-point-coincidence-one-loop}) is that they seem to imply that for some large or small values of $k$ or for some large enough time, the one-loop corrections become large enough to harm the perturbative expansion. This result seems to be at odds with the common intuition that, as a consequence of the approximate symmetry under time-translation and rescaling of the inflationary space time, every mode goes through the same history. The fact that the logarithm is proportional to a slow-roll parameter might suggest that it is precisely the deviation from de-Sitter which is generating the effect. A closer look at the calculation reveals that this is not the case.  Furthermore such an effect would imply the existence of a non-trivial correlation between very different modes which we find hard to believe. 

An obvious problems with the result in (\ref{eq:two-point-zeta}) is that the argument of the logarithm is a ratio of a comoving scale $k$ and a physical scale $\mu$. This cannot be the case and it is the sign of a mistake in the calculation. In fact  by its very definition the scale factor of an FRW Universe is unobservable. It just corresponds to a rescaling of the spatial coordinates. This implies that every physically correct result should be invariant under the following rescaling by a real number $\lambda$:
\be\label{eq:symmetry}
a\rightarrow\lambda\, a\ ,\qquad x\rightarrow x/\lambda\ , \qquad k\rightarrow \lambda\, k\ .
\ee
If we express the result of (\ref{eq:two-point-zeta}) in real space, we have to do two momentum integrations by $d^3k\; d^3k'$ (in (\ref{eq:two-point-zeta}) we neglected the momentum $\delta$-function which of course is there because of translation invariance). At this point the prefactor of the $\log(k/\mu)$ is invariant under (\ref{eq:symmetry}) while the $\log(k/\mu)$ is not.

The minimal way the above expression could be corrected is by replacing the argument of the $\log$ with $k/(a(\bar t)\mu)$. But what value of $\bar t$ should we use? There are basically three options: $\bar t\sim t_{early}$, $\bar t\sim t$ or $\bar t\sim t_k$, where $t_{early}$ is some early time before the mode crosses the horizon,  $t$ is the time at which the correlation function is being calculated and $t_k$ is the time when the mode $k$ crossed the horizon. The first option is ruled out by adiabaticity: the mode in the past has a very high frequency and is in its adiabatic vacuum. It therefore has no memory of the early times. The second option would imply an additional time dependence in the two-point function which would result in {\it no} model of slow roll inflation being eternal. Furthermore after a very long time (but no time is too a long  in the case of eternal inflation!) it would also mean that the one loop corrections would become large,  comparable to the tree level result. This is at odds with the common intuition of an approximate time-translation and rescaling symmetry of the inflationary space-time as well as with the expectation that very different modes should be only trivially correlated. More importantly as we will soon see if this were to be the case it would create problems not only for eternal inflation but for inflation in general!  

 If instead the true result is the third one then since $k/a(t_k)\sim H$ the logarithm would just become $\log(H/\mu)$, a result that would make a lot of physical sense. This can be understood by recollecting how logarithms appear in scattering amplitudes. They appear in the form of $\log(E/\mu)$ where now $E$ is the invariant energy of the collision. This logarithm takes into account the difference between the theory defined at the renormalization scale $\mu$ and at the scattering energy $E$.  The logarithmic corrections from the one-loop calculation are not large once the renormalization scale is chosen to be close to the scale at which the experiment is done, which led to the invention of the renormalization group~\cite{GellMann:1954fq,Wilson:1974mb}. We can reach a similar conclusion here: the energies probed by the interactions during inflation are of order $H$ and therefore the logarithms are small if we renormalize the theory at that scale. 
  
 Before moving on to the actual computation, we would like to stress a second motivation for studying loop corrections that we also find very important. There is another disturbing consequence that would arise if the way of making the result in (\ref{eq:two-point-zeta}) correct was to replace the argument of the logarithm with $k/(a(t)\mu)$. The theory studied by Weinberg and giving rise to (\ref{eq:two-point-zeta}) is a theory of the form
\begin{equation}
S=\int d^4x\; \sqrt{-g}\left[\frac{1}{2}(\partial\phi)^2-V(\phi)+\sum_{i=1}^{N}\frac{1}{2}g^{\mu\nu}
\partial_\mu\sigma_i\partial_\nu\sigma_i\right]\ ,
\label{Lagrangian2}
\end{equation}
 where there are $N$ massless scalar fields $\sigma_i$ interacting gravitationally with the inflaton, sourcing the fluctuations of the metric $\zeta$ through the vacuum fluctuations of their stress tensor $T_{(\sigma)\mu\nu}$.  Since these spectator fields are massless, only their gradients are important  and therefore the vacuum fluctuations of $T_{(\sigma)\mu\nu}$ are expected to become uncorrelated for distances longer than the horizon scale. If the correct answer were $\log(k/(a(t)\mu))$ it would mean that these fluctuations are capable of sourcing a $\zeta$ fluctuation even on scales much longer than the horizon. This would be very surprising and would create serious problems for the theory of inflation. In fact the predictivity of inflation relies on the fact that the perturbation $\zeta$ is expected to be constant on scales much longer than the horizon independently of what happens on scales of order of the horizon. This is very important because for some epochs such as reheating or a GUT phase transition (if this exist), we have almost no idea of what  happens on scales of the order or shorter than the horizon. In principle large fluctuations  on horizon scales can exist during these epochs. They are correlated only over horizon scale distances and so one expects they cannot affect a $\zeta$ mode of much longer wavelength. Instead a time dependence from the one loop corrections  would imply  the contrary. Notice that the situation can become really worrisome. The effect of the $\sigma$ fields, even assuming a time dependence, would be very small in standard inflationary scenarios. This is so because there is not enough time to overcome the tremendous suppression of order $H^2/\mpl^2$. Basically  the $\sigma$ fields are free fields and therefore do not have large fluctuations. The situation might be completely different  for fluctuations at the time of reheating or at a GUT phase transition where the relative fluctuations might well be of order one even on Hubble scales. This could induce a change of $\zeta$ of order one in a Hubble time: a huge effect that would undermine our capability of making predictions out of the inflationary phase without having a detailed understanding of all the epochs in between us and reheating. The same effects would probably also alter the scale invariance of the primordial density perturbations.
 
The purpose of this paper is to show that indeed there is no time-dependence nor deviation from scale-invariance induced by these interactions and that the one-loop logarithmic correction to the two-point function is of the form $\log(H/\mu)$. Achieving this will be  a rather challenging and sometimes technical task and for this reason we will give several independent derivations, which will also help us in  building intuition.

After reviewing in sec.~\ref{sec:loop_computation} the formalism for computing loop corrections, in sec.~\ref{sec:dotpicube} we will start  by studying a different and simpler theory given by~\cite{Cheung:2007st}
\bea\label{eq:action_intro}
S=\int d^4x\; a^3\left[-\dot H \mpl^2 \left(\dot\pi^2-\frac{1}{a^2}(\d_i\pi)^2\right)+\frac{2}{3}c_3 M^4\left(2\dot\pi^3+3\dot\pi^4-3\frac{1}{a^2}\dot\pi^2(\d_i\pi)^2\right)\right]\ ,
\eea
 where $\pi$ is the Goldstone boson of time translation, that are broken during inflation, and is related to $\zeta$ by the simple relationship $\zeta=-H\pi$. We will comment later on how this theory is derived. Here we just would like to point out that the scale $M$ which controls the size of the self-interactions can make them parametrically larger than gravitational. In fact, its current upper limit is just observational: the cubic self-interaction induces a three-point function in the CMB and is thus constrained by WMAP ~\cite{Senatore:2009gt}. Therefore we can study loop corrections induced by this self-interactions without including gravity. We think this is technically a simpler example.  
  
 We will calculate loop corrections to the two-point function in this model. We will perform the calculation with two kinds of regularization. The first, in sec.~\ref{sec:pidotcube_cutoff}, will be imposing a fixed physical cutoff in frequency and momentum.  We will find some power law divergencies, finite terms, and most importantly logarithmic divergencies. As usual, the form of the logarithmic divergencies is computable in the infrared even without performing an explicit renormalization, and it represents the running of the inflationary two-point function.  After we remove the power law divergencies through renormalization the logarithmic corrections can become parametrically much larger than the remaining finite terms  if the renormalization scale is taken to be very different than the energy typical of the inflationary processes which is  Hubble, up to the point of  potentially harming the perturbative expansion of weakly coupled theories. This is a situation historically known as the `large logarithms'.  Then, in sec.~\ref{sec:dim_reg_calculation}, we will perform the same calculation in dimensional regularization (dim. reg.) finding the same result for the logarithms. 
 
 Dimensional regularization was the scheme used in all the literature that followed Weinberg's first paper. We point out explicitly where we believe a mistake was made and perform  the correct calculation. Still working in this regularization, in sec.~\ref{sec:adimensionalization}, we will be able to show by making the loop integral dimensionless that even without explicitly doing the calculation one concludes that if a logarithm is present it has to be of the form $\log(H/\mu)$.
Finally, in sec.~\ref{sec:renormalization}, we will perform the renormalization of the two point function by identifying and solving for the proper local counterterms.

In this process we will be able to address another issue that arises in these one-loop calculations. Because of the symmetries of the space time, we will perform the loop phase space integrals using a partial Fourier basis where we keep time real but use spatial momentum. In doing these integrals it has been noticed that because of the nature of the interactions, the integrals in time converge even when the external time is taken to infinity. It is only the integrals in momentum that have UV divergencies, which are supposed to be reabsorbed by local counterterms.  The calculation in dim. reg. becomes simpler if the integrals are done in this order. However, as it was noticed by Weinberg \cite{Weinberg:2005vy,Weinberg:2006ac}, the momentum integrals that are left  can be more divergent than the ones one would get if the time integral was done up to finite time or if the order of integration had been switched. This implies that there is no guarantee that all the divergencies, including the logarithmic ones, obtained with the outlined procedure  are reabsorbable by local counterterms. It is in fact possible that they would disappear if one where to switch the  order of integration at the cost of inducing a time dependence of the one-loop result. This issue had been simply mentioned but not solved in the literature. We are able to address and solve this issue in several ways. Of course, the more direct way is by explicitly performing the renormalization of the correlation function, as we do in sec.~\ref{sec:renormalization}. The second way is by performing the calculation regularizing the integrals with a cutoff, in which case we first integrate in momentum and then in time finding the same result as in dimensional regularization. This means that the  divergency that we find has to be reabsorbed by a counterterm. Third in sec.~\ref{sec:dim_reg_finite_time} we  actually  perform the dim. reg. calculation keeping the external time finite finding again the same result. 

At this point extending our calculation to the theory of (\ref{Lagrangian2}) is quite simple. We do it in sec.~\ref{sec:sigma_fields}, finding again a logarithm of the form $\log(H/\mu)$. For this theory, in sec.~\ref{sec:theorem} we prove that there cannot be any time dependence of the two point function  in yet a different way that we also then generalize to the case of the $\pi$ self-interactions. This last proof can be thought of as computing the loop corrections by first doing the momentum integrals and then doing the time integral. In this way, we can assume that all the UV divergencies have been reabsorbed by a counterterm. This is so because in this way of doing the calculation, we are using directly the physical (renormalized) quantities defined at each time. In this method, computing loop corrections becomes equivalent to solving perturbatively the equations of motion for the various operators. It turns out this alternative method is not powerful enough for us to determine the exact result of the loop calculation  but it allows us to show in a very simple and physical way that the $\zeta$ two-point function  is time-independent. In particular, the reader not interested in the technical calculations but on the main result could read directly sec.~\ref{sec:theorem} to see that the $\zeta$ correlation function is time-independent and then make the integrals dimensionless  as in sec.~\ref{sec:adimensionalization} to see that the logarithm, if it exists, is of the form $\log(H/\mu)$. 

We find this last method of dealing with loop corrections particularly physical and intuitive. In section~\ref{sec:massive}, we use this intuition  to argue how our results extend to higher loops and different interactions. In particular we will claim that contrary to what was reported in the literature \cite{Weinberg:2006ac,literature_time} the $\zeta$ two point function does not acquire any time dependence even in the case where the spectator $\sigma$ fields have any form of stable potential~\footnote{This of course includes and agrees with the main conclusions of \cite{Weinberg:2006ac,literature_time} about the absence of an exponential dependence in time in the two-point function.}. It is conceivable that our results can be extended to other time dependent effects found in the literature when studying the zero mode in different theories~\cite{literature_time_different}. We will summarize our results in sec.~\ref{sec:conclusions}.

\section{One-loop corrections\label{sec:loop_computation}} 

Let us write the metric in the standard ADM parametrization:
\be
ds^2=-N^2dt^2+ h_{ij}\left(dx^i+N^i dt\right)\left(dx^j+N^j dt\right)\ .
\ee 

We can fix the gauge by choosing the inflaton $\phi$ to be uniform on equal time slices: $\delta\phi=0$ and by imposing the spatial metric $h_{ij}$ to be of the form:
\be\label{eq:zeta_gauge}
h_{ij}=a(t)^2e^{2\zeta}\delta_{ij}.
\ee

In inflation we are interested in computing late time expectation values of the metric fluctuation $\zeta$. This is due to the fact that we expect that for wavenumbers much outside of the horizon it is constant in time and more importantly independent of the local, often unknown, physical processes that happens during the various epochs of the history of the Universe. Here we are going to compute the one-loop corrections to its two-point function. This includes checking for its constancy in time. The expectation value for $\zeta$ is given by~\cite{Maldacena:2002vr,Weinberg:2005vy,Adshead:2008gk}
\be\label{eq:generic_expectation}
\langle\zeta^2(t)\rangle=\langle U_{int}(t,-\infty_+)^{\dag} \zeta^2(t) U_{int}(t,-\infty_+) \rangle\ ,
\ee
where 
\be
U_{int}(t,-\infty_+)=T e^{-i \int_{-\infty_+}^tdt'\;  H_{int}(t')}\ ,
\ee
where $H_{int}$ is the interaction Hamiltonian in the interaction picture, $T$ denotes time-ordering, and $-\infty_+$ means that the integral is performed on an analytically rotated contour $t'\rightarrow t' (1+i \epsilon)$, which projects the free vacuum state onto the interacting vacuum state in the infinite past. The operator $\zeta^2$ on the right-hand side is meant to be the freely evolving operator in the Heisemberg picture. 
At one-loop, eq.~(\ref{eq:generic_expectation}) becomes
\bea\label{eq:diagram_expectation}\nonumber
\langle\zeta^2(t)\rangle_{\rm{1-loop}}&=&-2\, {\rm Re}\left[\int_{-\infty_-}^t dt_2\int_{-\infty_-}^{t_2} dt_1\langle H^{(3)}_{int}(t_1) H^{(3)}_{int}(t_2)\zeta^2(t)\rangle\right]-2\,{\rm Im}\left[\int_{-\infty_-}^t dt_1\langle H^{(4)}_{int}(t_1) \zeta^2(t)\rangle\right]\\
&&+\int_{-\infty_-}^t dt_1\int_{-\infty_+}^t dt_2 \langle H^{(3)}_{int}(t_1) \zeta^2(t)H^{(3)}_{int}(t_2)\rangle \ ,
\eea
where $H^{(3)}_{int}$ and $H^{(4)}_{int}$  represent the cubic and the quartic interaction Hamiltonian. We notice that there are two terms that involve two insertion of a cubic interaction, and one term that involves one insertion of a quartic interaction.

We are going to apply eq.~(\ref{eq:diagram_expectation}) to two different physical Lagrangians that represent two different models of inflation. The first is a model of single field inflation where the inflaton is self-interacting with a strength that is parametrically larger than gravitational, and the second is a model where the inflaton is gravitationally interacting with $N$ free massless scalar fields. The fact that there are $N$ identical scalar fields makes the effect coming from them as they run in the loop parametrically larger than the analogous effect from loops of gravitons.

Before actually proceeding, let us specify what is our target for the calculation. The one-loop corrections in eq.~(\ref{eq:diagram_expectation}) will contain divergent pieces and finite pieces. It is expected that the divergent pieces will be reabsorbed by counterterms in the Lagrangian, which will affect also the finite terms of the result. In the present paper we will  not perform the explicit renormalization of the two Lagrangians we will study, but we will rather assume that this can be done~\footnote{In particular, for one of the theories, we will explicitly verify that this is the case by identifying and computing the counterterms.}. For one kind of divergencies, the logarithmic ones, it is particularly easy to realize how they will be transformed by the renormalization procedure. This is so because a logarithmic divergency will appear in the form 
\be
C \log\left(\frac{\Lambda}{\rm some\ physical\ mass\ scale}\right)\ ,
\ee
where $C$ is the coefficient of the logarithmic divergency, $\Lambda$ is the cutoff of the calculation,  and in the denominator we have some physical mass scale. Upon renormalization, the above expression will simply be transformed by replacing the cutoff $\Lambda$ with the renormalization scale $\mu$:
\be
C\ \log\left(\frac{\Lambda}{\rm some\ physical\ mass\ scale}\right) \qquad \rightarrow \qquad C\ \log\left(\frac{\mu}{\rm some\ physical\ mass\ scale}\right) \ .
\ee
This means that even without explicitly renormalizing the theory, we can compute the coefficient $C$ of the logarithmic running. Notice however that in principle we are unable to compute precisely what the mass scale which accompanies the renormalization scale is as we can always redefine it by a multiplicative constant at the cost of simply changing the finite terms. These finite terms are fixed once and for all by imposing the two-point function to have a certain value at some energy scale. They are part of the definition of the theory. The logarithmic divergency tells us instead how the amplitude of the process changes as we change the renormalization scale: it represents the running, in this case, of the two-point function and it can be large if the renormalization scale is chosen to be very different from the relevant energy scale of the process.  We will argue  that there is only one physical mass scale that can possible accompany the renormalization scale in this situation, while at the same time leaving the remaining numerical finite terms of order one. Several calculations of the coefficient $C$ and of the mass scale that accompanies the renormalization scale have been done since Weinberg in \cite{Weinberg:2005vy}, finding that the form of the log divergency is 
\be
C \log\left(\frac{k}{\Lambda}\right)\ ,
\ee
where $k$ is the comoving momentum of the $\zeta$ modes of which we compute the expectation value. We find this result not to be physical, as the comoving wavenumber $k$ is {\it not} a  physical quantity. Its value can be changed  by simply a constant rescaling of the scale factor $a$, which is not physical. As we pointed out in the introduction, this result could have had disastrous consequences for slow-roll eternal inflation, and more in general for inflation as well. The main point of this paper will be to show that all of those calculation are not correct, and that the mass scale accompanying the renormalization scale is the Hubble scale $H$:
\be
C\ \log\left(\frac{k}{\mu}\right) \qquad \rightarrow \qquad C\ \log\left(\frac{H}{\mu}\right) \ ,
\ee
proving in the same time that there cannot be any time dependence in the $\zeta$ correlation function. 

\section{Large $\dot\pi^3$ self-interaction\label{sec:dotpicube}}

In \cite{Cheung:2007st} an effective field theory for inflation was developed that made it possible to explore in full generality all the possible self-interactions of the inflaton, and in particular to clearly see that the inflaton can have large self-interactions without spoiling the background quasi de-Sitter solution. It was shown that the self-interactions can be parametrically larger than the one mediated by gravity and therefore we can concentrate on those without worrying about metric fluctuations. This simplifies the calculation a great deal. The Effective Lagrangian is a function of the field $\pi$ which represents the Goldstone boson of time translations which are spontaneously broken during inflation. It is related to the $\zeta$ metric fluctuation at linear level by the simple relation
\be
\zeta=-H\pi\ .
\ee
Notice that $\pi$ is a scalar field with dimensions of inverse mass. Here we do not present the derivation of this Lagrangian, that can be found in \cite{Cheung:2007st}, and we simply take a limit of its parameter space and use it for computing loop corrections. There are in general two interactions of comparable strength, $\dot\pi^3$ and $\dot\pi(\d_i\pi)^2$, each one accompanied by quartic interactions due to symmetry reasons. Upon tuning \cite{Cheung:2007st,Senatore:2009gt}, we can make the term containing $\dot\pi^3$ parametrically larger then the one containing $\dot\pi(\d_i\pi)^2$. This allows us to reduce the number of vertices, and therefore  to simplify the calculation. In this paper we are trying to assess conceptual and qualitative features of the loop corrections to the inflaton two point function and we can therefore study a tuned theory. Following the notation of \cite{Cheung:2007sv}, the action is of the form
\bea\label{eq:action}
S=\int d^4x\; a^3\left[-\dot H \mpl^2 \left(\dot\pi^2-\frac{1}{a^2}(\d_i\pi)^2\right)+\frac{2}{3}c_3 M^4\left(2\dot\pi^3+3\dot\pi^4-3\frac{1}{a^2}\dot\pi^2(\d_i\pi)^2\right)\right]\ . 
\eea
Here $c_3$ is a dimensionless parameter, while $M$ has dimensions of mass.  The interaction picture Hamiltonian is of the form
\be
H_{int}=\int d^3x\;a^3\left[-\frac{4}{3}c_3 M^4 \dot\pi^3-2c_3 M^4\left(1+2\frac{c_3 M^4}{\dot H\mpl^2}\right)\dot\pi^4+2c_3 M^4\frac{1}{a^2}\dot\pi^2(\d_i\pi)^2\right]\ .
\ee
We notice that we have a very simple interaction Hamiltonian at cubic level: only one interaction of the form $\dot\pi^3$.

We are now ready to begin the computation. We will perform it in two different regularization schemes: the first is by putting a sharp cutoff in frequency and momentum space, and the second by dimensional regularization. 

\subsection{Regularization by Frequency and Momentum Cutoff\label{sec:pidotcube_cutoff}}

We start by computing the two contributions with two insertions of $H^{(3)}_{int}$, and in particular the one of the form:
\bea\label{eq:diagr1}
\langle\zeta^2(t)\rangle_{\rm{1-loop,\,A}}=-2 H^2 {\rm Re}\left[\int_{-\infty_-}^t dt_2\int_{-\infty_-}^{t_2} dt_1\langle H^{(3)}_{int}(t_1) H^{(3)}_{int}(t_2)\pi^2(t)\rangle\right]\ ,
\eea
that we refer to as contribution $A$.
After some simple algebra we are left with:
\bea\label{eq:first_diagram}
&&\langle\zeta_{\vec k}(t)\zeta_{\vec k'}(t)\rangle_{\rm{1-loop,\,A}}=- (2\pi)^3 \delta^{(3)}(\vec k+\vec k')\\ \nonumber
&&\times\;128\, c_3^2 M^8 H^2\;{\rm Re}\left[ \int_{-\infty_-}^{t} dt_2\; a(t_2)^3 \int_{-\infty_-}^{t_2} dt_1\;  a(t_1)^3 \int \frac{d^3 k_1}{(2\pi)^3}\int \frac{d^3 k_2}{(2\pi)^3} (2\pi)^3 \delta^{(3)}\left(\vec k+\vec k_1+\vec k_2\right)\right. \\ \nonumber
&&\times\; \left.\dot\pi^{cl}_k(t_1)\pi^{cl\, *}_k(t)\;\dot\pi^{cl}_{k'}(t_2)\pi^{cl\, *}_{k'}(t)\  \dot\pi^{cl}_{k_1}(t_1) \dot\pi^{cl\, *}_{k_1}(t_2)\; \dot\pi^{cl}_{k_2}(t_1) \dot\pi^{cl\, *}_{k_2}(t_2)\right] \ .
\eea

Notice that there are other possible contractions of the operators, giving rise to diagrams that are not one-particle irreducible. However they involve derivatives of the wavefunctions at $k=0$ which vanish.
In order to get the coefficient of the one-loop logarithmic divergency, we realize that we can simply regulate the above integral with a cutoff in frequency and momentum space. Though this regulator is not diff. invariant we can exploit the fact that the logarithmic divergency receives equal contributions from all energy scales and therefore its coefficient is independent of the UV regulator. The coefficient of the logarithm obtained in this way is the correct one at one-loop but not at higher  loops as in this case it depends on the finite terms of the lower loop calculation which in turn depends on the regularization and renormalization procedure. 

We therefore regulate the  momentum integral with the following replacement:
\be
\int d^3k \qquad \rightarrow \qquad \int^{\Lambda a(t_1)} d^3k\ ,
\ee
where $\Lambda$ is a fixed {\it physical} cutoff, and therefore it appears cutting off the integral in comoving momentum $k$ accompanied with a factor of the scale factor $a$, so that $k_{\rm physical}=k/a(t)<\Lambda$. Notice that the scale factor $a$ is evaluated at time $t_1$. This can be understood in the following way. If we look back at eq.~(\ref{eq:generic_expectation}), we see that it involves the interaction picture evolution operator $U$ acting on the vacuum. In the interaction picture, we can think as the operators evolving with the free Hamiltonian, while the state evolving with the interaction Hamiltonian. The diagram we are computing corresponds to letting the vacuum state on the left (and then also the one on the right once we take the real part) evolve up to time $t$ with two insertions of the interaction Hamiltionian $H^{(3)}_{int}$, the first at time $t_1$ and the second at time $t_2$, with $t_1< t_2$. Momenta which are higher than the cutoff at the earlier time $t_1$,  do not contribute to the vertex at $t_1$, even in the case that by the time $t_2$ they have redshifted inside the cutoff. In the case there is no interaction at $t_1$, then there can no be a subsequent interaction at time $t_2$ and therefore we have to put the cutoff as $\Lambda\, a(t_1)$.
We have also to regulate the integral in $t_1$. This is analogously regulated with the same physical cutoff by replacing the $t_1$ integral in the following way:
\be
\int^{t_2}dt_1\qquad\rightarrow\qquad \int^{t_2-\frac{1}{\Lambda}}dt_1\ .
\ee

At this point, the calculation becomes quite straightforward. Before actually doing it let us first gain some intuition by doing the calculation in a non-expanding and then in a slowly expanding Universe. 

\subsubsection*{$\bullet$ Non-expanding Universe}
We start by completely neglecting the expansion of the Universe and take the wavefunctions to be:
\be\label{eq:classical_wave}
\pi^{cl}_k=\frac{1}{\left(-2 \dot H \mpl^2\right)^{1/2}}\frac{1}{k^{1/2}}e^{-i k t}\ , 
\ee
where we have set the scale factor $a$ to be equal to one. Notice that $\dot H$ enters here  just to canonically normalize the field, and does not play any role in our discussion. A straightforward calculation gives:
\bea 
&&\langle\zeta_{\vec k}(t)\zeta_{\vec k'}(t)\rangle_{\rm{1-loop,\,A,\, No\ Expansion}}\sim (2\pi)^3 \delta^{(3)}(\vec k+\vec k')\frac{c_3^2 H^2 M^8}{\dot H^4\mpl^8}\frac{1}{k^3}k^6\left(\log\left(\frac{\Lambda}{k}\right)+L\right)\, ,
\eea
where $L$ is a divergent constant and where we have neglected numerical factors. The result is clearly not scale invariant. Since the $\dot\pi^3$ interaction is of dimension six, the only way it gives rise to a logarithmic divergency is by pulling out some powers of the external momentum $k$. Notice that as $k\rightarrow 0$, the one-loop correction goes to zero, as the interaction we are dealing with has at least one derivative acting on the external leg.

\subsubsection*{$\bullet$ First corrections from the expanding Universe: $k\eta\gg1$}
Let us now include the first corrections from the space expansion and use the wavefunctions in de-Sitter space, taken in their high energy limit $k\eta\gg1$, where $\eta$ is conformal time:
\be\label{eq:classical_wave}
\pi^{cl}=i\frac{H}{\left(-2 \dot H \mpl^2\right)^{1/2}}\frac{1}{k^{3/2}}\left(1+i k \eta\right)e^{-i k \eta}\simeq-\frac{H}{\left(-2 \dot H \mpl^2\right)^{1/2}}\frac{(k\eta)}{k^{3/2}}  e^{-i k \eta} \ ,
\ee
where in the second passage we have taken the $k\eta\gg 1$ limit. In this case the expansion is very slow for the modes being considered. The result of the integration gives:
\bea\nonumber\label{eq:slow_loop}
&&\langle\zeta_{\vec k}(t)\zeta_{\vec k'}(t)\rangle_{\rm{1-loop,\,A,\, k\eta\gg 1}}\sim (2\pi)^3 \delta^{(3)}(\vec k+\vec k')\frac{c_3^2 H^8 M^8}{\dot H^4\mpl^8}\frac{(k\eta)^6}{k^3}\left(\log\left(\frac{k/a(\eta) }{\Lambda}\right)+C\right)\ , \\
\eea
We see that, apart for the $1/k^3$ term and the $\delta$-function, every other $k$ is accompanied by an $\eta$. In this way the expression, once Fourier transformed to real space, is correctly invariant under the rescaling $a\rightarrow \lambda\, a,\; k\rightarrow \lambda\, k,\; x\rightarrow x/\lambda$, which is a symmetry of the problem. Notice that again, in the limit $k\eta\rightarrow0$, the one-loop correction goes to zero. This is due to the fact that in order for this operator to be able to produce a logarithmic divergency, some powers of the external momentum have to be pulled out of the integral.  Symmetry arguments dictate that each additional power of momentum must be accompanied by a factor of $\eta$. This forces the expression, in the limit $\eta\rightarrow 0$, to go to zero. Further the structure of the logarithmic divergency has changed: now the factor of $k$ is accompanied by a scale factor evaluated at time $\eta$. We are beginning to see the relevant effect of the expanding Universe. We  can anticipate what the effect of including the expansion of the Universe even at later times ($k\eta\sim -1$) will be. As the expansion of the Universe goes on, and $k\eta$ becomes more and more infrared and eventually of order one, the amplitudes of the modes freeze and are evaluated when $k\eta\sim- 1$. This implies that the result becomes scale invariant, and that the logarithmic divergency takes the form of 
\be\label{eq:div}
\log\left(\frac{H}{\Lambda}\right)\ .
\ee
De-Sitter space has provided an infrared cutoff to the loop corrections. At the tree-level, quantum effects on scales larger than $1/H$ are larger than what they would be for the same scales in Minkowsky space. The same happens here for the loop corrections: on large scale, quantum effects are parametrically larger then what they would be in Minkowsky space. 

Notice two further comments. First, at least in principle, the term in (\ref{eq:div}) is a local term, and therefore one could imagine that a local counterterm proportional to $\log(H/\mu)$ could remove it completely. However this is not the case because for modes that are still inside the horizon the divergency is of the form of $\log(k/(a(t)\mu))$ which is not a local term and therefore cannot be removed by a local counterterm. The counterterm should be the same for all the modes, and therefore we conclude the $\log(H/\mu)$ is something that cannot be removed by a local counterterm. Second, the way the logarithm in (\ref{eq:slow_loop}) becomes (\ref{eq:div}) as the mode goes outside of the horizon tells us that the $H$ in the logarithm should be interpreted as the energy at which the process is evaluated for modes that have exited the horizon. This implies that we expect the constant $C$ to be of order one once $\mu$ is chosen to be of order $H$.

\subsubsection*{$\bullet$ Calculation in the Inflationary space}
 We are now ready to do the calculation in an inflationary Universe. 
We use the following classical wavefunction, expressed in conformal time:
\be\label{eq:classical_wave}
\pi^{cl}_k=i\frac{H}{\left(-2 \dot H \mpl^2\right)^{1/2}}\frac{1}{k^{3/2}}\left(1+i k \eta\right)e^{-i k \eta}\ . 
\ee
We would like to stress the technical point that since the momentum cutoff is time-dependent once expressed in comoving coordinates one has to perform the momentum integrals {\it first}  and the time integrals {\it second}. As we will see in the next subsection, in dimensional regularization we will perform the integrals in the opposite order. Taking the $t\rightarrow+\infty$ limit a straightforward calculation leads to:
\bea \label{eq:one-loop-result-A}
&&\langle\zeta_{\vec k}(t)\zeta_{\vec k'}(t)\rangle_{\rm{1-loop,\,A,\ t\rightarrow+\infty}}= -(2\pi)^3 \delta^{(3)}(\vec k+\vec k')\frac{1}{k^3}\frac{2}{15\pi^2}\frac{c_3^2 H^8 M^8}{\dot H^4 \mpl^8}\log\left(\frac{H}{\Lambda}\right)\ ,
\eea
where we have neglected all the finite terms and the power divergencies. Some details of this calculation are given in App.~\ref{app:cutoff-integrals}. The result is time-independent and scale invariant. The other contributions do not produce any other logarithmic divergency and we will be left with  $\log(H/\Lambda)$. 

Let us now  proceed with  the remaining terms. Let us analyze the additional one involving two insertions of $H^{(3)}$ which we will label with a $B$:
\bea\label{eq:diagr2}
&&\langle\zeta_{\vec k}(t)\zeta_{\vec k'}(t)\rangle_{\rm{1-loop,\; B}}=H^2\;\int_{-\infty_-}^t dt_1\int_{-\infty_+}^t dt_2\; \langle H^{(3)}_{int}(t_1) \pi_{\vec k}(t)\pi_{\vec k'}(t)H^{(3)}_{int}(t_2)\rangle \\ \nonumber
&& =- (2\pi)^3 \delta^{(3)}(\vec k+\vec k')\\ \nonumber
&&\times\;128\, c_3^2 M^8 H^2\;\int_{-\infty_-}^{t} dt_1\; a(t_1)^3 \int_{-\infty_+}^{t} dt_2\;  a(t_2)^3 \int \frac{d^3 k_1}{(2\pi)^3}\int \frac{d^3 k_2}{(2\pi)^3} (2\pi)^3 \delta^{(3)}\left(\vec k+\vec k_1+\vec k_2\right)  \\ \nonumber
&&\left.\times\; \dot\pi^{cl}_k(t_1)\pi^{cl\, *}_k(t)\;\dot\pi^{cl\, *}_{k'}(t_2)\pi^{cl}_{k'}(t)\  \dot\pi^{cl}_{k_1}(t_1) \dot\pi^{cl\, *}_{k_1}(t_2)\; \dot\pi^{cl}_{k_2}(t_1) \dot\pi^{cl\, *}_{k_2}(t_2)\right.\ .
\eea
This integral is convergent and gives
\bea 
&&\langle\zeta_{\vec k}(t)\zeta_{\vec k'}(t)\rangle_{\rm{1-loop,\,B,\ t\rightarrow+\infty}}= (2\pi)^3 \delta^{(3)}(\vec k+\vec k')\frac{1}{k^3}\frac{331}{7200\pi^2}\frac{c_3^2 H^3 M^8}{\dot H^4\mpl^8}\ .
\eea

Finally, we have to compute the contribution from the quartic interactions. It is quite easy to see that they do not give rise to logarithmic divergencies. Here for brevity we just concentrate on the $\dot\pi^4$ interaction, as this is the only one that will give an effect proportional to $c_3^2$ and so it could affect the logarithmic divergency we found in the first diagram. Keeping only the term proportional to $c_3^2$, the expression, that we call contribution $C$, reads:

\bea\label{eq:diagr3}\nonumber
&&\langle\zeta_{\vec k}(t)\zeta_{\vec k'}(t)\rangle_{\rm{1-loop,\; C}}=-2\; {\rm Im}\left[\int_{-\infty_-}^t dt_1\langle H^{(4)}_{int}(t_1) \pi_{\vec k}(t)\pi_{\vec k'}(t)\rangle\right]=96 (2\pi)^3\delta^{(3)}(\vec k+\vec k')\frac{c_3^2 H^2 M^8}{\dot H\mpl^2}\\
&&\times\;{\rm Im}\left[\int_{-\infty_-}^t dt_1\; a(t_1)^3\int^{\Lambda a(t_1)}\frac{ d^3k}{(2\pi)^3} \dot\pi^{cl}_k(t_1)\pi^{cl\, *}_k(t)\;\dot\pi^{cl}_{k'}(t_1)\pi^{cl\, *}_{k'} (t)\;  \dot\pi^{cl}_{k_1}(t_1) \dot\pi^{cl\, *}_{k_1}(t_1)\right]\ .
\eea
A straightforward integration shows that this contribution contains no logarithmic divergencies and goes as
\be
\langle\zeta_{\vec k}(t)\zeta_{\vec k'}(t)\rangle_{\rm{1-loop,\,C,\ t\rightarrow+\infty}} =(2\pi)^3 \delta^{(3)}(\vec k+\vec k')\frac{1}{k^3}\frac{c_3^2 H^4 M^8}{\dot H^4\mpl^8}{\cal{O}}\left(\Lambda^4,\,H^2\Lambda^2\right)\ .
\ee

Summarizing, we conclude that the $\zeta$ two-point function receives a logarithtmic correction of the form $\log(H/\Lambda)$:
\bea 
&&\langle\zeta_{\vec k}(t)\zeta_{\vec k'}(t)\rangle_{\rm{1-loop,\ t\rightarrow+\infty}}= (2\pi)^3 \delta^{(3)}(\vec k+\vec k')\frac{1}{k^3}\frac{2}{15\pi^2}\frac{c_3^2 H^8 M^8}{\dot H^4 \mpl^8}\left(\log\left(\frac{H}{\Lambda}\right)+C\right)\ ,
\eea
where $C$ is a numerical constant. We note that the unitarity bound $\Lambda_U$ of this theory is of order 
\be\label{eq:unitarity_cutoff}
\Lambda_U^4\sim \frac{\left(\dot H \mpl^2\right)^3}{M^8} \ ,
\ee
which means that the above one-loop correction scales as
\be
\langle\zeta^2\rangle_ {\rm{1-loop}}\sim\langle\zeta^2\rangle_ {\rm{tree}} \left(\frac{H}{\Lambda_U}\right)^4 \ .
\ee
As expected, the theory becomes strongly coupled when $H\sim\Lambda_U$. 

\subsection{Dimensional Regularization\label{sec:dim_reg_calculation}}

We now perform the same calculation as in the former subsection but in dimensional regularization. Dimensional regularization has the advantage of respecting diff. invariance. As we said before, this is not really necessary for computing the coefficient of the logarithmic running but it would be a great simplification if one wished to renormalize the theory and compute the finite terms of the loop corrections. Further, this is the regularization scheme  that has been applied in the  literature on the subject~\cite{Weinberg:2006ac,Adshead:2008gk,literature}, and we claim that it {\it has not} been applied correctly because several terms coming from the generalization of the expressions to $d$ dimensions have been omitted~\footnote{We will see that having forgotten those terms is equivalent to putting a cutoff in physical frequency-momentum space that grows exponentially in time.}. For these reasons we proceed to perform the same calculation in dimensional regularization (dim. reg.) where we make the integral convergent by changing the number of spatial dimensions.

Let us start with the diagram A, which becomes
\bea \label{eq:diag1dim}
&&\langle\zeta_{\vec k}(t)\zeta_{\vec k'}(t)\rangle_{\rm{1-loop,\,A}}=- (2\pi)^3 \delta^{(3+\delta)}(\vec k+\vec k')\; 128\, c_3^2 M^8 H^{2} \mu^{2\delta}\\ \nonumber
&&\times\; {\rm Re}\left[\int_{-\infty_-}^{t} dt_2\; a(t_2)^{3+\delta} \int_{-\infty_-}^{t_2} dt_1\;  a(t_1)^{3+\delta} \int \frac{d^{3+\delta} k_1}{(2\pi)^3}\int \frac{d^{3+\delta} k_2}{(2\pi)^3} (2\pi)^{3+\delta} \delta^{(3+\delta)}\left(\vec k+\vec k_1+\vec k_2\right)\right. \\ \nonumber
&&\qquad \times\; \left.\dot\pi^{cl}_k(t_1)\pi^{cl\, *}_k(t)\;\dot\pi^{cl}_{k'}(t_2)\pi^{cl\, *}_{k'}(t)\  \dot\pi^{cl}_{k_1}(t_1) \dot\pi^{cl\, *}_{k_1}(t_2)\; \dot\pi^{cl}_{k_2}(t_1) \dot\pi^{cl\, *}_{k_2}(t_2)\right] \ ,
\eea
where $\delta$ represents the difference between the number of spatial dimensions $d$ and 3 and $\mu$ is  the renormalization scale and it has been inserted in order to keep $\zeta$ dimensionless and $\pi$ with dimensions of time. Here we do not extend to $d$-dimensions  the numerical factors as they would only change the constant terms from the loop-corrections. At this point we have to express the wavefunction $\pi^{cl}$ in $d$-dimensions. This reads:
\be\label{eq:dimwave}
\pi^{cl,\,d}_k=-\frac{\sqrt{\pi}\,e^{i\pi \delta/4}\,H^{1+\delta/2}}{2\left(-\dot H \mpl^2\mu^\delta\right)^{1/2}}\frac{(-k\eta)^{(3+\delta)/2}}{ k^{(3+\delta)/2}}H^{(1)}_{(3+\delta)/2}(-k\eta)\ ,
\ee
where $H^{(1)}_\nu$ is the Hankel function of the first kind of index $\nu$.
We do not actually need to perform the integral in (\ref{eq:diag1dim}) with this wavefunction. In fact, the two wavefunctions which just depend on the external time $t$ and the external momentum $k$ can be brought outside of the integrals and taken to three dimensions and to the $\eta\rightarrow 0$ limit. They give:
\be\label{eq:waveto3}
\pi^{cl,\,d}_k(\eta\rightarrow0)= \frac{i}{(2\pi)^{1/2}}\frac{\,e^{i\pi \delta/4}\,H^{1+\delta/2}}{(-\dot H\mpl^2\mu^\delta)^{1/2}}\frac{1}{k^{(3+\delta)/2}}\qquad\rightarrow\qquad \frac{i}{(2\pi)^{1/2}}\frac{H}{(-\dot H\mpl^2)^{1/2}}\frac{1}{k^{3/2}}\ .
\ee
This is justified by the fact that every logarithm coming from the external wavefunctions will be canceled by the contribution from the counterterm which renormalizes this interaction. The same is true for the $\delta$-function of the external momenta:
\be
\delta^{(3+\delta)}(\vec k+\vec k')\qquad\rightarrow\qquad \delta^{(3)}(\vec k+\vec k') \ .
\ee
For the remaining wavefunctions inside the integrals we notice that we are interested in computing just the logarithmic running from the loop integral and we can therefore Taylor expand the above wavefunction around $\delta\rightarrow 0$, and keep only the linear term (as we will see, the loop integral has only one simple pole). Notice that the momentum in the wavefunction is still the full $d$-dimensional one which is a sufficient condition for keeping the integrals regularized. Taylor expanding (\ref{eq:dimwave}) around $\delta\rightarrow0$ we obtain:
\bea\label{eq:pi_expanded}
\pi^{cl,\,d}_k&\simeq & i\frac{H}{\left(-2 \dot H \mpl^2\mu^\delta\right)^{1/2}}\frac{1}{k^{3/2}}\left(1+i k \eta\right)e^{-i k \eta}\, \times \\ \nonumber
&& \left(1+\delta \left(\frac{1}{2}\log\left(-H\eta\right)+\frac{1}{1+i k\eta}-\frac{(1-i k\eta)}{4(1+i k\eta)}e^{2i k\eta}\left(- 3\pi i+2  \,{\rm Ci}(2 k\eta)-2\, i \, {\rm Si}(2k\eta)\right)\right)\right.\\ \nonumber
&&\left.+{\cal{O}}\left(\delta^2\right)\right)\ , \\ \nonumber
\dot\pi^{cl,\,d}_k&\simeq &-i \frac{H^2}{(-2 \dot H\mpl^2\mu^\delta)^{1/2}} k^{1/2}\eta^2 e^{-i k\eta}\times\\ \nonumber
&&\left(1+\delta\left(\frac{1}{2}\log\left(-H\eta\right)-\frac{1}{4}e^{2i k\eta}\left(-3\pi i+2  \,{\rm Ci}(2 k\eta)-2 \,i\, {\rm Si}(2k\eta)\right)+{\cal{O}}\left(\delta^2\right)\right)\right)\ ,
\eea
where
\be
{\rm Ci}(x)=\int^{x}_0 dx'\; \frac{\cos(x')-1}{x'}+\log(x)+\gamma \ ,\qquad \quad{\rm Si}(x)=\int^{x}_0 dx'\;\frac{\sin(x')}{x'}\ ,
\ee
and $\gamma$ is the Eulero-Mascheroni constant. We can now begin to do the integrals in (\ref{eq:diag1dim}) by starting from the ones in time.

\subsubsection*{$\bullet$ Contributions at order $\delta^0$}
We start by performing the time integrals in (\ref{eq:diag1dim}) for the component which is explicitly independent of $\delta$ (notice that $k$ is still the $d$-dimensional momentum). This was the only contribution included in the literature. We get:
\bea\label{eq:diagram1temp1}
&&\langle\zeta_{\vec k}(t)\zeta_{\vec k'}(t)\rangle_{\rm{1-loop,\,A,\, t\rightarrow +\infty,\,\delta^0}}=- (2\pi)^3 \delta^{(3)}(\vec k+\vec k') \frac{9}{16\pi^3} \frac{c_3^2 H^8 M^8}{\dot H^4 \mpl^{8}} \\ \nonumber
&&\times\; \frac{1}{k^7}\int \frac{d^{3+\delta} k_1}{(2\pi)^3\mu^\delta}\int \frac{d^{3+\delta} k_2}{(2\pi)^3} (2\pi)^{3+\delta} \delta^{(3+\delta)}\left(\vec k+\vec k_1+\vec k_2\right) \frac{k_1 k_2 \left(3\left(k1+k_2\right)^2+9\left(k_1+k_2\right)k+8 k^2\right)}{\left(k+k_1+k_2\right)^3}\ ,
\eea
where the subscript $\delta^0$ reminds us that we are taking only the terms in the wavefunctions of order $\delta^0$. Notice that in this case, contrary to what we did in the case of a regularization with a sharp cutoff in frequency and momentum space, we perform the time integral up to $t\rightarrow +\infty$ first, and then we perform the momentum integral. We will come back later in sec.~\ref{sec:dim_reg_finite_time} and sec.~\ref{sec:theorem}  to prove that there are no subtleties associated with this. By dimensional analysis, the remaining momentum integral will give a result of the form
\be\label{eq:Fdef}
k^4\left(\frac{k}{\mu}\right)^{\delta}F(\delta)\  ,
\ee
Because of the ultraviolet divergencies $F(\delta)$ has a singularities in the limit $\delta\rightarrow 0$ of the form:
\be
F(\delta)=\frac{F_0}{\delta}+F_1\ .
\ee
In the limit $\delta\rightarrow 0$ it leads to: 
\be
k^4 \left(\log(k/\mu)+L\right) \ ,
\ee
where $L$ is a divergent constant. It is straightforward to carry out the momentum integral and find:
\bea\label{eq:diagram1temp1res}
&&\langle\zeta_{\vec k}(t)\zeta_{\vec k'}(t)\rangle_{\rm{1-loop,\,A,\, t\rightarrow +\infty,\,\delta^0}}=- (2\pi)^3 \delta^{(3)}(\vec k+\vec k')\frac{2}{15\pi^2} \frac{c_3^2 H^8 M^8}{\dot H^4 \mpl^{8}} \frac{1}{k^3}\times\log\left(\frac{k}{\mu}\right) \ .
\eea

We now turn to consider the contribution from taking in each wavefunction the terms that are linear in $\delta$. These terms have been neglected in the literature. 

\subsubsection*{$\bullet$ Contribution proportional to \bf $\delta\log(-H\eta)$}
Again, as we are linearizing in $\delta$, we take the $\delta$-correction only from one wavefunction at the time. We start by considering the term in each wavefunction which contains the $\log(-H\eta)$:
\bea
\delta\pi^{cl,\,d}_k\simeq i\frac{H}{\left(-2 \dot H \mpl^2\right)^{1/2}}\frac{1}{k^3}\left(1+i k \eta\right)e^{-i k \eta}\, \times \delta \frac{1}{2}\log\left(-H\eta\right)\ .
\eea
Notice that this part of the Taylor expansion of (\ref{eq:pi_expanded}), taken as isolated, does not converge for $\eta\rightarrow 0$, as the $\log(-H\eta)$ diverges. However we are now going to notice that the time integrals are dominated by $\eta_1\sim\eta_2\sim\eta_k$ where $\eta_k$ is the time of horizon crossing for the $k$ mode ($k\eta_k\sim-1$), which implies that for $\delta$ sufficiently small, we can consider this isolated part of the expansion in (\ref{eq:pi_expanded}). 
Once we perform the integral over time, the effect of this additional logarithm will be  to multiply the expression in (\ref{eq:diagram1temp1}) by $\log(-c H\eta_k)$, where $c$ is some order one constant. This is so because the integrals in time in three dimensions read
\be\label{eq:time_integrals}
\int^0_{-\infty} d \eta_2\int^{\eta_2}_{-\infty} d\eta_1\; \eta_1^2\,\eta_2^2\; e^{(k+k_1+k_2)\eta_1} e^{(k-k_1-k_2)\eta_2}\ ,
\ee
after performing the contour rotation, and it is straightforward to see that they are dominated by the times $\eta_1\sim\eta_2\sim 1/k$ which is where the exponential suppression terminates. This is not changed when we add to the integral above a term of the form $\log(-H\eta_{1,2})$, and therefore the result of the time integrations will be just to replace $\log(-H\eta_{1,2})$ with $\log(-c H \eta_k)$, where $c$ is some order one constant.
At this point, the momentum integration is the same as before. In the limit $\delta\rightarrow 0$ this term will contribute to the logarigthmic divergency by multipling the coefficient of order $1/\delta$  of $F(\delta)$ in eq.~(\ref{eq:Fdef}). This is the same term that in the former contribution was multiplying the factor $\log(k/\mu)$. There are six factors of $\delta \log(-H\eta_k c)/2$ coming from the six wavefunctions integrated in the loops. Notice that there is also two analogous terms coming from Taylor expanding the factors of $a(\eta_{1,2})^\delta$ in the measure of integration around $\delta=0$. In the $\delta\rightarrow  0$ limit, they also multiply the coefficient of order $1/\delta$ of $F(\delta)$, with the only difference that this time they are proportional to minus two factors of $\delta \log(-c H\eta_k)$. Once we take the $\delta\rightarrow 0$ limit, all of these eight terms give the same result as in (\ref{eq:diagram1temp1res}) with the replacement:
\be
\log(k/\mu)\qquad\rightarrow\qquad \log(-H\eta_k )\ ,
\ee
that is
\bea\label{eq:diagram1temp2res}
&&\langle\zeta_{\vec k}(t)\zeta_{\vec k'}(t)\rangle_{\rm{1-loop,\,A,\, t\rightarrow +\infty,\,\delta \log}}=- (2\pi)^3 \delta^{(3)}(\vec k+\vec k')\frac{2}{15\pi^2} \frac{c_3^2 H^8 M^8}{\dot H^4 \mpl^{8}} \frac{1}{k^3}\times \log(-H\eta_k )\ .
\eea
where we have neglected the divergent term and the order one constant $c$ as it is irrelevant,  and where the $\delta \log$ subscript represents the fact that this is the contribution from the terms linear in $\delta \log(-H\eta)$. 

\subsubsection*{$\bullet$ Remaining $\delta-$components}
We are now left to consider the contribution from taking in one of the wavefunctions the remaining term proportional to $\delta$. Notice that this term is dimensionless and only a function of $k\eta_{1,2}$ and $k_{1,2}\eta_{1,2}$, depending on which wavefunction we are dealing with. So, if we were to perform the time integral, we would obtain an expression similar to (\ref{eq:diagram1temp1}). Then the remaining momentum integral, by dimensional analysis,  would give a result proportional to
\be
\delta\; k^4 \left(\frac{k}{\mu}\right)^\delta \tilde F(\delta) ,
\ee
where $\tilde F(\delta)$ is dimensionless and divergent for $\delta\rightarrow 0$. In order for a logarithmic running to come from this term, it would be necessary for $\tilde F$ to have a double pole for $\delta\rightarrow 0$. However, we saw in eq.~(\ref{eq:diagram1temp1res}) that the $F(\delta)$ defined in eq.~(\ref{eq:Fdef}) had only a single pole, and it is quite straightforward to realize that the additional terms proportional to $\delta$ that transform $F$ into $\tilde F$ do not increase the degree of divergency of $F$. In fact, it is quite tedious but straightforward to show that upon inclusion of the remaining $\delta$-correction to the wavefunction in the time integrals, these are still dominated by $\eta_1\sim\eta_2\sim 1/k$.  At this point, we look at the part of the $\delta$-corrected wavefunction that we are considering, which now can be thought of as depending on the ratio $k_{1,2}/k$ (in the other case we are just left with a number of order one), and concentrate only on those terms that can induce a logarighmic running. These are only those ones that do not scale as a simple power law in $k_{1,2}/k$.  The part we are considering of the $\delta$-corrected wavefunction has a logarighmic dependence only in the IR limit $k_{1,2}/k\rightarrow 0$ as $\log(k_{1,2}/k)$.  However, in this limit the integral we started with in the limit of $\delta\rightarrow 0$ was power law convergent, and therefore the additional infrared $\log(k_{1,2}/k)$ does not induce an additional logarithmic divergency. We conclude that $\tilde F(\delta)$ has no double pole as $\delta\rightarrow 0$, and that therefore no additional logarithms appear when considering this part of the $\delta$-correction to the wavefunction.  

We stress that these contributions proportional to $\delta$ were forgotten in the literature \cite{Weinberg:2006ac,Adshead:2008gk,literature} and led to parametrically incorrect results.
 
Finally, we comment on the decision we took in eq.~(\ref{eq:waveto3}) to take the external wavefunctions directly in $d=3$ dimensions. We can notice that if we had kept them in generic dimensions and we would have then expanded in $\delta$ as we did for the other wavefunctions, then naively there would be terms that, multiplying $F_0/\delta$ from the loop integral would give rise to additional logarithm of the form $\log(k)$ or $\log(H/\mu)$. However, if the divergencies are to be renormalized, than there is a counterterm that cancels the divergent term $F_0/\delta$  of the overall integral. This counterterm multiplies the same external waveunctions as in eq.~(\ref{eq:waveto3}), canceling in this way all the additional logarithmic divergency that we would have had if we had kept the external wavefunctions in $d$ dimensions~\footnote{We will perform explicitly the renormalization in sec.~\ref{sec:renormalization}. However, we can anticipate that schematically the counterterm will be of the form 
\be\label{eq:Hcounter}
H_{int}^{{\rm counter\ term}}\sim \int d^{3+\delta}x\;\mu^\delta\, a^{3+\delta}(t) \frac{F_0}{\delta}\frac{\dot H \mpl^2}{\Lambda^4_U} (\d^3\pi)^2 \ ,
\ee
where the number of derivatives comes from imposing that the counterterm has the same powers of the cutoff $\Lambda_U$ as the terms it renormalizes, as it is standard in non-renormalizable effective field theories, and where we have not been careful in distinguishing spatial or time derivatives, as in general all will appear in the counterterms. By treating (\ref{eq:Hcounter}) perturbatively in (\ref{eq:generic_expectation}), we will obtain terms schematically of the form
\be
\langle\zeta^2\rangle_{{\rm counter\ term}}\sim \delta^{(3+\delta)}(\vec k+\vec k') \pi^{cl\;*}_k(t)\pi^{cl\;*}_k(t)\frac{F_0}{\delta}\frac{\dot H \mpl^2}{\Lambda_U^4}\int^{\eta}d\eta'\; \mu^{\delta}\, a^{4+\delta}(\eta') \left(\frac{\d^3}{a^3}\pi^{cl}_k(\eta')\right)^2 \ .
\ee
The first two wavefunctions on the left are the same external wavefunctions we get in the loop. Taylor expansion of the integrand in $\delta\rightarrow 0$ (and using eq.~(\ref{eq:pi_expanded}) since the integral is dominated by $k \eta' \sim -1$), shows that no additional logarithmically divergent terms linear in $\delta$ are generated by the integrand. This implies that the additional logarithmically divergent pieces that comes from considering the external wavefunctions and the $\delta$-function of the external momenta in $3+\delta$ dimensions get canceled by the counterterm.
 }. This justifies our rather intuitive approach of treating the wavefunctions that do not participate to the loop integral directly in three dimensions.
   
Summarizing: in the calculation performed in dimensional regularization the effect of considering the terms proportional to $\delta$ in the measure and in the wavefunction has achieved the result of transforming the logarithmic divergency $\log(k/\mu)$ found in the  literature \cite{Weinberg:2006ac,Adshead:2008gk,literature} into
\be
\log(k/\mu)+\log(-H\eta_k )\sim\log(H/\mu)\ ,
\ee 
where we have used that $k\,\eta_k \sim -1$.
It is straightforward to see that the contributions we called $B$ and $C$ do not result in any logarithm and therefore we conclude that in dimensional regularization we also obtain: \bea\label{eq:diagram1resdim}
&&\langle\zeta_{\vec k}(t)\zeta_{\vec k'}(t)\rangle_{\rm{1-loop,\, t\rightarrow +\infty}}=- (2\pi)^3 \delta^{(3)}(\vec k+\vec k')\frac{2}{15\pi^2} \frac{c_3^2 H^8 M^8}{\dot H^4 \mpl^{8}} \frac{1}{k^3}\times\log\left(\frac{H}{\mu}\right) \, .
\eea

\subsubsection{$\log(H/\mu)$ without evaluating the loop integrals\label{sec:adimensionalization}}

We would like here to show that, even without doing the actual calculation in dim. reg., it is possible to see that the simple fact that the integrals are dimensionally regularized forces the logarithmic running to be of the form $\log(H/\mu)$. In order to this, it is however necessary to assume that the $\zeta$ two-point function does not depend on time. It is possible to show this without actually having to do any loop integral, as we will show in sec.~\ref{sec:theorem}, or, by actually doing the calculation in dim. reg. at finite time as we will do in sec.~\ref{sec:dim_reg_finite_time}. For the time being, we will assume this to be the case and we will therefore take the extremum of integration of the time integrals in the loops to plus infinity.

At this point, by proper redefinition of the variables of integration, it is possible to make the loop integrals dimensionless functions of $\delta$ only, and from this extract the form of the logarithmic running.  In fact we can take for example the loop term in (\ref{eq:diag1dim}) that we reproduce here for convenience:
\bea
&&\langle\zeta_{\vec k}(t)\zeta_{\vec k'}(t)\rangle_{\rm{1-loop,\,A}}=- (2\pi)^3 \delta^{(3)}(\vec k+\vec k')\; 128\, c_3^2 \frac{M^8 H^{4}\mu^\delta}{\dot H \mpl^2} \frac{1}{k^3} \\ \nonumber
&&\times\; \int_{-\infty_-}^{0} d\eta_2\; a(t_2)^{4+\delta} \int_{-\infty_-}^{\eta_2} d\eta_1\;  a(\eta_1)^{4+\delta} \int \frac{d^{3+\delta} k_1}{(2\pi)^3}\int \frac{d^{3+\delta} k_2}{(2\pi)^3} (2\pi)^{3+\delta} \delta^{(3+\delta)}\left(\vec k+\vec k_1+\vec k_2\right) \\ \nonumber
&&\times\; \dot\pi^{cl}_k(\eta_1)\;\dot\pi^{cl}_{k'}(\eta_2)\  \dot\pi^{cl}_{k_1}(\eta_1) \dot\pi^{cl\, *}_{k_1}(\eta_2)\; \dot\pi^{cl}_{k_2}(\eta_1) \dot\pi^{cl\, *}_{k_2}(\eta_2)  \ ,
\eea
where we have passed to conformal time,  taken the $\eta\rightarrow 0$ limit and we taken the external wavefunctions in three dimensions. We can make all the  quantities within the integral dimensionless by multiplying or dividing by the external momentum $k$. This is possible because the scale factor is unchanged in dim. reg:
\be
a(\eta_1)=-\frac{1}{H\eta_1}=-\left(\frac{k}{H}\right)\frac{1}{k\eta_1}\ ,
\ee
and because the wavefunctions in dim. reg. are just functions of $k\eta$ apart for multiplicative factors:
\bea\label{eq:dimwave_adim}
\pi^{cl,\,d}_{k_1}(\eta_1)&=&-\frac{\sqrt{\pi}H^{1+\delta/2}}{2\left(-\dot H \mpl^2\mu^\delta\right)^{1/2}}\frac{(-k_1\eta_1)^{(3+\delta)/2}}{ k_1^{(3+\delta)/2}}H^{(1)}_{(3+\delta)/2}(-k_1\eta_1)\\ \nonumber 
&=&-\frac{\sqrt{\pi}H^{1+\delta/2}}{2\left(-\dot H \mpl^2\mu^\delta\right)^{1/2}}\frac{\left(-\left(\frac{k_1}{k}\right)(k\eta_1)\right)^{(3+\delta)/2}}{ k^{(3+\delta)/2}(k_1/k)^{(3+\delta)/2}}H^{(1)}_{(3+\delta)/2}\left(-\left(\frac{k_1}{k}\right)(k\eta_1)\right)\\ \nonumber &=& -\frac{\sqrt{\pi}H^{1+\delta/2}}{2\left(-\dot H \mpl^2\mu^\delta\right)^{1/2}}\frac{1}{k^{(3+\delta)/2}}G\left(\frac{k_1}{k},k\eta_1\right)\ ,
\eea
where $G(k_1/k,k_1\eta)$ is a dimensionless function.
By extracting the relevant powers of $k$ and $H$, the remaining integral is just a dimensionless function of $\delta$. We obtain:
\bea
\langle\zeta_{\vec k}(t)\zeta_{\vec k'}(t)\rangle_{\rm{1-loop,\,A}}&=&- (2\pi)^3 \delta^{(3)}(\vec k+\vec k')\; 128\, c_3^2 \frac{M^8 H^{4}\mu^\delta}{(-\dot H \mpl^2)} \frac{1}{k^3} \times \frac{H^{4+\delta}}{(-\dot H \mpl^2)^3\mu^{3\delta}} I(\delta)\\ \nonumber 
&=&- (2\pi)^3 \delta^{(3)}(\vec k+\vec k')\; 128\, c_3^2 \frac{M^8 H^{8}}{\dot H^4 \mpl^8} \frac{1}{k^3} \times \left(\frac{H}{\mu}\right)^\delta I(\delta)\ ,
\eea
where $I(\delta)$ is a dimensionless function of $\delta$. In the limit $\delta\rightarrow 0$, the factor of $(H/\mu)^\delta$ can be Taylor expanded to give
\be
\left(\frac{H}{\mu}\right)^\delta\qquad\rightarrow\qquad 1+\delta \log\left(\frac{H}{\mu}\right)\ .
\ee
This quantity will multiply any pole coming from the integral $I(\delta)$ as $\delta\rightarrow 0$, giving rise to a logarithm of the form
\be
\log\left(\frac{H}{\mu}\right) \ .
\ee
This shows that {\it if} there is a logarithm as a result of this contribution {\it then} it is of the form $\log(H/\mu)$.  It is straightforward to see that the same result holds also for the other diagrams. This is an alternative proof of the fact that the running has the form $\log(H/\mu)$ that does not rely on us being able to actually perform the loop integrals. 

\subsubsection{Calculation in dimensional regularization at finite time\label{sec:dim_reg_finite_time}}

It is not difficult to generalize the calculation we did in sec.~\ref{sec:dim_reg_calculation} to finite external time. Apart from the fact that  the integrals becomes slightly more complicated the only subtlety is in how to deal with the corrections from the measure and the wavefunction that are proportional to $\delta$ (see eq.~(\ref{eq:pi_expanded})). In sec.~\ref{sec:dim_reg_calculation} we saw that, when the external time is taken to be plus infinity, the time integrals were dominated by $\eta_1\sim\eta_2\sim\eta_k$, where $\eta_k$ is the horizon crossing time of the mode $k$. This allowed us to replace the term $\delta\log(-H\eta_{1,2})$ in the wavefuntions and in the measure with $\log(-H\eta_k)$ and argue that the remaining $\delta$-corrections to the wavefunctions did not induce addional logarithmic divergencies. It is clear that if the external time $\eta$ is late enough so that the external mode $k$ is outside of the horizon (i.e. $-k\eta\ll1$), this still applies. In this regime the $\delta$-corrections can be treated exactly as in the case of infinite external time.

A straightforward calculation gives:
\bea\label{eq:diagram1resdim_finitetime}
\langle\zeta_{\vec k}(t)\zeta_{\vec k'}(t)\rangle_{{\rm1-loop,\ }k\eta\ll1}&=& -(2\pi)^3 \delta^{(3)}(\vec k+\vec k')\frac{2}{15\pi^2} \frac{c_3^2 H^8 M^8}{\dot H^4 \mpl^{8}} \frac{1}{k^3}\times\log\left(\frac{H}{\mu}\right)\\ \nonumber
&&\times\; {\rm Re}\left[-\frac{1}{4}(i+k\eta)^2 \left(4+k\eta\left(2 i+k\eta\right)\left(4+k\eta(6+i k\eta)\right)\right)\right] \, .
\eea

As we will discuss in detail later in sec.~\ref{sec:theorem} the loop integrals performed at finite external time are, at least in principle, more regular than the ones done at infinite external time. This is so because if we keep the external time finite the most ultraviolet $k$ modes are inside the horizon and therefore the integrals receive additional oscillations in the UV that improve their convergence. It is therefore possible that the logarithmic running that we found in the infinite external time limit disappears in the calculation done at finite time. It would be replaced by a time-dependent term that diverges as $\log(\eta)$ as we send the external time $\eta$ to zero.  We find that this is not the case. 
We will further explicitly verify this  in sec.~\ref{sec:renormalization} where we will find the counterterms that allow us to reabsorb the divergencies.  This proves that the $\zeta$ correlation function is time-independent at one-loop, and it also shows that the logarithmic running we computed in (\ref{eq:diagram1resdim_finitetime}) can indeed be computed without need of renormalizing the theory.


\subsection{Renormalization\label{sec:renormalization}}

We now proceed to the actual renormalization of the correlation function of $\zeta(t)^2$ for the case of the large $\dot\pi^3$ self-interactions. This theory is so simple, that this process is not complicated. Notice that this interaction is irrelevant, or equivalently non-renormalizable.

In non-renormalizable effective field theories, the divergencies are meant to be absorbed order by order in an expansion of energy over cutoff, or equivalently derivatives over cutoff. The effect we computed in (\ref{eq:diagram1resdim}) is from a dimension six operator (in canonical normalization), and therefore it is suppressed with respect to the tree-level result by $(H/\Lambda_U)^4$, where $\Lambda_U$ is the unitarity bound defined in eq.~(\ref{eq:unitarity_cutoff}). If we work in dim. reg. only the logarithmic divergencies appear and therefore we have to look for dimension eight quadratic operators. There are three dimension eight quadratic  operators compatible with the $\pi$ shift symmetry~\cite{Cheung:2007st}.  The counterterm Lagrangian is therefore given by:
\be
S_c=\int d^4x\; \mu^\delta\;a(t)^{3+\delta} \dot H\mpl^2 \frac{M^8}{\left(\dot H\mpl^2\right)^3}\left[C_1 (\d_t^3\pi)^2+C_2 \frac{1}{a^4}(\d_t \d_i^2\pi)^2+C_3 \frac{1}{a^2}(\d_t \d_t \d_i\pi)(\d_t \d_t \d_i\pi)\right]\ ,
\ee
where the $C_1,\,C_2,\,C_3$ are dimensionless numbers, $M^8/(\dot H \mpl^2)^3$ represents the cutoff suppression, and we have extracted a factor of $\dot H\mpl^2$ that represents the normalization term of the two derivative kinetic term~\footnote{In the unitary gauge where the $\pi$ Lagrangian is usually constructed \cite{Cheung:2007st}, these term arise for example by operators such as $(\d_t^2\delta g^{00})^2,\;(\d_t\d_i\delta g^{00})^2,\; (\d_t\delta K^{i}_i)^2$, where $K^{i}_i$ is related to the trace of the extrinsic curvature of uniform inflaton surfaces (see \cite{Cheung:2007st} for details.)}. 

Notice that the procedure would be different if we instead decided to perform the renormalization using the cutoff regularization. In this case, all the lower dimension operators would be generated with some power law divergency. These include the lowest dimensional one: the wavefunction renormalization. Notice however that the coefficient of the term $(\d_i\pi)^2$ is fixed by symmetries to be $\dot H \mpl^2$, due to the fact that $\pi$ is the Goldstone boson of time translations  \cite{Cheung:2007st}. It {\it cannot} be renormalized. This is quite an unusual non-renormalization theorem that affects this term. This means that the power law divergent terms will only renormalize the $\dot\pi^2$ term out of the two lowest dimension one. This effectively induces a speed of sound of the fluctuations $c_s$ that is different from one, unless it is tuned to be equal to one (and this is technically possible since we are dealing with a power law divergency) \cite{Cheung:2007st,Senatore:2009gt}. This was expected, as typical in effective field theories all allowed operators tend to be generated suppressed by the same scale unless we perform some tuning or unless there is a symmetry protecting it. In this case, there appears to be no symmetry protecting  $c_s=1$  and we decide to tune it. Notice that in this paper we are concentrating on the conceptual issues arising from the study of loop corrections to inflationary observable, and we consciously decide to study a consistent, though tuned, theory to simplify  the calculations.  

Connected to this, we notice that there are higher derivative quadratic operators of the form $(\d^2\pi)^2/\Lambda_U^2$ which would naturally give a correction to the $\zeta$ two-point function larger then the one in (\ref{eq:diagram1resdim}) by a factor of the order $(\Lambda_U/H)^2\gg 1$. Again, we are assuming that these terms are tuned away.

Coming back to reabsorbing the divergency of the  dim. reg. loop result of (\ref{eq:diagram1resdim}), the counterterms give a correction of the form
\bea\label{eq:diagram_counter}
\langle\zeta^2(t)\rangle_{\rm{counter\ term}}=-2H^2\,{\rm Im}\left[\int_{-\infty_-}^t dt_1\langle H^{(2)}_{int}(t_1) \pi^2(t)\rangle\right]\ ,
\eea
where $H^{(2)}_{int}=-L_c^{(2)}$, where $L_c^{(2)}$ is the counter term Lagrangian. This gives:
\bea\label{eq:diagram_counter_res}
&&\langle\zeta_k(t)\zeta_{k'}(t)\rangle_{\rm{counter\ term}}=(2\pi^3)\delta^{(3)}(\vec k+\vec k')\frac{1}{32\pi^3}\frac{H^8 M^8}{\dot H^4\mpl^8}\frac{1}{k^2}\\ \nonumber
&&\times \ \left[-\left(C_1+3C_2+C_3\right)\left(1+(k\eta)^2\right)-4\left(3C_1+C_3\right)(k\eta)^4+2(C_1+C_2+C_3)(k\eta)^6\right]\ .
\eea
These terms have to cancel the pole in $1/\delta$ coming in the loop integral. At finite time, this is  given by expression (\ref{eq:diagram1resdim_finitetime}) with $\log(H/\mu)$ taken to be equal to one. Notice that the finite time results depends on $(k\eta)^4,\;(k\eta)^2,\;(k\eta)^1, (k\eta)^0$: imposing the cancellation of the poles leads to four equation in three unknowns: $C_{1,2,3}$. The solution still exists: 
\be\label{eq:counter_sol}
C_1=4\pi\, c_3^2\frac{1}{\delta}\ , \qquad C_2=\frac{28}{15}\pi\, c_3^2\frac{1}{\delta}\ , \qquad C_3=-\frac{16}{3}\pi\, c_3^2\frac{1}{\delta}\ , 
\ee
providing another non trivial check of our results. 
The pole $1/\delta$ represents the fact that these counterterm are actually divergent in three dimensions. To each one of the coefficients we can add a finite term that can be determined by imposing the two point function at some renormalization scale to be equal to some quantity. We do not perform this last step here, as it is not the main point of our work.

In fact, our result shows that the divergency we found in sec.~\ref{sec:dim_reg_calculation} {\it are} reabsorbable by local couterterms. This is yet a third independent proof of this (the first one was obtained by performing the momentum integration first in the loop integral regularized with a cutoff; the second one was by performing the dim. reg.  loop integrals at finite external time in sec.~\ref{sec:dim_reg_finite_time}; we will give a fourth proof of this in sec.~\ref{sec:theorem}). 

Though the finite terms in (\ref{eq:counter_sol}) could be determined only by performing the loop integrals in sec.~\ref{sec:dotpicube} paying attention to the finite terms  it is worth mentioning that as usual the effect we have computed can be much larger than the finite effects if we decided to renormalize the theory at an energy scale $\mu$ very far from $H$. This is actually what makes sensible to compute the running and not the finite terms.

\subsection{Summary of calculation with large $\dot\pi^3$ self-interaction}

The case where the inflaton has a large self-interaction of the form $\dot\pi^3$ has allowed us to  calculate one-loop corrections to the inflaton two-point function. We have performed the calculation in two different kind of regularizations: one with a sharp cutoff in momentum and frequency space, and the other in dimensional regularization. We have also performed the dim. reg. calculation at finite time, and we have also shown the form of the logarithmic running by simply making dimensionless the loop integrals. We have then performed the renormalization of the correlation function, by explicitly finding and solving for the relevant counterterms.

We have found that all the calculations agree. The result is:
 \bea\label{eq:diagram1resdim}
&&\langle\zeta_{\vec k}(t)\zeta_{\vec k'}(t)\rangle_{\rm{1-loop,\, t\rightarrow +\infty}}= -(2\pi)^3 \delta^{(3)}(\vec k+\vec k')\frac{2}{15\pi^2} \frac{c_3^2 H^8 M^8}{\dot H^4 \mpl^{8}} \frac{1}{k^3}\times\log\left(\frac{H}{\mu}\right) \, .
\eea
As we had anticipated the logarithm is  $\log(H/\mu)$. We find this to be a very sensible result. First of all, contrary to the $\log(k/\mu)$ found in the former literature, the real space version of (\ref{eq:diagram1resdim}) is symmetric under the rescaling 
\be
a\rightarrow \lambda\, a\ , \qquad x\rightarrow x/\lambda\ ,\qquad k\rightarrow \lambda\,k\ ,
\ee
 a symmetry of the problem. Second, it makes sense from a physical point of view: the Hubble scale is cutting off the infrared behavior that would otherwise be there in Minkowky space. The resulting logarithm is of the form $\log(H/\mu)$, which is similar to the form found in scattering amplitudes: $\log(E/\mu)$ where $E$ is the center of mass energy and $\mu$ is the renormalization scale. The energy probed by the interactions during inflation is of order $H$.

\section{Gravitational interactions with $N$ massless scalar fields\label{sec:sigma_fields}}

We now turn to the inflationary theory that Weinberg originally studied in \cite{Weinberg:2005vy}. This is a theory where an inflaton with a standard kinetic term is rolling down a flat potential and is interacting gravitationally with $N$ massless scalar fields. The Lagrangian is of the form:

\begin{equation}\label{action_spectator}
S=\int d^4x\; \sqrt{-g}\left[\frac{1}{2}(\partial\phi)^2-V(\phi)+\sum_{n=1}^{N}\frac{1}{2}g^{\mu\nu}
\partial_\mu\sigma_n\partial_\nu\sigma_n\right]\ .
\end{equation}
Though we are considering purely gravitational interactions, the contribution from the $N$ massless scalar fields running in the loops will be enhanced by a factor of $N$ with respect to the analogous interactions coming from the graviton and the inflaton running in the loop. For this reason we can avoid the complication of letting those  run in the loops and concentrate on the $\sigma$ scalar fields~\footnote{The calculation done originally by Weinberg has a rather minor numerical mistake due to an inconsistent implementation of the $i\,\epsilon$ prescription. This was noticed and fixed in \cite{Adshead:2008gk}. Both \cite{Weinberg:2005vy,Adshead:2008gk} performed their calculations in dimensional regularization and did not include the contributions proportional to $\delta$ from the scale factors in the measure of integration and from the $d$-dimensional wavefunctions and therefore obtained an incorrect  logarithm of the form $\log(k/\mu)$.}.

We do not need to redo all the calculation. It is straightforward to include in the results of \cite{Weinberg:2005vy,Adshead:2008gk} the correction coming from the terms proportional to $\delta$. In fact  the only relevant terms proportional to $\delta$  were those of the form $\log(-H\eta)$ coming either from the wavefunction or from the measure of integration. After performing the time integrals, which converge and are dominated by $\eta_{1,2}\sim \eta_k$, those terms become of the form $\log(-H\eta_k)$. The contributions with two insertions of the cubing interaction Hamiltonian are of the form:
\be\label{eq:general_correction}
k^D \left(\frac{k}{\mu}\right)^\delta F(\delta)\times (1+\delta \log(-H\eta_k))\ ,
\ee
where $D$ is the correct number of dimension for the integral in three dimensions, while the overall factor of $k^\delta$ comes from the two measures of integration times a $(3+\delta)$ dimensional $\delta$-function. As before, $F(\delta)$ will be a dimensionless quantity of the form 
\be
F(\delta)=\frac{F_0}{\delta}+F_1\ ,
\ee
for $\delta\rightarrow 0$. The coefficient of the term $\delta \log(-H\eta_k)$ is equal to one because this contributions involves   six wavefunctions and two measures of integration. Eq.~(\ref{eq:general_correction}) can be expanded for small $\delta$ to give:
\be
k^4 \left(1+\delta \log\left(\frac{k}{\mu}\right)+\delta \log(-H\eta_k)\right)\left(\frac{F_0}{\delta}+F_1\right)\simeq k^4 \left(1+\delta \log\left(\frac{H}{\mu}\right)\right)\left(\frac{F_0}{\delta}+F_1\right)\ ,
\ee
which means that the coefficient of the logarithmic divergency is of the form
\be
\log\left(\frac{H}{\mu}\right)\ .
\ee
This analysis can be easily extended to the diagrams involving an insertion of the quartic interaction Hamiltonian. In this case the integral in time  converges and is dominated by the time of the horizon crossing for the $k$-mode. The result after the momentum integration will be of the form:
\be
k^D \left(\frac{k}{\mu}\right)^\delta F(\delta)\times (1+\delta \log(-H\eta_k))\ ,
\ee
which is of the same form as (\ref{eq:general_correction}). The factor in front of $\delta \log(-H\eta_k)$ is equal to one and this time it comes from the fact that there are four wavefunctions contributing as $2\delta \log(-H\eta_k)$ and one measure of integration contributing as $-\delta\log(-H\eta_k)$. \cite{Adshead:2008gk} shows that in the case of $N$ spectator scalar fields, $F(\delta)$ does not have a pole at $\delta\rightarrow 0$, and therefore no logarithm can come from this term. 


Summarizing, in the case of $N$ spectator scalar fields the form of the logarithm is $\log(H/\mu)$. By applying our correction to the results of \cite{Weinberg:2005vy,Adshead:2008gk} we find: 
\bea
&&\langle\zeta_{\vec k}(t)\zeta_{\vec k'}(t)\rangle_{\rm{1-loop,\, t\rightarrow +\infty}}= -(2\pi)^3 \delta^{(3)}(\vec k+\vec k') \frac{\pi}{6}N\frac{\dot H}{H^2}\frac{H^2}{\mpl^2}\frac{1}{k^3}\log\left(\frac{H}{\mu}\right)\ .
\eea

\subsubsection{Tadpole diagrams and non 1PI diagrams\label{sec:tadpole_cancellation}}

In computing one loop corrections to correlation functions of $\zeta$ there are important tadpole diagrams that have been so far neglected. These tadpole diagrams are important not only for the one-point function, but also, by attaching them to propagators, for all correlation functions. Because of translation invariance, the external line attached to a tadpole diagram has to have zero wavenumber.  In the former example ($\dot\pi^3$ interaction) these diagrams were therefore zero because the field $\pi$ had an exact shift symmetry so that every $\pi$ had a derivative acting on it. 
For the  $N$ spectator scalar fields this is not case.  The tadpole diagrams have to be included. The expectation value of $\zeta$ in this theory is given by:
\bea\label{eq:tadpole_diagram}
&&\langle\zeta_{\vec k}(t)\rangle_{\rm{1-loop}}= -2 {\rm Im}\left[\int^t_{\infty-}dt_1 \langle H^{(3) }_{int}\zeta_{\vec k}(t)\rangle\right]\\ \nonumber
&&=-(2\pi)^3 \delta^{(3)}(\vec k){\rm Im}\left[\int^t_{\infty-}dt_1 a(t_1)^3 \zeta^{cl\, *}_k(t)\left(\zeta^{cl}_k(t_1)\frac{1}{2}\left(2\langle\rho_\sigma(t_1)\rangle_0-3\langle p_\sigma(t_1)\rangle_0\right)+\frac{\dot\zeta^{cl}(t_1)}{H}\langle\rho_\sigma(t_1)\rangle_0\right)\right]\ ,
\eea
where we defined
\bea\nonumber
\langle\rho_\sigma(t)\rangle_0=\langle\frac{\dot\sigma^2(\vec x,t)}{2}+\frac{(\d_i\sigma)^2(\vec x,t)}{2}\rangle\ , \\
\langle p_\sigma(t)\rangle_0=\langle\frac{\dot\sigma^2(\vec x,t)}{2}-\frac{(\d_i\sigma)^2(\vec x,t)}{6}\rangle\ ,
\eea
and where the expectation value cannot depend on $\vec x$ because of translation invariance. At this order, $H^{(3)}_{int}=-L^{(3)}_{int}$ and is given by:
\be\label{eq:action_zetasigmasigma}
{\cal{L}}_{\zeta\sigma\sigma}=\int d^3x\; a^3\sum_{n=1}^N\left[-\frac{1}{2}\left(\zeta+\frac{\dot\zeta}{ H}\right) \frac{\d_i\sigma_n\d_i\sigma_n}{a^2}+\d_i\left(\frac{\zeta}{H}+\frac{\dot H}{H^2}\frac{1}{\d^2}\dot\zeta\right)\dot\sigma_n\d_i\sigma_n+\left(\frac{3}{2}\zeta-\frac{\dot\zeta}{2 H}\right) \dot\sigma_n^2\right]\ . 
\ee
We have assumed that the expressions for $\langle\rho_\sigma(t)\rangle_0$ and $\langle p_\sigma(t)\rangle_0$, which are UV divergent, have been renormalized and have been made finite (at most dependent on the renormalization scale). Their actual value is not important.
The wavefunctions of $\zeta$ and $\sigma$ in three dimensions are:
\bea\label{eq:zeta_wave}
&&\zeta^{cl}_k=i\frac{H^2}{\left(-2 \dot H \mpl^2\right)^{1/2}}\frac{1}{k^{3/2}}\left(1+i k \eta\right)e^{-i k \eta}\ , \\ \nonumber
&&\sigma^{cl}_k=i \frac{H^2}{k^{3/2}}\left(1+i k \eta\right)e^{-i k \eta} \ .
\eea

A diagrammatic representation of this calculation is given in Fig.~\ref{fig:tadpole}, where the distinction between dashed and continuous lines will be explained in sec.~\ref{sec:alternative_diagrams}. 
\begin{figure}
\begin{center}
\includegraphics[width=4cm]{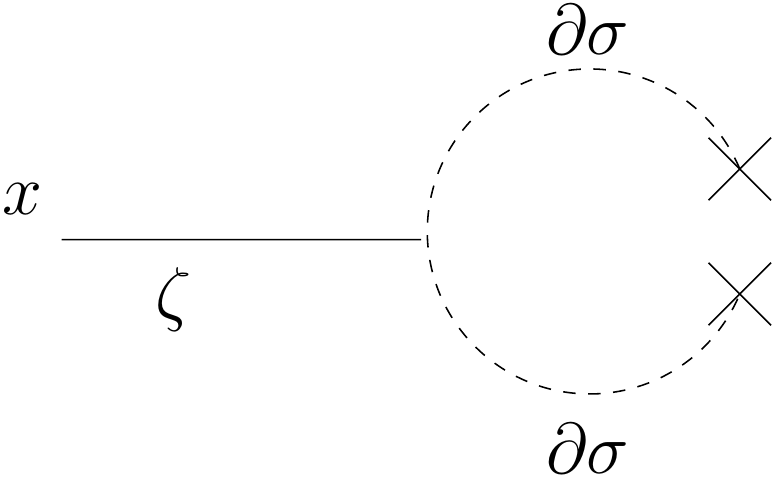}
\caption{\label{fig:tadpole} \small Tadpole diagram. Dashed lines represent correlation functions, continuos lines represents Green's functions. We will explain the origin of this notation later in sec.~\ref{sec:alternative_diagrams}.}
\end{center}
\end{figure}
The terms inside the Imaginary part in (\ref{eq:tadpole_diagram}) are not real and therefore the diagram is not zero. However, the wavefunction in (\ref{eq:zeta_wave}) diverges for $k=0$, which means that the tadpole diagram is actually infinite even after having renormalized $\langle\rho_\sigma(t)\rangle_0$ and $\langle p_\sigma(t)\rangle_0$. This is a problem that arises because  eq.~(\ref{eq:zeta_gauge}) does not fix the gauge at $k=0$. For tree-level calculations this is not important, as the zero mode of $\zeta$ is not observable. However, the tadpole diagram can be attached to a $\zeta$-propagator and affect at one-loop also correlation functions with $k\neq0$. Notice that the trilinear Lagrangian in $\zeta$, ${\cal{L}}_{\zeta^3}$ that can be found in \cite{Maldacena:2002vr} does contain vertexes of the form $\zeta (\d\zeta)^2$ and $\zeta^2\dot\zeta$ which allow for a $k=0$ mode to be attached to a propagator and give a non-zero (and actually infinite) result. 

However this problem can be dealt with in a rather straightforward way by ensuring that the zero mode of $\zeta$ is zero. This can be done by realizing that the tadpole is proportional to the expectation value of the stress-tensor of the $\sigma$ fields in the unperturbed metric. Since translation invariance forces the expectation value to depend only on $t$, this expectation value does nothing other than changing the background solution. The correct way to deal with the $\zeta$ zero-mode is therefore to define $\zeta$ as in  equation (\ref{eq:zeta_gauge}), but where in this case the background quantities (and in particular the scale factor) satisfy the following equations:

\bea\label{eq:new_background}
&&3 \mpl^2 H^2=\frac{1}{2}\dot\phi(t)^2+V\left(\phi(t)\right)+\langle\rho_\sigma(t)\rangle_0\ ,\\ \nonumber
&&\mpl^2\left(3H^2+2\dot H\right)=-\frac{1}{2}\dot\phi(t)^2+V\left(\phi(t)\right)-\langle p_\sigma(t)\rangle_0\ .
\eea
Formally, this corresponds to manipulating the Lagrangian of (\ref{action_spectator}) in the following way:
\begin{eqnarray}\label{eq:tadpole_action}
S&=&\int d^4x\; \sqrt{-g}\left[\left(\frac{1}{2}(\partial\varphi)^2-V(\phi)+\sum_{i=1}^{N_f}\frac{1}{2}g^{\mu
\nu}\langle\partial_\mu\sigma_i\partial_\nu\sigma_i\rangle\right)\right.\\ \nonumber
&&\qquad\qquad\left.+\left(\sum_{i=1}^{N_f}\frac{1}{2}g^{\mu\nu}\partial_\mu\sigma_i\partial_\nu\sigma_i-\sum_{i=1}^
{N_f}\frac{1}{2}g^{\mu\nu}\langle\partial_\mu\sigma_i\partial_\nu\sigma_i\rangle\right)\right]\ .
\end{eqnarray}
The background solution comes from imposing that the first line of (\ref{eq:tadpole_action}) starts quadratically in $\zeta$ (which just means solving the unperturbed equation of motions coming from the first line). Instead, the second term in the second line ensures that the contribution coming from contracting the two $\sigma$'s in the first term in the second line (i.e. every tadpole diagram or every subdiagram containing a tadpole subdiagram) is cancelled. This is represented pictorially in Fig.~\ref{fig:tadpole_cancellation} for the one-point and the two-point function.

\begin{figure}
\begin{center}
\includegraphics[width=10cm,]{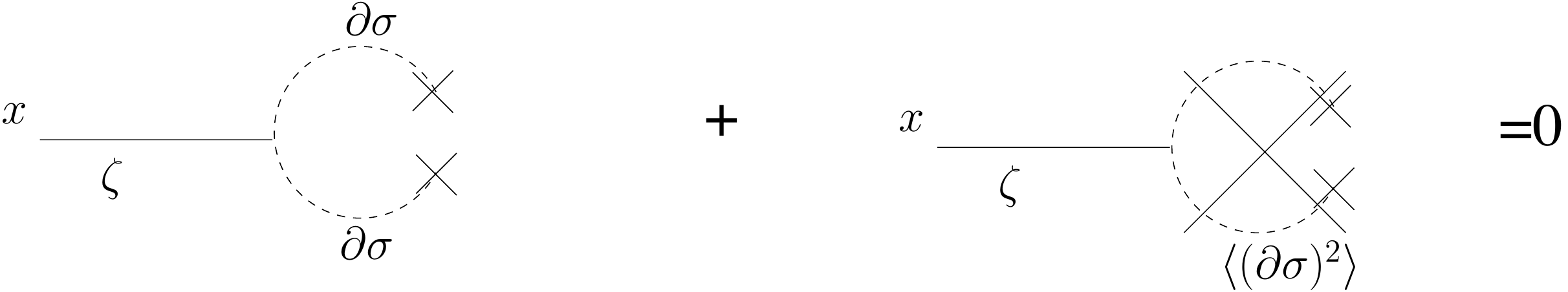}\\[0.5 in]
\includegraphics[width=14cm,]{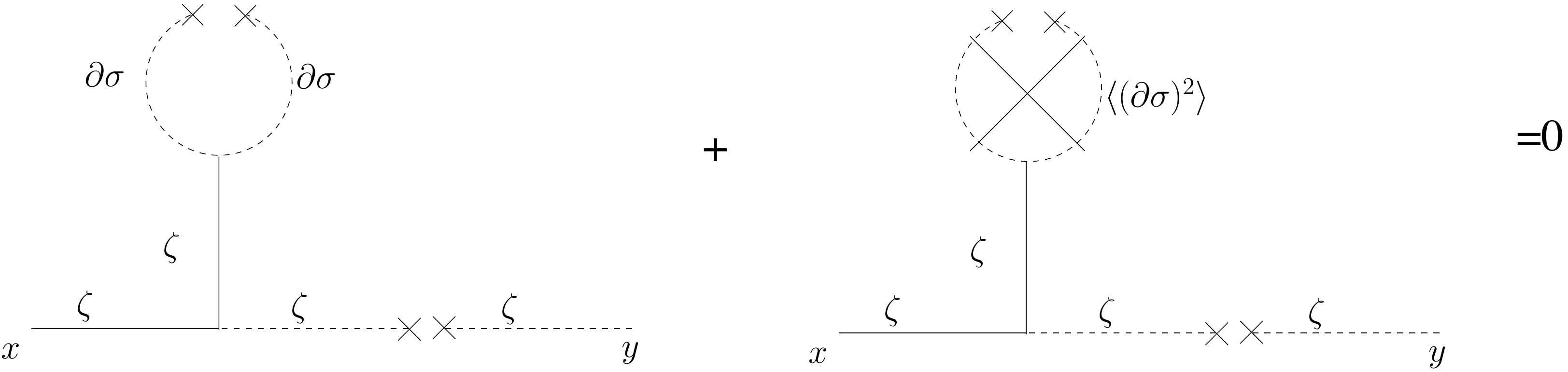}
\caption{\label{fig:tadpole_cancellation} \small Upper: Cancellation of the tadpole diagram. Lower: Cancellation of the tadpole subdiagram in a two-point function diagram. Dashed lines represent correlation functions, continuos lines represents Green's functions. We will explain the origin of this notation later in sec.~\ref{sec:alternative_diagrams}}
\end{center}
\end{figure}

\section{Time independence of $\zeta$ out of the horizon\label{sec:theorem}}

In the former sections we showed that the one-loop corrections to the two point function of $\zeta$ give rise only to logarithmic running of the form $\log(H/\mu)$, where $\mu$ is the renormalization scale. As part of the calculation we showed that loop corrections do not give rise to any additional time dependence of the two-point function of $\zeta_k$, which therefore stays constant once the mode $k$ is outside of the horizon. There is an important subtlety that was stressed by Weinberg in his original work~\cite{Weinberg:2005vy,Weinberg:2006ac}. 
It is associated with the fact that in order to compute the logarithmic running we did not need to renormalize explicitly the theory. In other words, we had simply to assume that all the divergences could be reabsorbed by  diff. invariant local counter terms in the Lagrangian. It is possible that this might not be the case for the following reason. If we take the calculation we did in dimensional regularization in sec.~\ref{sec:dim_reg_calculation} we notice that {\it first} we performed the time-integrals in the loops by sending the external time to plus infinity and then {\it second }we performed the momentum integral finding a divergency that led to a logarithmic running.  It is possible that this divergency might not be absorbable by a diff. invariant counter term in the Lagrangian.  This could happen if in doing the time integrals instead of sending the external time to infinity we were to keep it finite and the divergency only appeared as we sent this time to infinity. We are unable to say anything about this unless we perform explicitly the renormalization.


We have not ignored this issue completely in the former sections. For example when we did the regularization with a cut-off  we {\it first} performed the momentum integrals at finite external time and {\it second} we performed the time integrals and in doing this we were able to take the external time to infinity only at the very end. Since we found the same result as in dim. reg. this proved that the $\zeta$ correlation function could not depend on time and that the divergency we found should be re-absorbable by a counter term.

The only issue with this second calculation is that the cutoff in frequency and momentum space is not diff. invariant, which might lead us to suspect either that the logarithmic divergency that we found is not correct, or that there are additional time-dependent factors that were lost because of this. We believe that this is not the case, as the logarithmic running is an infrared quantity that should not be sensitive to the details of the UV regularization, particularly in our case. 

In order to further address this point, in sec.~\ref{sec:dim_reg_finite_time} we performed the calculation in the  dim. reg. for finite external time.   We found that the divergency was unaltered. This showed that the divergency was real and had to be re-absorbed by a counter-term.


The calculation at finite external time we just mentioned was done  only for the $\dot\pi^3$ theory. It is pretty clear that it can be extended to the case of the $N$ spectator $\sigma$ fields, but we have not done this explicitly. Here we will offer a proof of the constancy of $\zeta$ out of the horizon directly for the theory with $N$ spectator $\sigma$ fields. We will then generalize it to the theory with the large $\dot\pi^3$ self-interaction. We find that this additional proof improves  our intuition on loop corrections to inflationary correlation and it will further guide us in speculating on how our results are generalized to higher loops and to different interactions, which we will do in sec.~\ref{sec:massive}. This way of doing the calculation can be thought of as doing the loop calculation by first doing the momentum integrals and then doing the time integrals. Since if one does the calculation in this order there are no subtleties with the possible appearance of fake UV divergencies we can assume that the quantities we will deal with after the momentum integral are directly the renormalized, physical ones. This is why this method of solving the loop will be so intuitive: it will reduce to solving the linear equations of $\zeta$ in Fourier space working directly with the renormalized stress tensor. Though very intuitive this method will be not be powerful enough to show that there is logarithmic running in the two point function and calculate its  coefficient. But it will make very clear why the $\zeta$ correlation function is constant in time and scale invariant once the mode is outside of the horizon.

As we did in the introduction, we would like to stress that it would be a tragedy for inflation if the correlation function of $\zeta$ or of some other related operator did not become constant in time after horizon crossing because of some loop interactions involving modes of Hubble size. If this were to be the case, than we could expect the same to happen for interactions during the unknown part of the history of the Universe from the end of inflation to nucleosynthesis during which the modes relevant for observation are well outside of the horizon. In fact, as it will become clear later in this section, the quantum nature of the loops is not really relevant here: any type of fluctuations can lead to the same effect, and it is therefore extremely important to show that fluctuations on scales smaller or comparable to the horizon cannot influence modes outside the horizon. It is the constancy of $\zeta$ during the unknown parts of the history of the Universe that makes inflation, or any other theory of the early Universe, predictive.

{\bf Summary of the section.}  In order to prove the constancy of $\zeta$  we will introduce a alternative way of looking at loop corrections to correlation functions. This will take some time and therefore we would like to anticipate here the main logic and results. Concentrating on the theory with $N$ massless spectator scalar fields $\sigma$, let us start by noticing that the operator $\zeta_k$ at time $t$ is given by an 
expression of the following form:
\begin{equation} 
\label{eq:generic1}
\zeta_k(t)=\int_{-\infty}^t dt'\ G^{R;\mu\nu}_{\zeta}(k,t,t') T_{\sigma,\zeta;\mu\nu}(k,t') \ .
\end{equation}
Here $T_{(\sigma)\mu\nu}$ is the stress tensor of the $\sigma$ fields:
\begin{equation}
T_{(\sigma)\zeta;\mu\nu}(k,t)=\int d^3q\ \left(-\partial_{\mu}\sigma_{\zeta(k)}(q,t')\partial_{\nu}\sigma_{\zeta(k)}(k-
q,t')+g_{\mu\nu}\left(\zeta(k)\right)\,\partial_{\alpha}\sigma_{\zeta(k)}(q,t')\partial^{\alpha}\sigma_{\zeta(k)}
(k-q,t') \right) \ ,
\end{equation}
and the subscript $_\zeta$ represent the fact that in some cases the operators $\sigma$ are to be computed in the presence of a $\zeta$ fluctuations. $G^{R;\mu\nu}_{\zeta}(k,t,t')$ is the retarded Green's function for $\zeta$ associated to the $T_{(\sigma)\mu \nu}$. Since the $\sigma$'s interact with the inflaton only through gravitational interactions, it is $T_{(\sigma)\mu\nu}$ that sources $\zeta$. The above expression just comes form inverting the linearized Einstein Tensor (which is invertible once we fix the gauge), and in general it will be a complicated, apparently non-local expression. We will give the explicit form for $G_\zeta^R$ later in this section. Since we are interest in proving that correlation functions of $\zeta_k$ become constant when the mode $k$ is outside of the horizon, we will be interested in considering $t$ as a late time after horizon crossing as possible, for example the end of inflation. If $t^\star$ is some time after which the mode $k$ is outside of the horizon such that:
\be\label{eq:tstar}
\frac{k}{a(t^\star) H}\equiv\epsilon_{out}\ll 1\ ,
\ee
we can rewrite (\ref {eq:generic1}) as:
\begin{equation} 
\label{eq:generic2}
\zeta_k(t)=\int_{-\infty}^{t^\star} dt'\ G^{R;\mu\nu}_{\zeta}(k,t,t') T_{\sigma,\zeta;\mu\nu}(k,t')+\int_{t^\star}^t dt'\ G^
{R;\mu\nu}_{\zeta}(k,t,t') T_{\sigma,\zeta;\mu\nu}(k,t') \ ;
\end{equation}
The first term on the right-hand side picks up the contribution to $\zeta_k(t)$ generated by the sources that acted up to the time $t^\star$, i.e. while the $k$ mode is inside the horizon and as it goes out of the horizon, while the second term represents the contribution to $\zeta_k$ when $k$ is already outside of the horizon. We will analyze each of the two terms separately and we will show that they contribute to the $\zeta$ two point function in a way that is scale invariant and time independent. Let us anticipate the main result here. 

The contribution from the first term on the right-hand side can be thougth of as the free evolution of $\zeta$ as determined by some initial condition at time $t^\star$. Because the free evolution of the  $\zeta_k$ makes it freeze after horizon crossing this means that this contribution will become time-independent after horizon crossing. We will argue that this implies that the contribution is also scale invariant~\footnote{We will work initially in the approximation of de-Sitter space, and we will then generalize to deviations from exact de-Sitter.}.  

The contribution from the second term on the right-hand side of (\ref{eq:generic2}) is instead a bit more complicated to deal with. It represents how the fluctuations in the $T_{(\sigma)\mu\nu}$ source the $\zeta_k$ mode when this is well outside of the horizon. Since the $\sigma$'s have no potential term, their fluctuations are relevant only for the $\sigma$ modes smaller then the horizon. Still, the superposition of two short scale modes can lead to a long scale mode, and therefore affect $\zeta_k$. 

Concentrating on the $\zeta$ two point function, there are two kind of contributions from these terms. 
The first, that we will call cut-in-the-middle diagrams, represents how the free two-point function $T_{(\sigma)\mu\nu}$ affects the $\zeta_k$ two-point function. We will see that once the mode $k$ is well outside of the horizon, the various Hubble patches spanned by one wavelength of $\zeta_k$ become uncorrelated. This means that the $T_{(\sigma)\mu\nu}$ fluctuations quickly average out and are not able to provide the coherent effect that would be necessary to source $\zeta_k$ on large scales. $\zeta_k$ becomes therefore time-independent once outside of the horizon, and again the symmetry under rescaling of the scale factor $a$ will force the contribution from these terms to be scale invariant.

The second kind of contribution from the right-hand side of (\ref{eq:generic2}) comes from computing the expectation value of $T_{(\sigma)\mu\nu}$ in the presence of a background $\zeta_k$. The background $\zeta_k$ can be correlated with a freely evolving $\zeta_k$ to obtain a two-point function. We will refer to this contribution as the cut-in-the-side diagrams. The reason why this contribution shuts down as the $k$ modes go outside of the horizon is that the perturbation to the $T_{(\sigma)\mu\nu}$ due to the presence of a background $\zeta$ goes to zero. This happens because the background $\zeta$ at this order in perturbation theory evolves freely. When a free $\zeta$ mode is outside of the horizon it becomes constant and a simple redefinition of the scale factor $a$ which is locally unobservable. This means that the $\zeta$ mode cannot induce any physical perturbation to  $T_{(\sigma)\mu\nu}$. This implies that the effect of this term is again time independent, and due to the usual symmetry under rescaling of $a$ it is scale invariant.

This will conclude the altenative proof that $\zeta$ is time-independent and scale invariant  out of the horizon for the case of the $N$ spectators $\sigma$ fields. Finally, we will be able to comment on slow roll corrections to our calculation and extend this particular proof to the case of the large $\dot\pi^3$ self-interaction.

\subsection{An alternative diagrammatic expansion for loop corrections\label{sec:alternative_diagrams}}

We start the second proof of the constancy of $\zeta_k$ when it is outside of the horizon by presenting an alternative way of organizing the calculation.   This approach was originally developed in \cite{Musso:2006pt} for a restricted set of theories, and it was noted in \cite{Adshead:2008gk} that the derivation was not consistent with the $i\,\epsilon$ prescription for choosing the interacting vacuum in the past. Here we will generalize the approach of \cite{Musso:2006pt} to more generic theories and we will show how the correct $i\,\epsilon$ prescription can be implemented.

For concreteness let us specialize to the $\zeta$ two-point function. We start by taking expression (\ref{eq:generic_expectation}) and inserting the unit operator
\be
U_{int}(t,-\bar T) U^{-1}_{int}(t,-\bar T)\ , \qquad U_{int}(t,-\bar T)=T e^{-i \int_{-\bar T}^tdt'\;  H_{int}(t')}\ ,
\ee
between the two $\zeta$'s, to obtain
\be\label{eq:generic_expectation_new}
\langle\zeta^2(t)\rangle=\langle\left( U^{-1}_{int}(t,-\infty_-) \zeta(t) U_{int}(t,-\bar T) \right)\left(U^{-1}_{int}(t,-\bar T) \zeta(t)U_{int}(t,-\infty_+) \right)\rangle\ ,
\ee
Here $\bar T$ is some arbitrary early time in the past, and we stress that ${\it no}$ rotation of the contour of integration is performed for this operator. Ignoring for a moment the issue of the $i\, \epsilon$ prescription, we have written the expectation of the operator $\zeta(t)^2$ as the product of the two freely evolved $\zeta(t)$'s each evolved with the interaction picture time evolution operator $U_{int}$. In other words, the $\zeta(t)^2$ correlation function is simply given by the correlation function of the evolved $\zeta(t)$'s. A closer look at (\ref{eq:generic_expectation_new}) might let us think that this picture does not quite work because of the difference in the extremes and the contour of integration of the time integrals in the various evolutors. In order to make it clear that these differences do not play any significant role, we can deform the contour of integration for the two external $U_{int}$ to obtain:
\be
\langle\zeta^2(t)\rangle=\langle U^{-1}_{int}(-\bar T,-\infty_-)\left(U^{-1}_{int}(t,-\bar T) \zeta(t) U_{int}(t,-\bar T) \right)\left(U^{-1}_{int}(t,-\bar T) \zeta(t)U_{int}(t,-\bar T)\right)U_{int}(-\bar T,-\infty_+) \rangle\ .
\ee
The two most external $U_{int}$ have the function of projecting the free vacuum into the interacting vacuum, while the rest of the $U$'s evolve each $\zeta$ from the time $-\bar T$ to the time $t$. Notice that, in this approach, we cannot take the time $-\bar T$ to $-\infty$ from the beginning of the calculation because we would lose the capability of projecting on to the true vacuum. We will comment shortly of the physical meaning of this point.

Going back to eq.~(\ref{eq:generic_expectation_new}), we can Taylor expand in $H_{int}$ to obtain
\bea\label{eq:generic_expectation_new_expanded}
&&\langle\zeta^2(t)\rangle=\\ \nonumber
&&=\langle\left(\sum_{N=0}^\infty i^N\int^t dt_N\int^{t_N}dt_{N-1}\ldots \int^{t_2}dt_1 \left[H_{int}(t_1),\left[H_{int}(t_2),\ldots  \left[H_{int}(t_N),\zeta(t)\right]^{-\infty_-}_{-\bar T}\ldots\right]^{-\infty_-}_{-\bar T}\right]^{-\infty_-}_{-\bar T}\right)\\ \nonumber 
&&\times \left(\sum_{N=0}^\infty i^N\int^t dt'_N\int^{t_N}dt'_{N-1}\ldots \int^{t_2}dt'_1 \left[H_{int}(t'_1),\left[H_{int}(t'_2),\ldots  \left[H_{int}(t'_N),\zeta(t)\right]^{-\infty_-}_{-\bar T}\ldots\right]^{-\infty_-}_{-\bar T}\right]^{-\infty_-}_{-\bar T}\right)^\dag\rangle\ 
\eea
where 
\be
\int^{t}dt' \left[H_{int}(t'),\zeta(t)\right]^{-\infty_-}_{-\bar T}\equiv\int^{t}_{-\infty_-}dt'\;H_{int}(t')\zeta(t)-\int^{t}_{-\bar T}dt'\;\zeta(t)H_{int}(t')\ .
\ee
Expanding (\ref{eq:generic_expectation_new_expanded}) up to second order in $H_{int}$, we obtain
\be
\langle\zeta^2(t)\rangle=\langle\zeta^2(t)\rangle_{CIS}+\langle\zeta^2(t)\rangle_{CIM}\ ,
\ee
where we  have defined
\bea\nonumber\label{eq:CIMandCIS}\nonumber
\langle\zeta^2(t)\rangle_{CIS}&=&-2\,{\rm Re}\left[\left( \int^{t}dt_{2} \int^{t_2}dt_1 \langle\left[H_{int}^{(3)}(t_1),\left[H_{int}^{(3)}(t_2),\zeta(t)\right]^{-\infty_-}_{-\bar T}\right]^{-\infty_-}_{-\bar T}\right)\zeta(t)\rangle\right.\\  \label{eq:CIS_diagram}
&&\left.\qquad \qquad + i \left( \int^{t}dt_{1} \langle\left[H^{(4)}_{int}(t_1),\zeta(t)\right]^{-\infty_-}_{-\bar T}\right)\zeta(t)\rangle\right]\ , \nonumber \\  \label{eq:CIM_diagram}
\langle\zeta^2(t)\rangle_{CIM}&=&-\left( \int^{t}dt_1\langle \left[H_{int}^{(3)}(t_1),\zeta(t)\right]^{-\infty_-}_{-\bar T}\right)\left( \int^{t}dt'_1 \left[H_{int}^{(3)}(t'_1),\zeta(t)\right]^{-\infty_-}_{-\bar T}\rangle\right)^\dag
\eea
The subscript $_{CIS}$ denotes what we call cut-in-the-side diagram, while $_{CIM}$ denotes cut-in-the-middle diagram. If we remind ourselves that the $\zeta$ retarded Green's function is given by
\be
G^R_\zeta(x,x')=i\theta(t-t')\left[\zeta(x),\zeta(x')\right]\ ,
\ee
where the commutator on the right is taken on the free fields, we can identify the structure we anticipated in eq.~(\ref{eq:generic1}), with the additional subtlety that each term of the commutator is taken on a different time path (see eq.~(\ref{eq:CIMandCIS})). Alternatively, as in eq.~(\ref{eq:generic_expectation_new}), one can deform the contour of integration for the integral that goes up to $-\infty$, in such a way that it goes to $-\bar T$ on the real axis, and then goes from there to $-\infty$ on the imaginary axis. In this way, we can treat the integration from $-\infty$ to $-\bar T$ on the imaginary axis as the projection of the state at time $-\bar T$ onto the interacting vacuum, and we can interpret the diagrams in (\ref{eq:CIMandCIS}) as representing the evolution of the operators from time $-\bar T$ to time $t$ starting in the interacting vacuum. This subtlety will be largely irrelevant in proving the theorem. 

By specializing to the case of the $N$ spectator $\sigma$ fields, we can see that there are two ways in which the equation solution for $\zeta$ can be perturbed by interactions of the form $\zeta\sigma^2$. The first, which corresponds to the $CIM$ diagrams, is by considering vacuum fluctuations of two $\sigma$ fields, that combined in a $T_{(\sigma),\mu\nu}$ generate a $\zeta$ mode which then propagates freely.  The correlation of two of these terms is the $CIM$ diagram represented in Fig.~\ref{LeftRight2}. The notation is such that a dashed line corresponds to a free field, while a continuous line represents a retarded Green's function. The crosses represent correlation of free fields (two crosses have to be contracted together in order for a diagram not to be zero).
\begin{figure}
\begin{center}
\includegraphics[width=7cm]{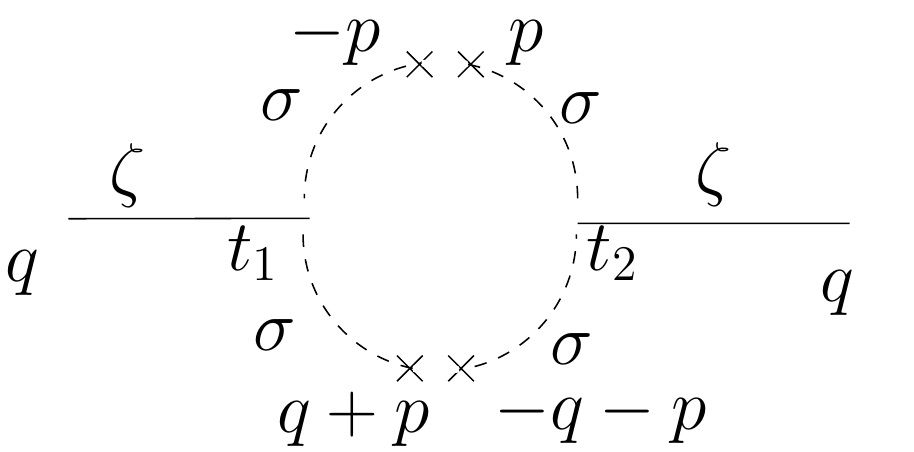}
\caption{\label{LeftRight2} \small\it Cut-in-the-middle diagrams. Continuos lines represents Green's functions, dashed lines represent free fields, and crosses represent correlations of free fields. Two crosses have to be contracted together in order for the diagram not to be zero.}
\end{center}
\end{figure}

The $CIS$ diagram represents instead the correction to the $\zeta$ correlation function due to the perturbation to the evolution of a primordial $\zeta$ fluctuation caused by the interaction with the $\sigma$ fields. It is represented in Fig.~\ref{LeftLeft2}. We see that an original $\zeta$ vacuum fluctuation interacts via the $\zeta\sigma^2$ interaction with a $\sigma$ vacuum fluctuation and modifies that fluctuation. This fluctuation evolves for some time and then interacts again  with another $\sigma$ vacuum fluctuation, generating a $\zeta$ fluctuation that then propagates freely up to time $t$. The correlation among the original $\sigma$ vacuum fluctations and of the original $\zeta$ fluctuation with another $\zeta$ fluctuation that instead propagates undisturbed up to present gives rise to the diagram. Notice that the subdiagram on the left of the last $\zeta$ Green's function is nothing but the expectation value of the $T_{(\sigma)\mu\nu}$ at time $t_1$ in the presence of a background $\zeta$ mode. 
\begin{figure}
\begin{center}
\includegraphics[width=10cm]{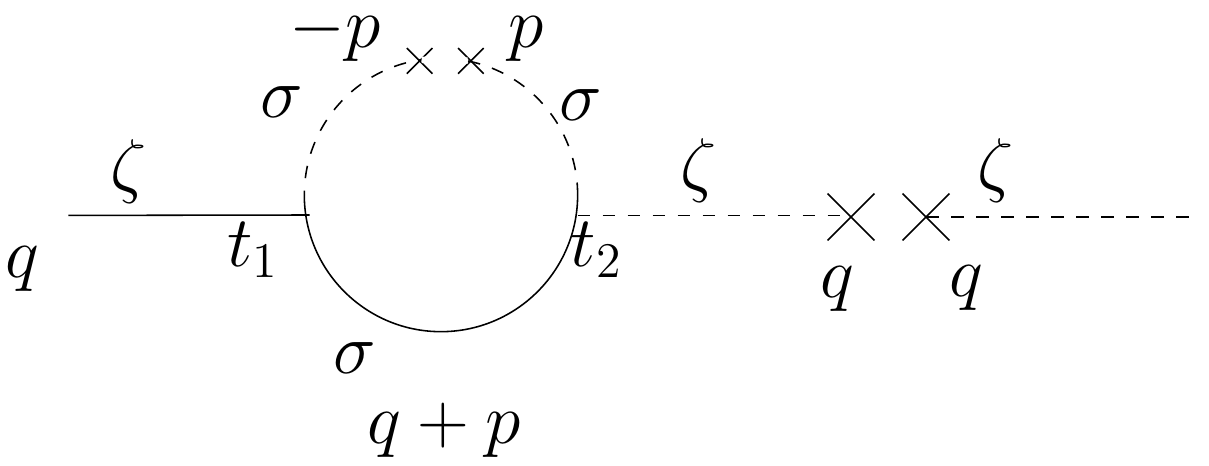}
\\[0.3 in]
\includegraphics[width=10cm]{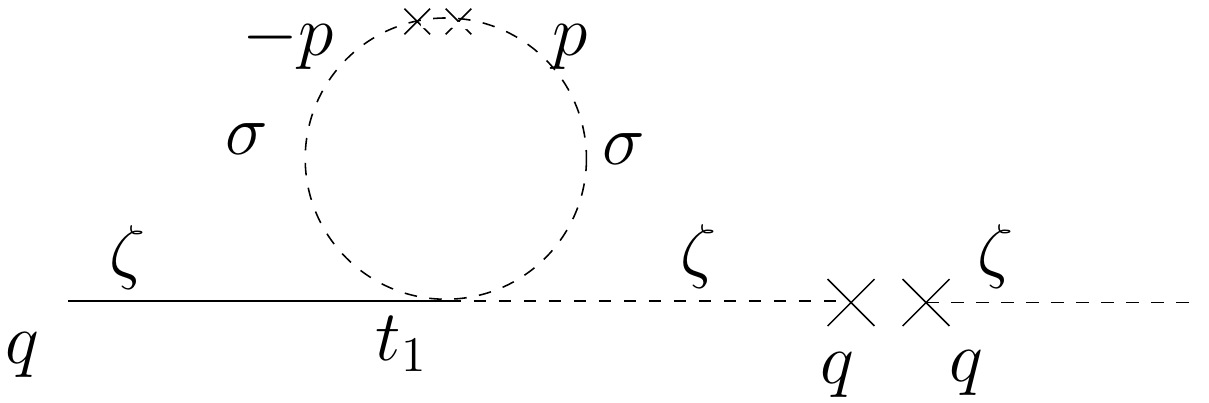}
\caption{\label{LeftLeft2} \small\it Cut-in-the-side diagrams. Continuos lines represents Green's functions, dashed lines represent free fields, and crosses represent correlations of free fields. Two crosses have to be contracted together in order for the diagram not to be zero.}
\end{center}
\end{figure}

The names cut-in-the-side diagram and cut-in-the-middle diagram are meant to point out the fact that in the $CIM$ diagrams both of the $\zeta$ operators undergo a perturbed evolution, while in the $CIS$ diagram it is only one of the two $\zeta$'s that is perturbed. Though here we have concentrated only on the 1PI diagrams with insertion of the cubic interaction Hamiltonian, it is straightforward to include also the quartic and the non 1PI diagrams. Clearly, the $CIM$ and the $CIS$ diagrams are physically well distinct: for example, the $CIS$ diagram is proportional to the primordial $\zeta$ vacuum fluctuations, and goes to zero in the limit that there are no $\zeta$ primordial fluctuations. On the other hand, the $CIM$ diagrams are independent of the primordial $\zeta$ fluctuation. In generic situations, these two diagrams can give parametrically distinct results. However in the case we are considering here where the $\sigma$ and the $\zeta$ fields are massless and undergo quantum vacuum fluctuations the two terms are comparable. In fact in the interacting theory the $\sigma$ and $\zeta$ fields  are mixed and both diagrams give equal size contributions. Equivalently although the overall result is independent of the time $T$ that  dependence only cancels in the sum of the $CIS$ and the $CIM$ diagrams. 

We will find the organization  scheme and the physical intuition provided by these diagrams extremely useful for proving the time-independence of the $\zeta_k$ correlation function once the mode $k$ goes very well outside of the horizon. Furthermore  in sec.~\ref{sec:massive} the intuition we will develop will allow us to generalize our results in various ways. 

\subsection{Early time contribution}

Let us now begin to analyze the two contributions from (\ref{eq:generic2}), and let us start from the first term on the right-hand side, that we can rewrite as 
\begin{equation} 
\label{eq:generic3}
\zeta_{k,1}(t)=\int_{-\infty}^{t^\star} dt'\ G^{R;\mu\nu}_{\zeta}(k,t,t') T_{(\sigma)\zeta;\mu\nu}(k,t')\ ,
\end{equation}
where the subscript $_1$ has been introduced to represent that we are talking of the contribution to $\zeta$ from the first term on the right-hand side of eq.~(\ref{eq:generic2}). Notice that since in this subsection we will automatically consider the sum of the $CIM$ and $CIS$ diagram, the subtlety about the two different time-integration paths in the Green's function in~(\ref{eq:CIMandCIS}) is irrelevant.  In~(\ref{eq:generic3}), the source is allowed to act only for the time up to $t^\star$ when the mode $k$ begins to be well outside of the horizon, and therefore represents how $\zeta$ is affected by early interactions.  But exactly because of this it is easy to realize that as far as this contribution is concerned the source will just create a certain $\zeta_k(t^\star),\;\dot\zeta_k(t^\star)$ at time $t^\star$ which will serve as initial conditions for the subsequent free evolution. Therefore, this term is equivalent to study the time dependence of a freely evolved $\zeta(t)$ given some initial condition $\zeta_k(t^\star),\,\dot\zeta_k(t^\star)$. The solutions in exact de-Sitter space are~\footnote{We will generalize later our results to deviations from de-Sitter.}:
\be
\zeta_k(t)=\zeta^{cl}_{k,1}(t) a_{\vec k}+\zeta^{cl,*}_{k,1}(t) a_{\vec k}^\dag\ ,
\ee
where the wavefunction $\zeta^{cl}_1$ is given by:
\be \label{eq:zeta_early}
\zeta_{k,1}^{cl}(t)=A_1(\zeta^\star,\dot\zeta^\star,k) \frac{1}{k^{3/2}}\left(k \eta \cos(k\eta)-\sin(k\eta)\right)+A_2(\zeta^
\star,\dot\zeta^\star,k) \frac{1}{k^{3/2}}\left( \cos(k\eta)+k \eta\sin(k\eta)\right)
\ee
where $A_1(\zeta^\star,\dot\zeta^\star,k)\ ,$ and $A_2(\zeta^\star,\dot\zeta^\star,k) $ are the two integration constants determined by $\zeta_k^\star$ and $\dot\zeta^\star_k$ at time $t^\star$, and $\eta$ is conformal time. Independently of the value of $A_{1,2}$, the  solution for $\zeta_k$ goes to a constant value exponentially fast in cosmic time $t$.

The symmetries of de-Sitter space suggest that  we demand that the contribution to the variance of $\zeta$ in physical real space that is given by modes between two fixed physical scales (both smaller that $H^{-1}$) should be independent of time. Satisfying this constraint implies that the coefficients $A_{1,2}$ should be independent of $k$ which then results in a scale invariant spectrum for $\zeta$. We thus expect that the $k$ dependence of $A_{1,2}$ is slow-roll suppressed and that the power spectrum of $\zeta$ is very close to scale invariant.

\subsection{Late time contribution: cut-in-the-middle diagrams \label{sec:CIMdiagrams}}

Let us now analyze the contribution from the second part of eq.~(\ref{eq:generic2}), that we denote with the subscript~$_2$: 
\begin{equation} 
\label{eq:generic4}
\zeta_{k,2}(t)=\int_{t^\star}^{t} dt'\ G^{R;\mu\nu}_{\zeta}(k,t,t') T_{(\sigma)\zeta;\mu\nu}(k,t')\ .
\end{equation}
The contribution from these terms represents how the small scale $\sigma$'s affects $\zeta$ even when the mode $k$ is well outside of the horizon. It is worth stressing that $T_{(\sigma)\zeta}(k)$ is not zero in this regime. Two small scale $\sigma$ modes can combine to create a low-$k$ $T_{(\sigma)\zeta}(k)$.


Let us start with the $CIM$ diagrams, where  $T_{(\sigma)}$ is evaluated in the unperturbed metric. In this case it is straightforward to find the equation of motion relating the $\sigma$'s to $\zeta$ because the $\sigma$'s live in an unperturbed metric. Further, since, as it will become evident soon, the $CIM$ diagrams involve the self-correlation of four $\sigma$'s, we can neglect all the terms of order  ${\cal{O}}\left(\zeta\sigma^2\right)$ from the equations of motion. The simplest way to find this equation is to use the action we wrote earlier in (\ref{eq:action_zetasigmasigma}) at order  ${\cal{O}}\left(\zeta\sigma^2\right)$  and to derive the equation of motion for $\zeta$. Alternatively, one can solve the $(0,0)$ and $(0,i)$ Einstein equations to find $N$ and $N^i$ at order $\zeta$ and $\sigma^2$ (but not at order $\zeta \sigma^2$), and plug these back into the continuity equation
\be
\nabla_\mu T^{\mu,0}=0\ .
\ee
One finds:
\bea\label{eq:constraints}
&&N=1+\frac{\dot\zeta}{H}+\frac{1}{2\mpl^2 H}\frac{1}{\d^2}\d^i\left(\dot\sigma\d_i\sigma\right)\ , \\ \nonumber
&&\d_iN^i=-\frac{\dot H}{H^2}\dot\zeta-\frac{1}{H}\frac{\d^2}{a^2}\zeta-\frac{1}{4\mpl^2H^2}\left(\dot\sigma^2+\frac{1}{a^2}(\d_i\sigma)^2\right)-\frac{(3H^2+\dot H)}{2\mpl^2 H^2}\frac{1}{\d^2}\d_i\left(\dot\sigma\d_i\sigma\right)\ ,
\eea
and the resulting equation for $\zeta$ is
\bea\label{eq:zeta}
&&\ddot\zeta+\left(3H-2\frac{\dot H}{H}+\frac{\ddot H}{\dot H}\right)\dot\zeta-\frac{\d^2}{a^2}\zeta-\frac{1}{4\mpl^2}\left(\dot\sigma^2 +\frac{(\d_i\sigma)^2}{a^2}\right)\\ \nonumber 
&&+\frac{1}{\mpl^2}\left(\frac{\d_t}{2}+\frac{3}{2}H-\frac{\dot H}{H}+\frac{\ddot H}{2\dot H}\right)\frac{1}{\d^2}\d^i\left(\dot\sigma\d_i\sigma\right)=0\ .
\eea
Notice that in all of the above expressions (\ref{eq:constraints}) and (\ref{eq:zeta}) the terms of ${{\cal{O}}(\sigma^2)}$ should be understood as having the zero mode subtracted, a result of the tadpole subtraction procedure. 

The above equation can also be written in terms of $T_{(\sigma)\mu\nu}$:
\bea\label{eq:zeta_Tmunu}
&&\ddot\zeta+\left(3H-2\frac{\dot H}{H}+\frac{\ddot H}{\dot H}\right)\dot\zeta-\frac{\d^2}{a^2}\zeta-\frac{1}{2\mpl^2}T_{(\sigma)00}+\frac{1}{\mpl^2}\left(\frac{\d_t}{2}+\frac{3}{2}H-\frac{\dot H}{H}+\frac{\ddot H}{2\dot H}\right)\left(\frac{1}{\d^2}\d_iT_{(\sigma)0i}\right)=0\ .\nonumber \\ &&\ \ \quad
\eea

We are interested in solving this equation when the mode $k$ of $\zeta$ is well outside the horizon. We can therefore expand in powers of the external momentum (keeping track of  the difference between $\d_i(\dot\sigma\d^i\sigma)$ and $\d_i\dot\sigma \d^i\sigma$), and keep only the leading terms. We obtain:

\bea\label{eq:zeta_late}
&&\ddot\zeta+3H\dot\zeta-\frac{1}{4\mpl^2}\left(\dot\sigma^2 +\frac{(\d_i\sigma)^2}{a^2}\right)+\frac{3H}{2\mpl^2}\frac{1}{\d^2}\d^i\left(\dot\sigma\d_i\sigma\right)=0
\eea

Notice that at late time there are two sources for $\zeta$: $T_\sigma^{00}$, which represents the energy density of the $\sigma$'s, and $\frac{1}{\d^2}\d_iT_\sigma^{0i}$, which represents a non-local term proportional to the divergency of the velocity of the $\sigma$'s.

Eq.~(\ref{eq:zeta_late}) can be integrated to give:
\be\label{eq:zeta_out_horizon}
\zeta_2(k,t)=\int_{t^\star}^t dt' \frac{1}{6\mpl^2 H}\left(1-\frac{a(t')^3}{a(t)^3}\right) \left.\left(-\frac{1}{2}\left(\dot\sigma_\zeta^2 +\frac{(\d_i\sigma_\zeta)^2}{a^2}\right)+3H\frac{1}{\d^2}\d^i\left(\dot\sigma_\zeta\d_i\sigma_\zeta\right)\right)\right|_{k}\ ,
\ee
where we have used the fact that we are interested in the contribution of the source from $t^\star$ onwards. Notice how this expression is indeed of the form of eq.~(\ref{eq:generic4}).

We are now ready to begin the computation of the $CIM$ diagram.   We have:
\bea\label{eq:cut-in-the-middle}
&&\langle\zeta_{2,k}(t)\zeta_{2,k'}(t)\rangle_{CIM}=\int^t_{t^\star} dt_1\int^t_{t^\star} dt_2\; \left( \frac{1}{6\mpl^2 H}\right)^2\left(1-\frac{a(t_1)^3}{a(t)^3}\right)\left(1-\frac{a(t_2)^3}{a(t)^3}\right) \times\\ \nonumber
&& \langle\left.\left(-\frac{1}{2}\left(\dot\sigma^2 +\frac{(\d_i\sigma)^2}{a^2}\right)+3H\frac{1}{\d^2}\d^i\left(\dot\sigma\d_i\sigma\right)\right)(t_1)\right|_{k} \left.\left(-\frac{1}{2}\left(\dot\sigma^2 +\frac{(\d_i\sigma)^2}{a^2}\right)+3H\frac{1}{\d^2}\d^i\left(\dot\sigma\d_i\sigma\right)\right)(t_2)\right|_{k'}\rangle\ .
\eea

There are six correlation functions to analyze. Let us start with the one involving
\be
\langle(\dot\sigma^2)_k(t)(\dot\sigma^2)_{k'}(t')\rangle\ ,
\ee
We notice that this operator is divergent, and here and in the rest of the section, the operators appearing in the correlation functions should be understood to be regularized. Notice that here we are not interested on the numerical value of these operators, but just on how they affect the late time behavior of $\zeta$. The fact that they are regularized will be enough to show that they cannot induce a time dependence on $\zeta$~\footnote{Notice that we can even assume that these quantities have been renormalized: contrary to the case where we integrate first in time up to plus infinity and then in momentum, here we are doing the integration in opposite order, and therefore all UV divergencies are true UV divergencies of the theory and have to be reabsorbable either by a counterterm or by a field redefinition.}. At this point we can also notice that $\dot\sigma^2$ has a tadpole in de-Sitter space which renormalizes  the background Universe and its contribution to $\zeta$ is cancelled as explained  in sec.~\ref{sec:tadpole_cancellation}. We can therefore concentrate only on the connected part of the correlation $\langle(\dot\sigma^2)_k(t)(\dot\sigma^2)_{k'}(t')\rangle$.

We can rewrite $\langle(\dot\sigma^2)_k(t)(\dot\sigma^2)_{k'}(t')\rangle$ as
\bea
\langle(\dot\sigma^2)_k(t)(\dot\sigma^2)_{k'}(t')\rangle&=&\int d^3x\int d^3x' e^{i(\vec k\cdot \vec x+\vec k'\cdot \vec x')}\langle(\dot\sigma^2)(\vec x,t))(\dot\sigma^2)(\vec x',t')\rangle\\ \nonumber
&=&\delta^{(3)}(\vec k+\vec k')\int d^3x_- e^{2 i \vec k\cdot \vec x_-}\langle(\dot\sigma^2)(\vec x_-,t)(\dot\sigma^2)(0,t')\rangle\ ,
\eea
where $\vec x_-=|\vec x-\vec x'|/2$. Since we are interested in small $k$'s if this integral converges even for $k=0$ then we can set the oscillating exponential to one, and do the integral. In this way we will obtain a result which is exact up to corrections of order $k/(a(t^\star)H)\ll1$.
We are therefore led to evaluate
\be\label{eq:sigmasigmalate}
\langle(\dot\sigma^2)_{k\simeq0}(t)(\dot\sigma^2)_{k'\simeq0}(t')\rangle=\delta^{(3)}(\vec k+\vec k')\int d^3x_-\langle(\dot\sigma^2)(\vec x_-,t)(\dot\sigma^2)(0,t')\rangle\ .
\ee
In order to do this, we take the correlation function $\langle(\dot\sigma)(\vec x_-,t)(\dot\sigma^2)(0,t')\rangle$ which can be constructed from $\langle\sigma(\vec x_-,t)\sigma(0,t')\rangle$, and take the large distance limit. The resulting expression is
\be\label{eq:approximate}
\langle\sigma(\vec x_-,t)\sigma(0,t')\rangle\sim H^2\log(\Lambda_{IR}\Delta x^2)\ ,\qquad H \Delta x\gg 1 \ .
\ee
where $\Delta x^2= -(\eta-\eta')^2+x_-^2$ and $\Lambda_{IR}$ is the IR cutoff.
Notice that at equal times this is the standard logarithmic dependence of massless scalar fields in de-Sitter space. If we define
\be
I(t,t')\equiv\int d^3x_- \langle(\dot\sigma^2)(\vec x_-,t)(\dot\sigma^2)(0,t')\rangle\ ,
\ee
then we have that at large distances 
\be
\langle\dot\sigma^2(\vec x_-,t)(\dot\sigma^2)(0,t)\rangle\sim \frac{H^4}{a(t)^4x_-^4} 
\ee
which tells us that $I(t,t)$ is convergent in the infrared. Since the expression is regularized (or even renormalized), then it converges even in the UV. This implies that $I(t,t)$ is finite. Notice that in order for $I(t,t)$ to be compatible with the symmetry
\be
a\rightarrow\lambda\, a\ ,\qquad x\rightarrow x/\lambda\ , \qquad k\rightarrow \lambda\, k\ ,
\ee
it must be proportional to $1/a(t)^3$.  Dimensional analysis fixes the powers of $H$, and we conclude that $I(t,t')$ goes as
\be\label{eq:Iexpr}
I(t,t)\sim  \frac{c_{UV} H^{5}}{a(t)^3}\ .
\ee
where $c_{UV}$ is a number (expected to be or order one) that depends on the regularization and the renormalization procedure, and that controls what is the strength of the correlations of $\dot\sigma^2$. 

Eq.~(\ref{eq:Iexpr}) tells us that the correlation function of $\dot\sigma^2$ is Poisson like distributed, {\it i.e.} its power spectrum is $k$ independent. This is enough for us to be able to  show that the contribution to $\zeta$ from the cut-in-the-middle diagram is time independent and therefore scale invariant.

In fact by using  that
\be
\left| I\left(t,t'\right)\right|\leq\sqrt{\left| I\left(t,t\right)I\left(t',t'\right)\right|}
\ee
we can write the contribution of this term to eq.~(\ref{eq:cut-in-the-middle}) as follows
\bea\nonumber
&&\langle\zeta_{2,k}(t)\zeta_{2,k'}(t)\rangle=
\delta^{(3)}(k+k')\int_{t^\star}^t dt_1\int_{t^\star}^t dt_2 \left( \frac{1}{24\mpl^2 H}\right)^2\left(1-\frac{a(t_1)^3}{a(t)^3}\right)\left(1-\frac{a(t_2)^3}{a(t)^3}\right)I(t_1,t_2) \\ \nonumber
&&\leq\delta^{(3)}(k+k')\int_{t^\star}^t dt_1\int_{t^\star}^t dt_2 \left( \frac{1}{24\mpl^2 H}\right)^2\left(I(t_1,t_1)I(t_2,t_2)\right)^{1/2} \\ \nonumber
&&\sim\delta^{(3)}(k+k')\left( \frac{1}{\mpl^2 H}\right)^2 c_{UV} H^5\int_{t^\star}^t dt_1\int_{t^\star}^t dt_2\frac{1}{a(t_1)^{3/2}a(t_2)^{3/2}}\\
&&\sim\delta^{(3)}(k+k') \frac{c_{UV} H}{\mpl^4}  e^{-3H t^\star}\ ,
\eea
where in the last passage we have neglected numerical factors. Eq.~(\ref{eq:tstar}) implies
\be
t^\star=\frac{1}{H}\log\left(\frac{k}{H\epsilon_{out}}\right)\ ,
\ee
so we obtain:
\be\label{eq:cut-middle-temp}
\langle\zeta_{2,k}(t)\zeta_{2,k'}(t)\rangle_{CIM}\lesssim \delta^{(3)}(k+k')\frac{1}{k^3}\frac{H^4}{\mpl^4}c_{UV}\,\epsilon_{out}^3 N  \ ,
\ee
where we have reinserted the factor of $N$ associated to the number of $\sigma$ fields.
We see that the contribution from  $\langle(\dot\sigma^2)_k(t)(\dot\sigma^2)_{k'}(t')\rangle$ to the cut-in-the-middle diagrams is time-independent and scale invariant. 

It is worth giving some physical interpretation of this result. It is quite easy to understand why the result is time-independent. In order for the $CIM$ diagram to be able to induce a time dependence on $\zeta_k$, it is necessary for the self correlation of $T_{(\sigma)\mu\nu}$ to be coherent on a wavelength of order $a(t)/k$ as the external time is taken to infinity. However, since the $\sigma$ fields are massless, the relevant fluctuations are only at most of order Hubble in size, and therefore $T_{(\sigma)\mu\nu}$ becomes quickly uncorrelated as the mode $k$ spans many Hubble regions. This is why the main effect from the $CIM$ diagram is peaked at the smallest time $t^\star$ when the mode $k$ spans the smallest number of independent Hubble patches. 
Notice in fact how the signal scales like $\epsilon_{out}^3$ which is approximately the inverse of the number of independent Hubble patches in a box of radius $a(t^*)/k$. Once the  correlation function is time independent the symmetries of the problem forces it to be scale invariant. If we go back to real space, we have
\be
\langle\zeta_{2}(\vec x, t)^2\rangle_{CIM}\lesssim \frac{H^4}{\mpl^4}c_{UV}N\,\epsilon_{out}^3 \int d^3k\; \frac{1}{k^3}
\ee
which is invariant under rescaling of $k$.

It is easy to see that the contributions  from all the other terms in eq.~(\ref{eq:cut-in-the-middle})  behave in a very similar way, and results in a time-independent and scale-invariant correlation function.  We compute explicitly the contribution from the most interesting additional terms in App.~\ref{app:CIMcorrelations}. We  conclude that the contribution from the $CIM$ diagrams are time-independent and scale-invariant and we proceed to  study of the cut-in-the-side diagrams.

\subsection{Late time contribution: cut-in-the-side diagrams\label{sec:CISdiagrams}}

We now proceed to the study of the contribution from the $CIS$ diagrams. These diagrams require to compute the expectation value of the operator stress-energy tensor  $T_{(\sigma)\zeta;\mu\nu}$ of the $\sigma$'s due to the interaction with a $\zeta$ fluctuation at some earlier time. Let us represent this as 
\be\label{eq:tmunuexpt}
\langle T_{(\sigma)\mu\nu}(\vec x,t)\rangle_{\zeta{_k}}\ .
\ee 
Due to the divergencies involved in this calculation the expression of  $T_{(\sigma)\zeta;\mu\nu}$ needs to be regularized. As in the former subsection we think of $T_{(\sigma)\mu\nu}$ directly as the regularized, or even renormalized, stress tensor at that time. The one-loop $CIS$ diagram is represented in Fig.~\ref{LeftLeft2} where the vertex labeled by $t_2$ represents the perturbation to a  $\sigma$ mode due to the interaction with the primordial $\zeta$ mode.  There are three important things to notice that will make the proof possible even without having to explicitly perform the computation of the expectation value in (\ref{eq:tmunuexpt}). The first is that, because of translation invariance, the wavenumber $k$ of the $\zeta$ that perturbs the operator $\langle T_{(\sigma)\mu\nu}\rangle_{\zeta{_k}}$ must be the same as the external $\zeta$ for which we are computing the expectation value, and therefore it is very outside the horizon for the times of interest here: $t>t^\star$. The second important thing to notice is that because the $\sigma$'s appear always with derivatives the $\sigma$ modes contributing to the finite part of  $\langle T_{(\sigma)\mu\nu}(\vec x,t)\rangle_{\zeta{_k}}$ have a momentum of order $H$ or larger. 
The third important point is that at this order in perturbation theory the mode $\zeta$ that perturbs $ T_{(\sigma)\mu\nu}(\vec x,t)$ is the free field that lives in the unperturbed metric. This implies that the initial $\zeta_k$ approaches a constant in time value as $k/(a(t)H)\rightarrow 0$ and that the induced metric approaches the unperturbed one, up to the simple rescaling of  the scale factor
\be
a(t)\rightarrow e^{\zeta_k}a(t)  \ .
\ee
The above points can be summarized by saying that we are left to compute how the expectation value of $\langle T_{(\sigma)\mu\nu}(\vec x,t)\rangle$ gets altered in the presence of a metric that  is just the unperturbed one with the simple rescaling of the scale factor, which is unobservable in the limit that it is spatially constant. We write $T_{(\sigma)\mu\nu}$ using the obervable quantities $\rho,\,p,\, v$ in the standard way as:
\be
\langle T^{\mu}_{(\sigma)\nu}\rangle_{\zeta_k}=\langle(\rho_\sigma+p_\sigma)u^\mu u_\nu+p_\sigma \delta^\mu_\nu \rangle_{\zeta_k}
\ee
where
\be
u^\mu=\left(\frac{1}{N_{\zeta_k}} ,\delta v^i_\sigma\right)
\ee
and
\bea
&&\langle\rho_\sigma\rangle=\langle\rho_{\sigma}(t)\rangle_{0}+\langle\delta\rho_{\sigma}\rangle_{\zeta_k}\ , \\ \nonumber
&&\langle p_\sigma\rangle=\langle p_{\sigma}(t)\rangle_0+\langle\delta p_{\sigma}\rangle_{\zeta_k}
\eea
where the subscript $\zeta$ represent that these are expectation values taken with respect to the $\sigma$ fields in the background of $\zeta$, while the subscript $_0$ represents that they are computed in an unperturbed metric. The fact that a super-horizon wavelength becomes locally unobservable implies that  for $k/\left(a(t) H\right) \ll1 $:
\bea\label{eq:assumption}
&&\langle\delta\rho_{\sigma}(t)\rangle_{\zeta_k}\sim \langle\delta p_{\sigma}(t)\rangle_{\zeta_k}\sim  {\cal{O}}\left(\langle\rho_\sigma(t)\rangle_0\frac{k}{a(t) H}\zeta_k(t)\right)\ , \\ \nonumber
&& \langle \delta v_{\sigma}^i(t)\rangle_{\zeta_k}\sim {\cal{O}}\left(\frac{k}{a(t) H}\zeta_k(t)\right)\ .
\eea
There is one subtlety in this argument that we would like to put into evidence: when we compute $\langle T_{(\sigma)\mu\nu}(\vec x,t)\rangle$, the insertion of the $\zeta_k$ perturbation as represented in Fig.~\ref{LeftLeft2} can be at such an early time $t_2\ll t^\star$ that the mode $\zeta_k$ is inside the horizon. In this case, the mode $\zeta$ {\it is} observable, and in principle  could imprint an effect on the $T_{(\sigma)\mu\nu}$. However, in this regime the modes whose frequency is higher than order $H$ at $t\sim t^\star$ (which are the modes contributing to $T_{(\sigma)\mu\nu}$) are much more energetic than the one of $\zeta$, and therefore they just follow their adiabatic interacting vacuum up to $t\sim t^\star$~\footnote{Notice that this same kind of reasoning applies to contributions to the $\zeta$ correlation function at times after inflation, for example during the period of reheating or during a GUT phase transitions (if this exists). These are epochs during which we know very little of what is going on. Still, the time scale associated to  the thermal fluctuations for the case of reheating, or to bubble collisions in a GUT phase transition, is expected to be much faster than the $\zeta_k$ mode when this is inside the horizon. Obviously, these fluctuations stay in their adiabatic vacuum at those times.}. 
A related subtlety in the above equation is also the fact that we are declaring that $\langle\delta\rho_{\sigma}(t)\rangle_{\zeta_k},\ \langle\delta p_{\sigma}(t)\rangle_{\zeta_k}\,$,  and $ \langle \delta v_{\sigma}^i(t)\rangle_{\zeta_k}$ depend on $\zeta$ evaluated at the same $t$. This is of course not true, and in general the relationship will be non-local in time but given that at late times $\zeta_k(t)$ approaches a constant and that the expectation value is dominated by this regime we can approximate the dependence as being $\zeta_k(t)$.

It is easy to see that eq.~(\ref{eq:assumption}) is enough for proving that the correlation function of $\zeta$ becomes time-independent out of the horizon.  In order to do this, we write again the continuity equation for the full stress tensor
\be
\langle\nabla_\mu T^{\mu0}\rangle_{\zeta_k}=0
\ee
which gives
\bea\label{eq:zeta-CIS}
&&\ddot\zeta+\left(3H-2\frac{\dot H}{H}+\frac{\ddot H}{\dot H}\right)\dot \zeta+\frac{k^2}{a^2}\zeta\\ \nonumber
&&+\frac{\dot H  {\langle\dot p_\sigma\rangle_0}-\ddot H (\langle \rho_\sigma\rangle_0+\langle p_\sigma\rangle_0)}{2\mpl^2 \dot H^2}\dot\zeta+\frac{ H}{2\mpl^2 \dot H}\left(\langle\delta\dot\rho_{\sigma}\rangle_{\zeta_k}+3H\left(\langle\delta\rho_{\sigma}\rangle_{\zeta_k}+\langle\delta p_{\sigma}\rangle_{\zeta_k}\right)\right)\\ \nonumber 
&&+ \frac{H}{2\mpl^2} (\langle \rho_\sigma\rangle_0+\langle p_\sigma\rangle_0) \langle \frac{\d_i}{a} \delta v_{\sigma}^i(t)\rangle_{\zeta_k} +\left(6H^2+\frac{H \ddot H}{\dot H}\right)\delta N_{\zeta\sigma^2}+H\chi_{\zeta\sigma^2}+H\dot{\delta N}_{\zeta\sigma^2}=0
\eea
where we have defined:
\bea
&& N=1+\delta N_\zeta+\delta N_{\sigma^2}++\delta N_{\zeta\sigma^2}\ ,\\
&&\chi=\d_i N^i=\chi_\zeta+\chi_{\sigma^2}+\chi_{\zeta\sigma^2}\ ,
\eea
with $\delta N_\zeta$ being proportional to $\zeta$,  $\delta N_{\sigma^2}$ to $\sigma^2$ and  $\delta N_{\zeta\sigma^2}$ to $\zeta \sigma^2$, and similarly for $\chi$. Notice that in eq.~(\ref{eq:zeta-CIS}), $\delta N_{\sigma^2}$ and $\chi_{\sigma^2}$ do not appear. This is so because the we are taking the expectation value with respect to the $\sigma$ field, and the tadpole cancellation ensures that those terms cancel.

As usual, we are interested in eq.~(\ref{eq:zeta-CIS})  when the mode $k$ is well outside the horizon. In this regime, we can find solutions for $\delta N_{\zeta\sigma^2}$ and $\chi_{\zeta\sigma^2}$ at leading order in $k/(a(t)H)\ll 1$. By solving at this order the $(0,0)$ and the $(0,i)$ Einstein equations we find:
\bea
&& \delta N_{\zeta\sigma^2}\simeq -\frac{1}{2\mpl^2\left(3H^2+\dot H\right)}\langle\delta\rho_{\sigma}\rangle_{\zeta_k}\, \\ \nonumber 
&&\chi_{\zeta\sigma^2}\simeq {\cal{O}}\left(\frac{k^2}{a(t)^2H^2}\zeta\right)\ ,
\eea
Eq.~(\ref{eq:assumption}) implies that $\delta N_{\zeta\sigma^2}$ is of order $k/(a(t) H)\zeta_k$ and $\chi_{\zeta\sigma^2}$ is of higher order. Therefore, by looking back at equation (\ref{eq:zeta-CIS}),  in the long wavelength limit, we can see that the sources for $\zeta$ at late time decay, which is enough to see that the correlation function of $\zeta$ will not depend on time and is scale invariant. In fact by taking the leading terms in $k/(a H)\ll1$ and in slow-roll parameters in eq.~(\ref{eq:zeta-CIS}), and using (\ref{eq:assumption}), we obtain: 
\bea
&&\zeta_k(t)\sim \int_{t^\star}^t dt' \frac{1}{H}\left(1-\frac{a(t')^3}{a(t)^3}\right) \frac{H^2}{\mpl^2\dot H}H^4\zeta_k(t') \frac{k}{a(t')H}\sim \frac{H^5}{\mpl^2\dot H} \zeta_k(t)\int_{t^\star}^t dt'  \frac{k}{a(t')H}\\ \nonumber 
&&\sim  \frac{H^5}{\mpl^2\dot H} \zeta_k(t) \frac{k}{H} \frac{1}{H}e^{-Ht^\star}\sim
\frac{H^2}{\mpl^2\epsilon} \zeta_k(t) \epsilon_{out}\ ,
\eea
Here for simplicity we have dropped higher order corrections in the slow roll parameters. The result can be trivially extended to include those as well. Corrections suppressed by higher derivatives in eq.~(\ref{eq:assumption}) and (\ref{eq:zeta-CIS}) are down by further powers of $\epsilon_{out}\ll1$. For the cut-in-the-side diagrams we therefore obtain: 
\be
\langle\zeta_{2,k}(t)\zeta_{2,k'}(t)\rangle_{CIS}\sim 
\delta^{(3)}(k+k') \frac{1}{k^3} \frac{H^4}{\epsilon^2 \mpl^4}N\epsilon_{out}\ ,
\ee
where we have reinserted the number $N$ of spectator fields. This is a time-independent and scale-invariant result. 

So far in this section we have neglected almost completely the non 1PI diagrams that we studied in sec.~\ref{sec:tadpole_cancellation}. The tadpole cancellation procedure applies also in this case  where we perform the momentum integration first and the time integration second. Therefore, they are automatically zero once the background has been redefined according to (\ref{eq:new_background}). All the $CIS$ diagrams  give a time-independent scale-invariant result.

Finally, we need to study the case where a $\zeta$ generated by the first term of the right-hand side of~(\ref{eq:generic2}), that we called $\zeta_1$, correlates with a term generated by the second term on the right-hand side of~(\ref{eq:generic2}), that we called $\zeta_2$. By using the standard property of correlation functions, we have
\be
|\langle\zeta_{k\, , 1}(t)\zeta_{k'\, , 2}(t)\rangle|\lesssim \left(\langle\zeta_{k\, , 1}(t)\zeta_{k'\, , 1}(t)\rangle\right)^{1/2}  \left(\langle\zeta_{k\, , 2}(t)\zeta_{k'\, , 2}(t)\rangle\right)^{1/2}\ .
\ee
Since each of the terms inside the square root is time-independent and scale-invariant, we conclude that so is for the correlation between $\zeta_1$ and $\zeta_2$. 

This last result completes the analysis of all the contributions to the correlation of $\zeta$ at late time and therefore we conclude that the result is scale-invariant and time-independent.  The procedure to compute the loops that we used in sec.~\ref{sec:sigma_fields} is justified in yet another way.\\

\subsection{Slow-roll corrections\label{sec:slowroll_corr}}

So far in the above calculation we have not included all the slow roll corrections, for example those ones that come from the wavefunction of the fields as the spacetime deviates from de-Sitter. It is rather simple to include the first corrections due to the slow roll parameters as follows.
In the former subsection \ref{sec:theorem}, we have seen that $\zeta_k$ at some late time is the result of three distinct contributions: the free evolution of $\zeta_k$ from some initial condition at time $t_\star$ (which is in practice just a constant), the contribution which comes from the variance of $T_{(\sigma)\mu\nu}$ (which is exponentially peaked at $t_\star$), and finally the contribution from the perturbation to the  expectation value of $T_{(\sigma)\mu\nu}$ due to a primordial $\zeta$ fluctuation (which is again exponentially peaked at $t_\star$). 

The last two of these contributions are exponentially sensitive only to the time $t_\star$, and so, in the case of some time dependence of $H,\;\dot H,\ldots$, the result will be the same as the one we obtained, just replacing $H,\;\dot H,\ldots$ with $H(t_\star),\;\dot H(t_\star),\ldots\ $. 

The same is true also for the first contribution if $t_\star$ is chosen to be close to the time of horizon crossing so that the variation of the parameters of the spacetime between $t_\star$ and the time of horizon crossing is negligible. This is clearly possible given the slow variation of the parameters with respect to the exponential stretching of the modes. We therefore conclude that with the replacement 
\be
H,\;\dot H,\ldots\ \ \rightarrow \ \ H(t_\star),\;\dot H(t_\star),\ldots\ ,
\ee
we can incorporate the leading slow roll corrections.

  \subsection{$\dot\pi^3$ large self-interactions}

The above proof of the constancy of $\zeta$ was done for the case of $N$ spectator massless scalar fields. The same result holds for the case of the  $\dot\pi^3$ interactions. Considering for example the cubic interactions, everything in the above proof proceeds in the same way, apart from trivial combinatoric factors, if, of the three $\pi$'s in the vertex $\dot\pi^3$, we consider one to be the inflaton and the other two to be two spectator fields. In fact, in the above proof, there were only two crucial points. The first is that for the $CIM$ diagrams,the correlation of $\dot\sigma^2$ and the similar operators was not scale-invariant, but rather Poisson like. There is no difference in this case (notice that because here we could  ignore metric fluctuations the proof would be even simpler). The second important point of the above proof was that, in the $CIS$ diagrams, $\langle\delta \rho_\sigma\rangle_{\zeta_k},\ \langle\delta p_\sigma\rangle_{\zeta_k},\, \ldots$ were all going to zero as $k/(a(t)H)\rightarrow 0$. Clearly this will be the same also in this case, as the free mode $\pi^{cl}_k$ becomes a constant out of the horizon and $\pi$ is only derivatively coupled. It is again correct to assume that $\langle\dot\pi^2\rangle_{\zeta_k},\ ,\, \ldots$ are all going to zero as $k/(a(t)H)\rightarrow 0$. We therefore conclude that the correlation function of $\zeta$ is constant also in this case, justifying also in this alternative way the calculation of sec.~\ref{sec:dotpicube}.  Slow roll corrections can be included as before. 

\section{Other interactions and higher loops\label{sec:massive}}

Finally  we comment on how our results are expected to generalize to higher loop calculations and for more generic interactions.

The two kind of theories we have studied resulted in logarithms of the form $\log(H/\mu)$, where $\mu$ is the renormalization scale. Current bounds of the non-Gaussianities of the CMB that limit the size of the $\dot\pi^3$ self-interactions \cite{Senatore:2009gt} make these corrections extremely small. The one loop effects  change with the $k$-mode only slowly due to the slight dependence of the Hubble scale on the horizon-crossing time of the $k$-mode.  These slow-roll suppressed effects are in general very small but  it is possible to imagine that if this were not to be the case, then a procedure similar to the renormalization group could be developed and implemented. This lies beyond the scope of the current paper.

One common characteristic of the two interactions we had studied was that the fields running in the loop appeared in the Lagrangian with at least  one derivative. What happens when this is not the case, for example if the $\sigma$ fields have a (positive) mass term? This case has been studied and solved by Weinberg in \cite{Weinberg:2006ac}. In his first paper \cite{Weinberg:2005vy}, he noticed that if the mass term was treated perturbatively as a two-line vertex, and one used in the loops the wavefunctions of a massless scalar field in de-Sitter space, then the integral in time in the loops would not converge. This interaction might then give rise to a time-depedence of the $\zeta$ correlation function. In a following paper \cite{Weinberg:2006ac}, however, he solved the puzzle by noticing that summing up all the perturbative mass insertions into the propagator (i.e. using the wavefunction of a massive field in de-Sitter) resulted in a  wavefunction that decayed exponentially at late times and the time integrals  in the loop converged again. The apparent time-dependence of the $\zeta$ correlation function disappeared.

We would like to see how this effect appears in the arguments we presented in the previous section.   For massive $\sigma$'s there is a term proportional to $m^2 \sigma^2$  in the stress energy tensor. Let us imagine treating  the mass term perturbatively and repeating the steps of sec.~\ref{sec:theorem}. We expect to find a time dependence in the $\zeta$ correlation function. 

In the cut-in-the-side  ($CIS$) diagrams of sec.~\ref{sec:CISdiagrams}, we would have to compute the perturbation to the expectation value of the $\sigma$ stress tensor $T_{(\sigma)\mu\nu}$ due to a background free $\zeta$ mode. It is still true also in this case that a constant $\zeta$ is unobservable and that therefore the perturbations to $\rho_\sigma$ and $p_\sigma$ would be proportional to the first derivative of $\zeta$. This is enough to make the source in the $CIS$ diagrams shut down at late time, and therefore there is no induced time dependence from the $CIS$ diagrams in the case of massive $\sigma$'s either.

The cut-in-the-middle ($CIM$) diagrams of sec.~\ref{sec:CIMdiagrams} are different. Those diagrams involved the correlation function of $T_{(\sigma)\mu\nu}$. In the massless case, the various terms of the stress tensor involved always derivatives acting on the $\sigma$ fields, and this was enough to ensure that the correlation functions of the various terms of the stress tensor became Poisson like for distances longer than the horizon. This implied that these fluctuations were not able to source a $\zeta$ mode once the wavelength was much outside of the horizon. In the massive case things are different. The part of the stress tensor that is proportional to $m^2$ does not involve derivatives of $\sigma$ which means that the correlation function of this part of the stress tensor {\it is not} Poisson but  scale invariant. This part of $T_{(\sigma)\mu\nu}$ has correlations for very long distances  and therefore it is able to source a $\zeta$ mode regardless of how much outside of the horizon that mode is. Physically this is quite easy to understand. Let us consider a $\zeta$ fluctuation of wavenumber $k$. At the time when this fluctuation crosses the horizon there will be a fluctuations of $T_{(\sigma)\mu\nu}$ on the same wavelength generated by two $\sigma$ fluctuations one of which has a wavelength of order $1/k$, while the other has a longer wavelength and crossed the horizon at an earlier time. At this point if we are using the massless $\sigma$ wavefunctions the $\sigma$'s freeze and remain constant. They remain up in their potential  wherever they happened to land at horizon crossing. This results in a region of additional potential energy which now begins to expand as the Universe expands, without never decaying very much like a space-dependent cosmological constant where the space dependence is given by the wavenumber $k$ of the $T_{(\sigma)\mu\nu}$ fluctuation. Since the additional potential energy never decays it is able to coherently source the mode $\zeta_k$ and therefore it induces a time-dependent effect. Notice that fluctuations of $T_{(\sigma)\mu\nu}$ on the same scale produced at a later time by shorter scale $\sigma$'s are instead suppressed, because they average out as in the massless case. It is just the fluctuations of the $\sigma$ modes of wavelength up to order $k$ that matter.

It is clear that as a result of the approximations used the above calculation does not capture what will actually happen.  Once the mode is outside of the horizon the $\sigma$ field cannot remain up its potential and will move down and reach the minimum. In fact it will do so either exponentially fast in time undergoing oscillations (if $m>3H/2$) or very slowly  (if $m<3H/2$). We just do not see this effect above because we have treated the mass term perturbatively. Since the out of the horizon solution for a massive field scales as~\cite{Weinberg:2006ac}:
\be
a(t)^{\lambda_{\pm}}\ ,
\ee
where 
\be
\lambda_{\pm}=-\frac{3}{2}\pm\sqrt{\frac{9}{4}-\frac{m^2}{H^2}}\ ,
\ee
the effect of the mass term is perturbatively of order $\frac{m^2}{H^2}\,H t$. If one is interested in times when this correction is important then the only way to properly take it into account is to re-sum  all the mass insertions and use the wavefunctions of a massive field (see fig.~\ref{fig:massive_propagator}).
If one were to do so, then, after some long time in the case of small $m$ one would see that the source for the $\zeta$ mode shuts down and $\zeta$ becomes constant. Notice that for small masses this happens after a time
\be
t_c\sim\frac{1}{H}\frac{m^2}{H^2}\ .
\ee
Wether this time scale is long  or not depends on the problem. For example, in standard inflation, this time scale might be longer than the time to reheating and in this case the time dependence of $\zeta$ will be relevant. Notice however that the overall effect goes down as the energy density, as $m^2$, and therefore it becomes smaller and smaller in the limit $m\rightarrow 0$. This time scale $t_c$ is clearly short  if one wishes to study time-dependence of correlation functions  in eternal inflation or even in de-Sitter (which would require changing gauge and defining some relevant observable \cite{Creminelli:2008es}). In this case one can conclude that the $\zeta$ correlation function becomes constant in time  a time of order  $t_c$ after horizon crossing.

\begin{figure}
\begin{center}
\includegraphics[width=12cm,]{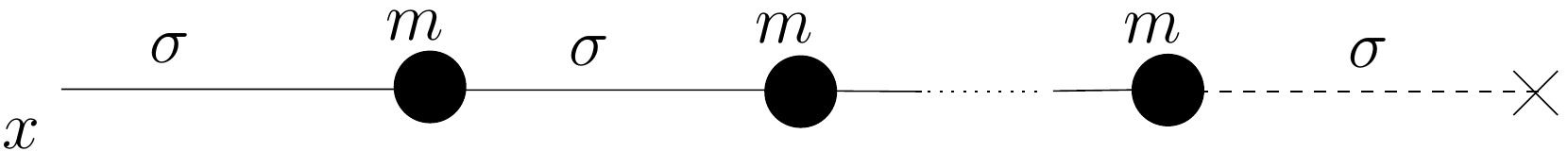}
\caption{\label{fig:massive_propagator} \small Mass insertions to obtain the massive propagator.}
\end{center}
\end{figure}

What happens if the $\sigma$ fields do not have a mass term, but for example a potential of the form $\lambda\sigma^4$ or even  higher dimensional potentials such as   $\sigma^6/\Lambda^2$, where the $\sigma=0$ point is a minimum? All these terms would enter if we were to do the calculation at more than one loop. Let us just concentrate on  $\lambda\sigma^4$, the other cases being just a trivial higher-loop generalization of this one. Clearly if one were to treat the $\lambda$ term perturbatively one could exactly reproduce the argument we just made for the massive case (extended this time to two loops) and find that the $CIM$ diagram produce a time dependence in the $\zeta$ correlation function. The solution is the same as in the massive case. In that case, we had to re-sum all the insertion of the mass term in the propagator (see fig.~\ref{fig:massive_propagator}), here we will have to re-sum all the insertion of the quartic interaction in the propagator (see fig.~\ref{fig:one_loop_propagator}). This will generate effectively a mass term for the $\sigma$ fields and will make the field decay and the induced time dependence of the $\zeta$ correlation function disappear eventually. As in the former case it is still true that the time at which a $\zeta_k$ mode becomes constant after horizon crossing might be very long. There is just one subtlety in the above argument. This is the fact that in principle a massive counter term could cancel the mass term induced by the quartic interaction (this would require a large amount of tuning, as we know from the Higgs particle in the Standard Model but still in principle it could happen) and therefore one would still have a massless field which would therefore induce a time dependence on $\zeta$ by jumping up its potential and staying there. We find it very hard to imagine that this cancellation is actually possibly in a quasi de-Sitter space as we expect that finite terms in the loop to inherit some time dependence from the time dependence of $H$ and the counterterm should not be able to reabsorb that. However even assuming that the mass can be tuned away there are other diagrams that are relevant and that make the field run down. In fact a particle in a quartic potential rolls down to the bottom even in the absence of any explicitly quadratic term. For example one could take the diagram in Fig.~\ref{fig:down_the_potential} where all $\sigma$'s in the dashed lines which represent primordial fluctuations get correlated with other primordial ones from the right-handed part of the CIM diagram~\footnote{This diagram can be thought of as representing the difference between $\sigma\sigma^2$ and $\sigma\langle\sigma^2\rangle$, where it is only the second term that can be potentially removed from the mass counterterm.}. This diagram will let the $\sigma$ field roll down its potential and therefore make $\zeta_k$ constant eventually though possibly a long time after it has crossed the horizon. 

Notice that similarly to the apparent time-dependence of $\zeta$ induced by these kinds of interactions, we expect there to be also IR divergencies associated with the momenta loop integrals. These have been found for example in \cite{literature,literature_time,Bartolo:2007ti,Riotto:2008mv, literature_time_inflaton} in the context of spectator scalar fields and of the standard slow rolling inflaton. If we consider for example the case of the massive spectator $\sigma$ fields we discussed at the beginning of the section, these IR divergencies are associated to the counting of all the $\sigma$ modes that crossed the horizon {\it before} the mode $k$ did. If one uses the massless wavefunctions, these modes that crossed the horizon arbitrarily in the past do not decay and all contribute to the correlation function, inducing an IR divergency.
As for the induced time-dependence of $\zeta$, we expect that the resummation of the diagrams we discussed will let all the spectator fields decay, effectively multiplying the wavefunction of the modes outside of the horizon by some power of $(k \eta)$, where $k$ is the wavenumber of the mode. In this way, the IR divergency is expected to disappear as well. Reference \cite{Riotto:2008mv}  offers an example, valid in some particular theories, of the resummation techniques of the diagrams and shows that upon resummation of the wavefunction  this particular kind of IR divergencies disappear. 


\begin{figure}
\begin{center}
\includegraphics[width=10.5cm,]{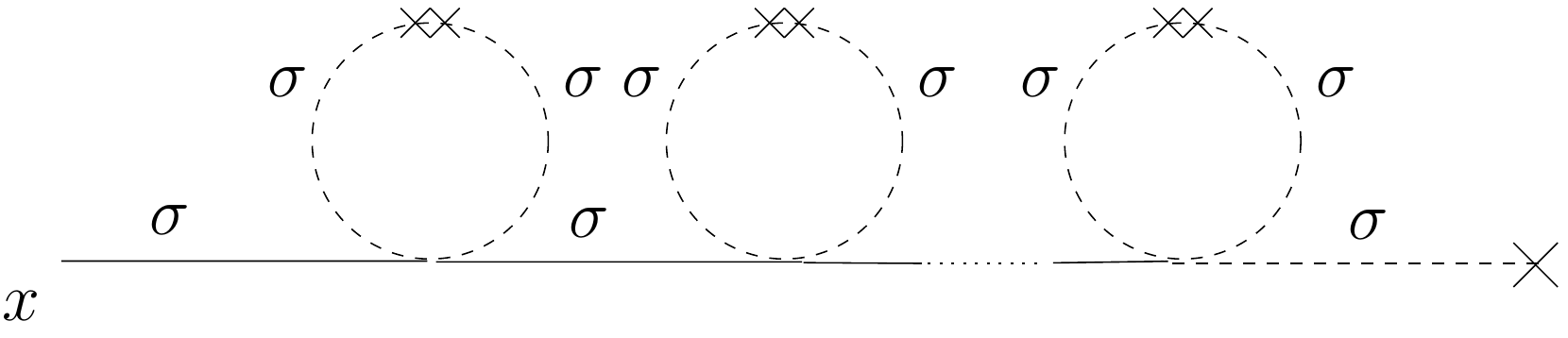}
\caption{\label{fig:one_loop_propagator} \small One-loop insertions to obtain the effectively massive propagator.}
\end{center}
\end{figure}

\begin{figure}
\begin{center}
\includegraphics[width=10cm,]{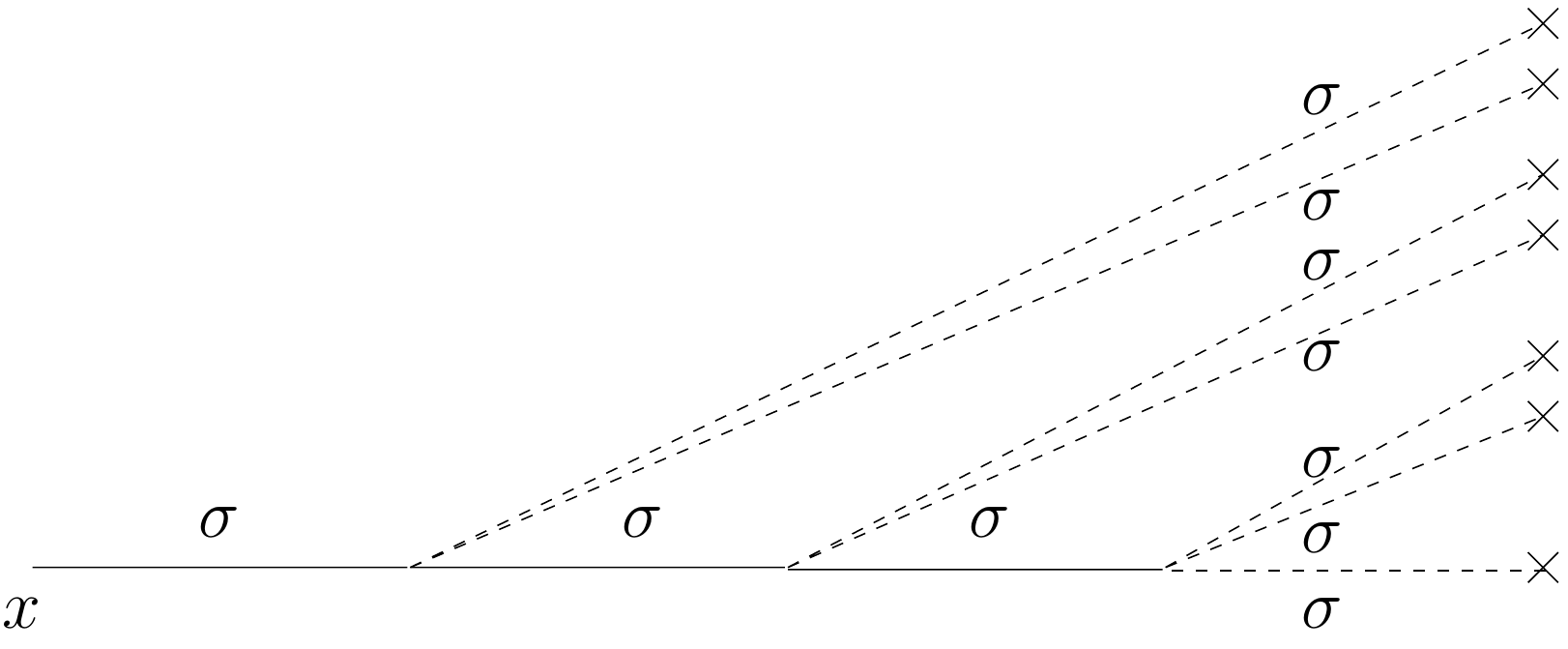}
\caption{\label{fig:down_the_potential} \small Left-handed part of one of the CIM diagram whose resummation will make $\sigma$ decay out of the horizon even in the absence of a mass term. Each of the $\sigma$'s in the dashed lines has to be correlated with a corresponding one on the right-handed side of the diagram.}
\end{center}
\end{figure}

There are other interactions that Weinberg pointed out would make the time integrals in the loop not converge and therefore apparently make $\zeta$ time-dependent~\footnote{ These are for example the ones that contain time derivatives of the fields, and that appear in the Lagrangian as multiplying in three dimensions a number of scale factors $a$ equal to one, after counting as minus two powers of $a$ every appearance of a time derivative \cite{Weinberg:2005vy}.}. Studying these interactions goes beyond the scope of the current paper. We just mention here that an example of a term of this kind is $\dot\zeta\sigma^n$ with $n$ being an integer. Such a term would for example arise when the $\sigma$ fields have a potential of the form $\sigma^n$. But because  the $\sigma$'s have a potential they would roll down to their minimum and would not induce a time dependence of $\zeta$ even after including the  $\dot\zeta\sigma^n$ vertex. We believe this is missed in Weinberg's theorem \cite{Weinberg:2005vy,Weinberg:2006ac} as in its proof every interaction is analyzed on its own  assuming that all fields are massless. In reality, it seems to us these kind of interactions are highly constrained by symmetries and come in groups. Therefore we think they should be studied together in specific setups.


Though we believe the above arguments to be very sound, strictly speaking we did not prove them. It would be interesting to directly verify them and also to explicitly find and study cases, if they exist in local theories, where the kind of interactions we discussed in the last paragraph appear and do induce a time dependent effect of the $\zeta$ correlation function even after the proper treatment of the mass term and of the related diagrams we discussed in Fig.s~\ref{fig:massive_propagator}, \ref{fig:one_loop_propagator}, and~\ref{fig:down_the_potential} has been taken into account.

\section{Conclusions\label{sec:conclusions}}

In this paper we have studied one-loop corrections to the two-point function of the curvature perturbation $\zeta$ in two kind of theories: one in which the inflation has large self-interactions, and one in which the inflaton interacts gravitationally with $N$ massless spectator scalar fields.

In both cases, we found that the one-loop corrections are time-independent and scale-invariant. Further, we have shown that there is a logarithm of the form 
\be
\langle\zeta_k^2\rangle_{\rm 1-loop}\sim\frac{1}{k^3}\times\beta\log\left(\frac{H}{\mu}\right)\ ,
\ee
where $H$ is the Hubble scale during inflation, $\mu$ is the renormalization scale, and $\beta$ can be thought of as being related to the $\beta$ function of the correlation function. This result is  physically sensible:  loop corrections for a weakly coupled theory are small if we choose the renormalization scale close to the energy scale of the process we are interested in. For inflation, this energy is Hubble.

Our results differ from the previous literature, which instead found that a logarithmic running of the form $\log(k/\mu)$. After arguing that symmetries forbid this result we were able to identify  the mistake in previous calculations. Dimensional regularization was applied incorrectly.  We also performed our calculation using a different regularization scheme finding the same result. We were also able to explicitly renormalize the simplest of the two theories we studied.

For the first time we studied the time dependence of the $\zeta$ correlation function finding that for the interactions we considered $\zeta$ becomes constant outside of the horizon even at one-loop level. This is the case once the background spacetime has been properly redefined in order to take into account tadople diagrams. These findings offer an alternative proof that the divergencies  {\it have} to be reabsorbed by local counterterms and are not instead infrared divergencies that appear as UV divergencies because of the particular way the calculation is performed. We achieved this not only by explicitly renormalizing the theory but also by performing the loop calculation at finite external time. We further developed an alternative method for computing loop corrections based on solving perturbatively the Heisemberg equations for the operators, which allowed us to work at every time directly with the renormalized physical quantities. In this approach the physical reason why $\zeta$ becomes constant outside of the horizon appears clearly. It is due to the combined effect of the fact that a constant $\zeta$ mode is unobservable and that the stress energy tensor which sources $\zeta$ becomes rapidly uncorrelated for scales longer than the horizon.

We are then able to use the intuition we developed to infer how our results generalize to different interactions and to higher loops. In particular, contrary to claims in the literature, we argue that no long term time dependence is generated by any form of spectator field, even if they have a potential term.

We find our result to be important not only because generally they increase our understanding of inflation and of de-Sitter space, but also specifically for two problems. The first is related to the fact that during the evolution of the Universe from the end of inflation up to now there are various epochs, such as reheating, where we do not know the details of the physical processes at play and thus we could imagine that there were very large fluctuations on the horizon scale. The theory of the spectators massless scalar fields can be thought of as a weakly coupled version of this, and it would have been worrisome if we had found a time-dependence on $\zeta$ on large scales. In fact, while the effect from the $\sigma$ fields would have been in any event small, the effect from fluctuations during reheating could have altered the value of $\zeta$ completely and the predictivity of inflation would have been lost. Our findings prove that this is not the case.

A second reason why we consider our findings relevant is connected to slow roll eternal inflation. If we trivially extend our finding to the inflaton two-point function at coincidence, the fact that the one loop corrections are time-independent and scale-invariant means that the smoothed two point function of the inflaton field at coincidence grows linearly with time. This means that the results of \cite{Creminelli:2008es,Dubovsky:2008rf} still apply after including the kind of loop corrections we computed, and that therefore, at least for what these interactions are concerned, standard slow roll inflation exists and there is a sharp phase transition as a function of the slope of the potential where eternal inflation begins.

We think our results motivate other interesting calculations. For example, it would be interesting to verify with an explicit calculation our arguments of sec.~\ref{sec:massive} about the fact that the resummation of higher loops makes the time dependence induced by spectator fields with a potential at a fixed perturbative order disappear, following and generalizing the calculation of \cite{Riotto:2008mv}, and understand the issue also in the case in which the inflaton and the graviton are let run in the loops. This would be particularly important for assessing the existence and the characterization of slow roll eternal inflation. 
Further, it would be interesting to study loop corrections and how the renormalization procedure works for the most generic theories of inflation. For single field inflation, this would amount to study the most general form of the Lagrangian presented in \cite{Cheung:2007st}, of which we studied only one particular limit in sec.~\ref{sec:dotpicube}.  There are also questions at a more fundamental level that our calculation might help answer. For example, in the dS/CFT correspondence~\cite{Strominger:2001pn,Witten:2001kn}, it is speculated that the time-dependence of correlation functions, which occurs at perturbative level in theories where spectators scalar fields have a potential term, should be mapped into renormalization group running for some operators in the CFT. It would be very interesting to see how this happens in a concrete way. Further, it is possible that the structure of the loop corrections that we found might be related, in string theory, to the density of single-string states, similarly to what happens in negatively curved spaces \cite{McGreevy:2006hk}. This is another direction worth exploring.

\bigskip
{\bf Acknowledgements:} We thank Sergei Dubovsky for intial collaboration in the project. 
We acknowledge David Gross for pointing out to us the possible relevance of  loop-corrections in eternal inflation. We thank Peter Adshead, Nima Arkani-Hamed, Paolo Creminelli, Thomas Dumitrescu, Richard Easther, Raphael Flauger, Zohar Komargodski, Eugene A. Lim, Juan Maldacena,  Alberto 
Nicolis, Steve Shenker, David Shih, Eva Silverstein, Filippo Vernizzi, Giovanni Villadoro and Steven Weinberg for many useful and stimulating comments.
This work was supported by NSF grants AST-0506556,  AST-0907969 and PHY-0855425, 
and fellowships from the David and Lucile Packard Foundation.

\appendix

\section{Loop integrals with momentum frequency cutoff \label{app:cutoff-integrals}}

Here we give some details of the calculation of the loop integrals that lead to eq.~(\ref{eq:one-loop-result-A}). The two momenta integrals can be done exploiting the identity:
\be
\int d^3k_1\;d^3k_2\;\delta^{(3)}(\vec k_1+\vec+k_2+\vec k)f(k_1,k_2,k)=\frac{2\pi}{k}\int_0^{\Lambda a(t_1)} dk_1\; k_1 \int_{|k_1-k|}^{k_1+k}dk_2\;k_2\, f(k_1,k_2,k).
\ee
The result of the two momenta integrations therefore splits into the sum of two terms: one in which we take the contributions of small $k_1$'s such that $k_1-k<0$, and one where we take the contribution from the large $k_1$'s such that $k_1-k>0$. The contribution from the large $k_1$'s gives:
\bea \label{eq:one-loop-integration-A}\nonumber
&&\!\!\!\!\!\!\!\!\!\!\!\!\langle\zeta_{\vec k}(t)\zeta_{\vec k'}(t)\rangle_{\rm{1-loop,\,A,\ t\rightarrow+\infty}}= (2\pi)^3 \delta^{(3)}(\vec k+\vec k')\frac{1}{k^3} \frac{c_3^2H^4M^8}{2\pi^2\dot H^4 \mpl^8} \int^0_{-\infty} d\eta_2\int^{\eta_2(1+\frac{H}{\Lambda})}_{-\infty} d\eta_1\;\frac{\eta_2^2}{\eta_1^2(\eta_1-\eta_2)^6} \\ \nonumber
&&\!\!\!\!\!\!\!\!\!\!\!\!\times \left\{H^4\eta_1^4\left[e^{2k\eta_1}\left(8-11 k(\eta_1-\eta_2)+7 k^2(\eta_1^2-\eta_2^2)\right)\right.\right.\\ \nonumber 
&&\!\!\!\!\!\!\!\!\!\!\!\!\qquad\quad\ \    \left.-\, 4 e^{2 k(2\eta_1-\eta_2)}\left(8-21 k(\eta_1-\eta_2)+27 k^2(\eta_1-\eta_2)^2-20k^3(\eta_1-\eta_2)^3+8 k^4(\eta_1-\eta_2)^4\right)\right]+\\ \nonumber
&&\!\!\!\!\!\!\!\!\!\!\!\!+e^{2\frac{\Lambda}{H}\left(\frac{\eta_2}{\eta_1}-1\right)}\left[e^{2 k\eta_1}\left(H^2\eta_1^2\left(H^2\eta_1^2+2\Lambda\, H\eta_1(\eta_1-\eta_2)+2\Lambda^2(\eta_1-\eta_2)^2\right)\left(8-5(\eta_1-\eta_2)+k^2\left(\eta_1-\eta_2\right)^2\right)\right)+
\right. \\ \nonumber
&&\!\!\!\!\!\!\!\!\!\!\!\!\left.\qquad\qquad\qquad\qquad\ \ +4\Lambda^3 H \eta_1(\eta_1-\eta_2)^3\left(2-k(\eta_1-\eta_2)\right)+2\Lambda^4(\eta_1-\eta_2)^4\right)+\\ \nonumber
&&\!\!\!\!\!\!\!\!\!\!\!\!\qquad\qquad\quad\   +e^{2 k\eta_2}\left(H^2\eta_1^2\left(H^2\eta_1^2+2\Lambda\, H\eta_1(\eta_1-\eta_2)+2\Lambda^2(\eta_1-\eta_2)^2\right)\left(8+5(\eta_1-\eta_2)+k^2\left(\eta_1-\eta_2\right)^2\right)\right)+\\ \nonumber
&&\!\!\!\!\!\!\!\!\!\!\!\! \left.\left.\left.\qquad\qquad\qquad\qquad\ \ +4\Lambda^3H\eta_1(\eta_1-\eta_2)^3\left(2+k(\eta_1-\eta_2)\right)+2\Lambda^4(\eta_1-
\eta_2)^4\right)\right]\right\}\ . \\
\eea
It is quite straightforward to realize that integrations of the terms in the third, fourth, fifth and sixth lines of (\ref{eq:one-loop-integration-A}) can not give rise to any logarithmic divergency. Because of the ${\rm Exp}\left(2\frac{\Lambda}{H}\left(\frac{\eta_2}{\eta_1}-1\right)\right)$ the integrand becomes exponentially small as soon as $|\eta_1-\eta_2|$ becomes larger than $\eta_2 H/\Lambda$, which is the minimum distance in conformal time allowed by our cutoff. This means that for these term the integral is peaked at the cutoff region, and it can not give rise to any logarithmic dependence.
Integrations of the terms in  the second and third line of (\ref{eq:one-loop-integration-A}) instead is supported on all scales, and a straightforward integration leads to our result in~(\ref{eq:one-loop-result-A}).

We do not reproduce here the contribution from the small $k_1$'s. It is quite straightforward and it gives rise only to finite terms.

\section{Additional correlation functions in the $CIM$ diagrams\label{app:CIMcorrelations}}

Here we complete the calculation of the correlation functions in (\ref{eq:cut-in-the-middle}) for the $CIM$ diagrams. It is easy to see that the contribution to  the $\zeta$ correlation from $\langle(\dot\sigma^2)_k(t)\left((\d_i\sigma)^2\right)_{k'}(t')\rangle$ and from $\langle\left((\d_i\sigma)^2\right)_k(t)\left((\d_j\sigma)^2\right)_{k'}(t')\rangle$ proceeds in very similar terms to the term studied in (\ref{eq:cut-middle-temp}), and results in a contribution parametrically equal. To analize the contribution from $\langle\left(\frac{1}{\d^2}\d^i\left(\dot\sigma\d_i\sigma\right)\right)_k(t)\left(\frac{1}{\d^2}\d^i\left(\dot\sigma\d_i\sigma\right)\right)_{k'}(t')\rangle$ we have to study
\bea\nonumber
&&\langle\left(\frac{1}{\d^2}\d^i\left(\dot\sigma\d_i\sigma\right)\right)_{k\simeq0}(t)\left(\frac{1}{\d^2}\d^i\left(\dot\sigma\d_i\sigma\right)\right)_{k'\simeq0}(t')\rangle\sim\delta^{(3)}(\vec k+\vec k')\frac{k^ik^j}{k^4}\int d^3x_-\; \langle\left(\dot\sigma\d_i\sigma\right)(\vec x_-,t)\left(\dot\sigma\d_j\sigma\right)(0,t')\rangle\\ \nonumber 
&&\lesssim \delta^{(3)}(\vec k+\vec k')\left(\frac{k^i k^l}{k^2} \int d^3x_-\; \langle\left(\dot\sigma\d_i\sigma\right)(\vec x_-,t)\left(\dot\sigma\d_l\sigma\right)(0,t)\rangle\right)^{1/2}\times\\ \nonumber
&&\qquad\times\left(\frac{k^j k^m}{k^2} \int d^3x_-\; \langle\left(\dot\sigma\d_j\sigma\right)(\vec x_-,t')\left(\dot\sigma\d_m\sigma\right)(0,t')\rangle\right)^{1/2}\\ \nonumber
&& \sim  \delta^{(3)}(\vec k+\vec k')\left(\frac{k^i k^l}{k^2} \int d^3x_-\; \langle\dot\sigma(\vec x_-,t)\dot\sigma(0,t)\rangle\langle\d_i\sigma(\vec x_-,t)\d_l\sigma(0,t)\rangle\right)^{1/2}\times \\ \nonumber 
&&\qquad\times\left(\frac{k^j k^m}{k^2}\int d^3x_-\; \langle\dot\sigma(\vec x_-,t')\dot\sigma(0,t')\rangle\langle\d_j\sigma(\vec x_-,t')\d_m\sigma(0,t')\rangle\right)^{1/2}\\
&&\sim \frac{1}{k^2}\frac{1}{ a(t)^{1/2}a(t')^{1/2}} c_{UV} H^5\ ,
\eea
where we have used (\ref{eq:approximate}) and in particular that $\langle\dot\sigma(\vec x_-,t)\d_i\sigma(0,t)\rangle=0$. Plugging back in the expression for $\langle\zeta\zeta\rangle_{CIM}$, and proceeding as in the former example, we obtain that this term contributes in the following way
\bea\nonumber
&&\langle\zeta_{2,k}(t)\zeta_{2,k'}(t)\rangle_{CIM}\lesssim 
\delta^{(3)}(k+k')\frac{c_{UV} H^5}{\mpl^4}\frac{1}{k^2}\int_{t^\star}^t dt_1\int_{t^\star}^t dt_2  \frac{1}{a(t)^{1/2}a(t')^{1/2}}\\ \nonumber
&&\sim  \delta^{(3)}(k+k')\frac{c_{UV} H^3}{\mpl^4}\frac{1}{k^2} e^{-H t^\star} \sim  \delta^{(3)}(k+k')\frac{H^4}{\mpl^4}\frac{1}{k^3} c_{UV}\,\epsilon_{out}\ .
\eea
We find that this contribution is also time independent and scale invariant.  Proceeding analogously with the correlation from $\langle\left(\dot\sigma^2\right)_k(t)\left(\frac{1}{\d^2}\d^i\left(\dot\sigma\d_i\sigma\right)\right)_{k'}(t')\rangle$ we find that
\bea
&&\langle\left(\dot\sigma^2\right)_{k\simeq0}(t)\left(\frac{1}{\d^2}\d^i\left(\dot\sigma\d_i\sigma\right)\right)_{k'\simeq0}(t')\rangle\lesssim\delta^{(3)}(\vec k+\vec k') \frac{1}{k}\frac{1}{ a(t)^{3/2}a(t')^{1/2}} c_{UV} H^5\ ,
\eea
which leads to 
\bea
&&\langle\zeta_{2,k}(t)\zeta_{2,k'}(t)\rangle_{CIM}\lesssim 
\delta^{(3)}(k+k')\frac{c_{UV} H^4}{\mpl^4}\frac{1}{k}\int_{t^\star}^t dt_1\int_{t^\star}^t dt_2  \frac{1}{a(t)^{3/2}a(t')^{1/2}}\\ \nonumber
&& \sim  \delta^{(3)}(k+k')\frac{H^4}{\mpl^4}\frac{1}{k^3} c_{UV}\,\epsilon_{out}^2\ .
\eea
which is again time-independent and scale-invariant.
It is straightforward to see that the contribution from $\langle\left(\left(\d_i\sigma\right)^2\right)_k(t)\left(\frac{1}{\d^2}\d^i\left(\dot\sigma\d_i\sigma\right)\right)_{k'}(t')\rangle$ gives a parametrically equal contribution. This last one exhausted all contributions to the late time correlation of $\zeta$ from the $CIM$ diagrams, and we therefore conclude that this contribution is time-independent and scale-invariant.


\begin{thebibliography}{nn}


\bibitem{Maldacena:2002vr}
  J.~M.~Maldacena,
  ``Non-Gaussian features of primordial fluctuations in single field
  inflationary models,''
  JHEP {\bf 0305} (2003) 013
  [arXiv:astro-ph/0210603].


\bibitem{Weinberg:2005vy}
  S.~Weinberg,
  ``Quantum contributions to cosmological correlations,''
  Phys.\ Rev.\  D {\bf 72} (2005) 043514
  [arXiv:hep-th/0506236].


\bibitem{Cheung:2007st}
  C.~Cheung, P.~Creminelli, A.~L.~Fitzpatrick, J.~Kaplan and L.~Senatore,
  ``The Effective Field Theory of Inflation,''
  JHEP {\bf 0803} (2008) 014
  [arXiv:0709.0293 [hep-th]].


\bibitem{Alishahiha:2004eh}
  M.~Alishahiha, E.~Silverstein and D.~Tong,
  ``DBI in the sky,''
  Phys.\ Rev.\  D {\bf 70} (2004) 123505
  [arXiv:hep-th/0404084].


\bibitem{ArkaniHamed:2003uz}
  N.~Arkani-Hamed, P.~Creminelli, S.~Mukohyama and M.~Zaldarriaga,
  ``Ghost Inflation,''
  JCAP {\bf 0404} (2004) 001
  [arXiv:hep-th/0312100].


\bibitem{Senatore:2004rj}
  L.~Senatore,
  ``Tilted ghost inflation,''
  Phys.\ Rev.\  D {\bf 71} (2005) 043512
  [arXiv:astro-ph/0406187].


\bibitem{Chen:2006nt}
  X.~Chen, M.~x.~Huang, S.~Kachru and G.~Shiu,
  ``Observational signatures and non-Gaussianities of general single field
  inflation,''
  JCAP {\bf 0701} (2007) 002
  [arXiv:hep-th/0605045].


\bibitem{Flauger:2009ab}
  R.~Flauger, L.~McAllister, E.~Pajer, A.~Westphal and G.~Xu,
  ``Oscillations in the CMB from Axion Monodromy Inflation,''
  arXiv:0907.2916 [hep-th].


\bibitem{Zaldarriaga:2003my}
  M.~Zaldarriaga,
  ``Non-Gaussianities in models with a varying inflaton decay rate,''
  Phys.\ Rev.\  D {\bf 69} (2004) 043508
  [arXiv:astro-ph/0306006].


\bibitem{Lyth:2002my}
  D.~H.~Lyth, C.~Ungarelli and D.~Wands,
  ``The primordial density perturbation in the curvaton scenario,''
  Phys.\ Rev.\  D {\bf 67} (2003) 023503
  [arXiv:astro-ph/0208055].


\bibitem{Green:2009ds}
  D.~Green, B.~Horn, L.~Senatore and E.~Silverstein,
  ``Trapped Inflation,''
  arXiv:0902.1006 [hep-th].

\bibitem{Barnaby:2009mc}
  N.~Barnaby, Z.~Huang, L.~Kofman and D.~Pogosyan,
  ``Cosmological Fluctuations from Infra-Red Cascading During Inflation,''
  Phys.\ Rev.\  D {\bf 80} (2009) 043501
  [arXiv:0902.0615 [hep-th]].








\bibitem{Smith:2009jr}
  K.~M.~Smith, L.~Senatore and M.~Zaldarriaga,
  ``Optimal limits on $f_{\rm NL}^{\rm local}$ from WMAP 5-year data,''
  JCAP {\bf 0909} (2009) 006
  [arXiv:0901.2572 [astro-ph]].

\bibitem{Senatore:2009gt}
  L.~Senatore, K.~M.~Smith and M.~Zaldarriaga,
  ``Non-Gaussianities in Single Field Inflation and their Optimal Limits from
  the WMAP 5-year Data,''
  arXiv:0905.3746 [astro-ph.CO].


\bibitem{Slosar:2008hx}
  A.~Slosar, C.~Hirata, U.~Seljak, S.~Ho and N.~Padmanabhan,
  ``Constraints on local primordial non-Gaussianity from large scale
  structure,''
  JCAP {\bf 0808} (2008) 031
  [arXiv:0805.3580 [astro-ph]].


\bibitem{eternal}
  A.~H.~Guth,
  ``The Inflationary Universe: A Possible Solution To The Horizon And Flatness
  Problems,''
  Phys.\ Rev.\  D {\bf 23} (1981) 347;
A.~D.~Linde, ``Nonsingular Regenerating Inflationary Universe," Cambridge University preprint Print-82-0554 (1982);
P.~J.~Steinhardt, ÒNatural Inflation,Ó in {\it The Very Early Universe}, ed. G.W. Gibbons, S.W. Hawking and S. Siklos, Cambridge University Press, (1983);
  A.~Vilenkin,
  ``The Birth Of Inflationary Universes,''
  Phys.\ Rev.\  D {\bf 27} (1983) 2848;
  A.~S.~Goncharov, A.~D.~Linde and V.~F.~Mukhanov,
  ``The Global Structure Of The Inflationary Universe,''
  Int.\ J.\ Mod.\ Phys.\  A {\bf 2} (1987) 561.


\bibitem{Weinberg:1987dv}
  S.~Weinberg,
  ``Anthropic Bound on the Cosmological Constant,''
  Phys.\ Rev.\ Lett.\  {\bf 59} (1987) 2607.



\bibitem{Creminelli:2008es}
  P.~Creminelli, S.~Dubovsky, A.~Nicolis, L.~Senatore and M.~Zaldarriaga,
  ``The Phase Transition to Slow-roll Eternal Inflation,''
  JHEP {\bf 0809} (2008) 036
  [arXiv:0802.1067 [hep-th]].



\bibitem{Dubovsky:2008rf}
  S.~Dubovsky, L.~Senatore and G.~Villadoro,
  ``The Volume of the Universe after Inflation and de Sitter Entropy,''
  JHEP {\bf 0904}, 118 (2009)
  [arXiv:0812.2246 [hep-th]].


\bibitem{ArkaniHamed:2007ky}
  N.~Arkani-Hamed, S.~Dubovsky, A.~Nicolis, E.~Trincherini and G.~Villadoro,
  ``A Measure of de Sitter Entropy and Eternal Inflation,''
  JHEP {\bf 0705} (2007) 055
  [arXiv:0704.1814 [hep-th]].










\bibitem{Adshead:2008gk}
  P.~Adshead, R.~Easther and E.~A.~Lim,
  ``Cosmology With Many Light Scalar Fields: Stochastic Inflation and Loop
  Corrections,''
  Phys.\ Rev.\  D {\bf 79} (2009) 063504
  [arXiv:0809.4008 [hep-th]].







\bibitem{Weinberg:2006ac}
  S.~Weinberg,
  ``Quantum contributions to cosmological correlations. II: Can these
  corrections become large?,''
  Phys.\ Rev.\  D {\bf 74} (2006) 023508
  [arXiv:hep-th/0605244].



\bibitem{Cheung:2007sv}
  C.~Cheung, A.~L.~Fitzpatrick, J.~Kaplan and L.~Senatore,
  ``On the consistency relation of the 3-point function in single field
  inflation,''
  JCAP {\bf 0802} (2008) 021
  [arXiv:0709.0295 [hep-th]].



\bibitem{literature}
  K.~Chaicherdsakul,
  ``Quantum cosmological correlations in an inflating universe: Can fermion and
  gauge fields loops give a scale free spectrum?,''
  Phys.\ Rev.\  D {\bf 75} (2007) 063522
  [arXiv:hep-th/0611352];
  D.~Seery,
  ``One-loop corrections to a scalar field during inflation,''
  JCAP {\bf 0711} (2007) 025
  [arXiv:0707.3377 [astro-ph]];
  E.~Dimastrogiovanni and N.~Bartolo,
  ``One-loop graviton corrections to the curvature perturbation from
  inflation,''
  JCAP {\bf 0811} (2008) 016
  [arXiv:0807.2790 [astro-ph]].
  P.~Adshead, R.~Easther and E.~A.~Lim,
  ``The 'in-in' Formalism and Cosmological Perturbations,''
  arXiv:0904.4207 [hep-th];
  X.~Gao and F.~Xu,
  ``Loop Corrections to Cosmological Perturbations in Multi-field Inflationary
  Models: I. Entropy Loops,''
  JCAP {\bf 0907} (2009) 042
  [arXiv:0905.0405 [hep-th]];
  D.~Campo,
  ``Quantum corrections during inflation and conservation of adiabatic
  perturbations,''
  arXiv:0908.3642 [hep-th].

  
\bibitem{GellMann:1954fq}
  M.~Gell-Mann and F.~E.~Low,
  ``Quantum electrodynamics at small distances,''
  Phys.\ Rev.\  {\bf 95} (1954) 1300.


\bibitem{Wilson:1974mb}
  K.~G.~Wilson,
  ``The Renormalization Group: Critical Phenomena And The Kondo Problem,''
  Rev.\ Mod.\ Phys.\  {\bf 47} (1975) 773.


  

\bibitem{literature_time}
 Our results might affect for example: 
  M.~van der Meulen and J.~Smit,
  ``Classical approximation to quantum cosmological correlations,''
  JCAP {\bf 0711}, 023 (2007)
  [arXiv:0707.0842 [hep-th]].
  They might have influence also on the literature on the zero mode of the perturbations:
  V.~K.~Onemli and R.~P.~Woodard,
  ``Super-acceleration from massless, minimally coupled $\phi^4$,''
  Class.\ Quant.\ Grav.\  {\bf 19}, 4607 (2002)
  [arXiv:gr-qc/0204065];
  V.~K.~Onemli and R.~P.~Woodard,
  ``Quantum effects can render $w < -1$ on cosmological scales,''
  Phys.\ Rev.\  D {\bf 70} (2004) 107301
  [arXiv:gr-qc/0406098];
  E.~O.~Kahya and V.~K.~Onemli,
  ``Quantum Stability of a $w < - 1$ Phase of Cosmic Acceleration,''
  Phys.\ Rev.\  D {\bf 76}, 043512 (2007)
  [arXiv:gr-qc/0612026];
  A.~M.~Polyakov,
  ``De Sitter Space and Eternity,''
  Nucl.\ Phys.\  B {\bf 797} (2008) 199
  [arXiv:0709.2899 [hep-th]].

  

\bibitem{literature_time_different}
  S.~P.~Miao and R.~P.~Woodard,
  ``The fermion self-energy during inflation,''
  Class.\ Quant.\ Grav.\  {\bf 23} (2006) 1721
  [arXiv:gr-qc/0511140], 
  S.~P.~Miao and R.~P.~Woodard,
  ``Leading log solution for inflationary Yukawa,''
  Phys.\ Rev.\  D {\bf 74} (2006) 044019
  [arXiv:gr-qc/0602110].



  
  




\bibitem{Musso:2006pt}
  M.~Musso,
  ``A new diagrammatic representation for correlation functions in the in-in
  formalism,''
  arXiv:hep-th/0611258.
  
\bibitem{Bartolo:2007ti}
  N.~Bartolo, S.~Matarrese, M.~Pietroni, A.~Riotto and D.~Seery,
  ``On the Physical Significance of Infra-red Corrections to Inflationary
  Observables,''
  JCAP {\bf 0801} (2008) 015
  [arXiv:0711.4263 [astro-ph]].

  
\bibitem{Riotto:2008mv}
  A.~Riotto and M.~S.~Sloth,
  ``On Resumming Inflationary Perturbations beyond One-loop,''
  JCAP {\bf 0804}, 030 (2008)
  [arXiv:0801.1845 [hep-ph]].


\bibitem{literature_time_inflaton}
    M.~S.~Sloth,
  ``On the one loop corrections to inflation and the CMB anisotropies,''
  Nucl.\ Phys.\  B {\bf 748} (2006) 149
  [arXiv:astro-ph/0604488];
  M.~S.~Sloth,
  ``On the one loop corrections to inflation. II: The consistency relation,''
  Nucl.\ Phys.\  B {\bf 775} (2007) 78
  [arXiv:hep-th/0612138];
  D.~Seery,
  ``One-loop corrections to the curvature perturbation from inflation,''
  JCAP {\bf 0802} (2008) 006
  [arXiv:0707.3378 [astro-ph]];
  D.~Seery,
  ``A parton picture of de Sitter space during slow-roll inflation,''
  JCAP {\bf 0905} (2009) 021
  [arXiv:0903.2788 [astro-ph.CO]];
  Y.~Urakawa and K.~i.~Maeda,
  ``One-loop Corrections to Scalar and Tensor Perturbations during Inflation in
  Stochastic Gravity,''
  Phys.\ Rev.\  D {\bf 78} (2008) 064004
  [arXiv:0801.0126 [hep-th]].


\bibitem{Strominger:2001pn}
  A.~Strominger,
  ``The dS/CFT correspondence,''
  JHEP {\bf 0110} (2001) 034
  [arXiv:hep-th/0106113].


\bibitem{Witten:2001kn}
  E.~Witten,
  ``Quantum gravity in de Sitter space,''
  arXiv:hep-th/0106109.

\bibitem{McGreevy:2006hk}
  J.~McGreevy, E.~Silverstein and D.~Starr,
  ``New dimensions for wound strings: The modular transformation of geometry to
  topology,''
  Phys.\ Rev.\  D {\bf 75} (2007) 044025
  [arXiv:hep-th/0612121];
  E.~Silverstein,
  ``Dimensional mutation and spacelike singularities,''
  Phys.\ Rev.\  D {\bf 73} (2006) 086004
  [arXiv:hep-th/0510044].


  
  \end{thebibliography}
\end{document}